\documentclass[english]{iopart}

\usepackage{graphicx}
\newcounter{fig}
\newcommand{\chit}{\tilde{\chi}}

\begin{document}

\title[Extreme series]{\Large
Experimental mathematics on the magnetic susceptibility of
the square lattice Ising model}

\author{ 
S. Boukraa$^\dag$, A. J. Guttmann$^\ddag$, S. Hassani$^\S$, I. Jensen$^\ddag$,
J.-M. Maillard$^{||}$, B. Nickel$^+$ and N. Zenine$^\S$}
\address{\dag LPTHIRM and D\'epartement d'A{\'e}ronautique,
 Universit\'e de Blida, Algeria}
\address{$\ddag$ ARC Centre of Excellence for 
Mathematics and Statistics of Complex Systems 
Department of Mathematics and Statistics,
The University of Melbourne, Victoria 3010, Australia}
\address{\S  Centre de Recherche Nucl\'eaire d'Alger, 
2 Bd. Frantz Fanon, BP 399, 16000 Alger, Algeria}
\address{$||$ LPTMC, Universit\'e de Paris, Tour 24,
 4\`eme \'etage, case 121, 
 4 Place Jussieu, 75252 Paris Cedex 05, France} 
\address{$+$ Department of Physics, University of 
Guelph, Guelph, Ontario N1G 2W1, Canada}
\ead{maillard@lptmc.jussieu.fr, maillard@lptl.jussieu.fr, tonyg@ms.unimelb.edu.au, 
 I.Jensen@ms.unimelb.edu.au, njzenine@yahoo.com, boukraa@mail.univ-blida.dz}

\begin{abstract}
We calculate very long low- and high-temperature series for the susceptibility 
$\chi$ of the square lattice Ising model as well as very long series for
the five-particle contribution $\, \chi^{(5)}$ and six-particle contribution 
$\, \chi^{(6)}$. These calculations have been made possible by the use of highly 
optimized polynomial time modular algorithms and a total of more than 150000 CPU 
hours on computer clusters. The series for $\chi$ (low- and high-temperature regime), 
$\, \chi^{(5)}$ and $\, \chi^{(6)}$ are now extended to 2000 terms.
In addition, for $\chi^{(5)}$,  10000 terms of the series are calculated {\it modulo} 
a single prime, and have been used to  find the linear ODE satisfied by $\chi^{(5)}$ 
{\it modulo} a prime.

A diff-Pad\'e analysis of the 2000 terms series for $\,\chi^{(5)}$ and $\chi^{(6)}$ 
confirms to a very high degree of confidence previous conjectures 
about the location and strength of the singularities 
of the $n$-particle components of the susceptibility, up to a small set of ``additional'' 
singularities. The exponents at {\em all} the singularities of the Fuchsian linear ODE 
of  $\chi^{(5)}$ and the (as yet unknown) ODE of $\chi^{(6)}$ are given: they are 
{\em all} rational numbers. We find the presence of  singularities at $w=1/2$ 
for the linear ODE of  $\chi^{(5)}$, and $w^2= 1/8$ for the ODE of  $\chi^{(6)}$, which 
are {\it not} singularities of the ``physical'' $\chi^{(5)}$ and $\chi^{(6)},$ that is 
to say the series-solutions of the ODE's which are analytic at $\, w\, = \, \, 0$.

Furthermore, analysis of the long series for $\,\chi^{(5)}$ (and $\,\chi^{(6)}$) combined with the 
corresponding long series for the full susceptibility $\chi$ yields
previously conjectured singularities in some $\chi^{(n)}$, $\,n\, \ge\, 7$. The exponents 
at all these singularities are also seen to be rational numbers.

We also present a mechanism of resummation of the logarithmic singularities of the 
$\,\chi^{(n)}$ leading to the known power-law critical behaviour occurring in the full 
$\chi$, and perform a power spectrum analysis giving strong arguments in favor 
of the existence of a natural boundary for the full susceptibility $\chi$.

\end{abstract} 

\vskip .5cm


\noindent {\bf AMS Classification scheme numbers}: 34M55, 
47E05, 81Qxx, 32G34, 34Lxx, 34Mxx, 14Kxx 


\section{Introduction \label{pre}} 

The magnetic susceptibility of the two-dimensional Ising model has attracted 
the attention of many of the mathematicians and physicists who work (or worked) 
in the area of equilibrium statistical mechanics for more than 60 years, ever 
since Onsager's celebrated solution of the free-energy of the two-dimensional 
Ising model~\cite{Onsager}.

Let $\, \sigma_{i,j}$ be the spin at lattice site $(i,j)$. The two-point 
correlation function is defined as 
\begin{eqnarray}
\label{CMNdef}
C(M,N)\,=\,\, \langle \sigma_{0,0} \sigma_{M,N} \rangle,
\end{eqnarray}
and the magnetic susceptibility is given by
\begin{eqnarray}
\label{chidef}
kT \cdot \chi \, = \,\,\,
\sum_{M} \sum_{N} \bigr( C(M,N)\, - {\mathcal M}^2 \bigl),
\end{eqnarray}
where the magnetisation ${\mathcal M}$ is zero for $T\,  >\,  T_c$ and, 
in the isotropic case, ${\mathcal M} = (1-s^{-4})^{1/8}$  for $T\,  <\,  T_c$ (in the scaling 
limit $\, M, \, N\, \rightarrow \, \infty$), where $s=\, \sinh(2J/kT)$ .
 
To this day, while we still lack a closed-form  solution for the susceptibility, 
we do have a great deal of associated knowledge. While there have been many 
breakthroughs in the study of this problem, we would like to mention five in particular.

Firstly, in 1976, Wu, McCoy, Tracy and Barouch~\cite{wu-mc-tr-ba-76} showed 
that the susceptibility can be expressed as an infinite sum of contributions,
known as {\em $n$-particle contributions}, so that the high-temperature 
susceptibility is given by
\begin{eqnarray}
\label{plus}
kT \cdot \chi_H(w) \, \, =  \, \, \, \sum \chi^{(2n+1)}(w) \,\, = \, \,\,
{{1} \over {s}} \cdot (1 - s^4)^{\frac{1}{4}} \cdot \sum \tilde{\chi}^{(2n+1)}(w)
\end{eqnarray}
and the low-temperature susceptibility is given by
\begin{eqnarray}
\label{chimoins}
kT \cdot \chi_L(w)\, \,  = \,\,\, \sum \chi^{(2n)}(w) \,\, = 
 \, \,\, (1 - 1/s^4)^{\frac{1}{4}} \cdot \sum \tilde{\chi}^{(2n)}(w)
\end{eqnarray}
in terms of the self-dual temperature variable $w=\frac{1}{2}s/(1+s^2).$  

As is now well known~\cite{wu-mc-tr-ba-76}, the $n$-particle contributions
have an integral representation and are given by the $(n-1)$-dimensional 
integrals~\cite{nickel-99,nickel-00,pal-tra-81,yamada-84},
\begin{eqnarray}
\label{chi3tild}
\tilde{\chi}^{(n)}(w)\,\,=\,\,\,\, {\frac{1}{n!}}  \cdot 
\Bigl( \prod_{j=1}^{n-1}\int_0^{2\pi} {\frac{d\phi_j}{2\pi}} \Bigr)  
\Bigl( \prod_{j=1}^{n} y_j \Bigr)  \cdot   R^{(n)} \cdot
\,\, \Bigl( G^{(n)} \Bigr)^2, 
\end{eqnarray}
where\footnote[3]{The Fermionic term $\,G^{(n)}$ has several 
representations~\cite{nickel-00}.} 
\begin{eqnarray}
\label{Gn}
G^{(n)}\,=\,\, \prod_{1\; \le\; i\;<\;j\;\le \;n} \, h_{ij}, \,  \quad 
h_{ij}\,=\,\,
{\frac{2\sin{((\phi_i-\phi_j)/2) \cdot \sqrt{x_i \, x_j}}}{1-x_ix_j}}, 
\end{eqnarray}
and
\begin{eqnarray}
\label{Rn}
R^{(n)} \, = \,\,\, \,  {\frac{1\,+\prod_{i=1}^{n}\, 
x_i}{1\,-\prod_{i=1}^{n}\, x_i}}, 
\end{eqnarray}
with
\begin{eqnarray}
\label{thex}
&&x_{i}\, =\,\,\,  \,  \frac{2w}{1-2w\cos (\phi _{i})\, 
+\sqrt{\left( 1-2w\cos (\phi_{i})\right)^{2}-4w^{2}}},  
\\
\label{they}
&&y_{i} \, = \, \, \,
\frac{2w}{\sqrt{\left(1\, -2 w\cos (\phi _{i})\right)^{2}\, -4w^{2}}}, 
\quad \quad  \quad  \quad \sum_{j=1}^n \phi_j=\, 0  
\end{eqnarray}
valid for small $w$ and, elsewhere, by analytic continuation. The variable $w$ 
corresponds to small values of $s$ {\em as well as to} large values of $s$.
In discussing analytic properties of $\chit^{(n)}$ we will often refer to the 
principal disc by which we mean $| s| \leq 1$ for $\chit^{(2n+1)}$ and 
$| s| \geq 1$ for $\chit^{(2n)}$. For simplicity in writing we will however 
always refer to the principal disc as  $| s| \leq 1$ for the generic $\chit^{(n)}$ 
and leave it to the reader to understand the correct high/low temperature distinction.

Next, in 1996, Guttmann and Enting~\cite{GE96} gave compelling arguments (though
not a proof) that the {\em anisotropic} Ising susceptibility  was in a different 
class of functions to that of most functions of exactly solved lattice models.
In particular, both the Ising free-energy and magnetisation are known to be 
holonomic functions (i.e. {\em differentiably finite} or {\em D-finite} functions), 
while the susceptibility, they argued, was not. This is an important clue as to why the 
susceptibility is, mathematically, a much more difficult problem than 
the free-energy or magnetisation.

In 1999 and 2000, Nickel~\cite{nickel-99, nickel-00} suggested that the {\em isotropic} 
Ising susceptibility possessed a {\em natural boundary on the unit circle} $|s|=1.$ 
While again not providing a rigorous proof, Nickel's arguments were most compelling. 
Note that functions with a natural boundary cannot be D-finite.

Then in 2001, Orrick, Nickel, Guttmann and Perk~\cite{Orrick} presented a 
{\em polynomial time} algorithm for the generation of the coefficients of the series 
expansion of the susceptibility, in fact in time O$(N^6)$ for a series of $N$ terms. 
From an algebraic-combinatorics viewpoint, a polynomial time algorithm is considered 
a solution. Naturally, a closed form solution is preferable, both for elegance, and 
so that the analytic structure can be readily extracted. Furthermore, Orrick 
{\it et al.}~\cite{Orrick} gave a detailed asymptotic analysis, based on a series
expansion of some $\, 323$  terms in both high- and low-temperature expansions, 
and most crucially, short distance correlation functions as series in 
$\tau\,=\, (1/s-s)/2$, the temperature deviation from the critical value.
Various exponents and amplitude parameters were extracted to unimaginable accuracy, 
typically 30 or more digits, and these answered most of the outstanding questions
as to the structure of the scaling functions.

Orrick {\it et al.}~\cite{Orrick} also addressed the question of the implications of 
the unit (complex $s$-plane) circle as a natural boundary~\cite{nickel-99, nickel-00}.  
Isolated singularities can obviously be ``seen" and indeed Orrick {\it et al.} needed 
to subtract or otherwise suppress the effect of a number of such singularities, including
a contribution from $\tilde{\chi}^{(6)}$, to achieve the accuracy they did in their 
critical point analysis.  They also presented a heuristic argument and calculation 
to show that the $\, n\, \rightarrow \, \infty$ accumulation of the  $\tilde{\chi}^{(n)}$ 
singularities implied that expansions in $\,\tau$  would only be {\em asymptotic}.  
Their prediction of the rate at which expansion coefficients diverge was qualitatively 
confirmed by numerical results on a toy model, subsequently called the diagonal 
Ising model by Boukraa {\it et al.}~\cite{Diag}.

In 2004, Zenine, Boukraa, Hassani and Maillard~\cite{ze-bo-ha-ma-04} made an important
step towards the understanding of the three-particle contribution $\tilde{\chi}^{(3)}$, 
based on a novel and powerful method for series expansions~\cite{ze-bo-ha-ma-05}, 
followed by an inspired computer search for the underlying linear ODE generating 
the series expansion.  They obtained the Fuchsian linear ODE for $\tilde{\chi}^{(3)}$.  
In 2005, the same group~\cite{ze-bo-ha-ma-05b} found the Fuchsian linear ODE for
$\tilde{ \chi}^{(4)}$ by similar methods. An important observation coming out of 
the $\tilde{\chi}^{(3)}$ work was that there were singularities that had not been 
predicted by Nickel~\cite{nickel-99,nickel-00} and that the  $\tilde{\chi}^{(n)}$
were in fact much more complicated functions than had been imagined.  This gave 
considerable urgency to finding new results for higher order  $\tilde{\chi}^{(n)}$.

Since finding the linear ODE for  $\tilde{\chi}^{(n)}$ with $n>\, 4$ is clearly 
a huge task, another strategy to get some kind of handle on the analytic structure 
of $\tilde{\chi}^{(n)}$ was considered. A set of simplified integrals were 
introduced~\cite{bo-ha-ma-ze-07,bo-ha-ma-ze-07b} by making the integrand of
$\tilde{\chi}^{(n)}$ simpler and/or by restricting the multiple integral
to an integral over a single variable. The hope was that these integrals will
preserve some (or hopefully all) of the singularity structure of the full problem. 
From these model integral investigations, a reasonably consistent picture
of the singularities emerged and a large set of candidate singularities for
those actually occurring in the linear ODE's of $\,\tilde{\chi}^{(n)}$ was obtained.

In this paper we return to the problem of finding  linear ODE's for $\, \tilde{\chi}^{(n)}$ 
(for $n>\, 4$) or, failing that, to obtaining long series that can be analysed by other 
means.  To save space and repetition, we will drop the important adjective {\em linear}
 before ODE in the following, but all mention of ODEs should be taken as a reference 
to a {\em linear Fuchsian ODE}, unless otherwise stated. A strong motivation for 
obtaining such long series  for the full susceptibility $\,\tilde{\chi}$, and the $n$-particle 
contributions $\,\tilde{\chi}^{(5)}$ and  $\,\tilde{\chi}^{(6)}$ 
is not to improve already known numerical calculations (critical exponents, amplitudes, ...),
but to shed some light on important, and not yet understood, physical problems such as 
the structure of the ODE's they satisfy, the implication of new singularities for the 
natural boundary,  heuristic asymptotics of Orrick {\it et al.}~\cite{Orrick}, and the 
mechanism of resummation of logarithmic singularities.

Indeed, in all previous analyses of the full  $\,{\chi}$ and the individual  
$\,{\chi}^{(n)}$, a point appeared that was left unresolved~\cite{ze-bo-ha-ma-05c}.
This issue is the power/log behaviour of each $\,\chi^{(n)}$ at the singular
points which is not the same as the behaviour of the full $\chi$. That is to say, 
the asymptotic behaviour of the full susceptibility ${\chi}$
is (with $\tau=\,\frac{1}{2}( 1/s-s)$)
\begin{eqnarray}
\label{tau1}
\chi \, \,\sim \,\,\, ct. \,\,\, |\tau|^{-7/4}\, \, + ct.\,\,
  + ct. \,\, |\tau|^{1/4}\, \, +{\rm O}(|\tau|\log|\tau|),
\end{eqnarray}
whereas: 
\begin{eqnarray}
\label{tau2}
\chi^{(n)}\,  \sim\, \, \, ct. \, \,  |\tau|^{-7/4}  \, 
+ ct. \,\, |\tau|^{1/4}(\log|\tau|)^{n-1}\, 
+{\rm O}(\log|\tau|^{n-2}),
\end{eqnarray}
where $\, ct.$ denotes  constants. In order to resolve the issue of how the 
individual terms of the form $ |\tau|^{1/4}(\log|\tau|)^{n-1}$ occurring in 
$\chi^{(n)}$ combine to give a constant in the full susceptibility, we require 
a better understanding of each  $\chi^{(n)}$ rather than getting the full 
$\,\chi$ to higher accuracy. With the complete solutions for $\tilde{\chi}^{(3)}$ and
$\tilde{\chi}^{(4)}$, long series expansions for the higher terms  ${\chi}^{(n)}$
and for the full  ${\chi}$ will allow us to completely resolve this issue.

Our achievements, with regard to series generation, are that we have extended the 
series for the full $\chi$,  $\tilde{\chi}^{(5)}$ and  $\tilde{\chi}^{(6)}$ to about 2000 
terms\footnote[3]{More precisely, for $\tilde{\chi}^{(6)}$ we have $\, 3260$ coefficients 
in $\, w$ or $\,1616$ non-zero terms (that is in $x=w^2$).}.  We also have series for  
$\tilde{\chi}^{(5)}$, {\em modulo a single prime}, to 10000 terms, and this has enabled us 
to find the associated linear ODE {\em modulo} a prime and get a complete picture of the 
singularities and exponents in that case. Most of the details of this $\tilde{\chi}^{(5)}$ 
linear ODE analysis will be given in a future paper. 

In this paper we report in Section \ref{chilong} on how the generation of the long series has 
been achieved and in Section 3 we describe an efficient method for finding the
associated linear ODE's and give some results from the linear ODE mod prime 
analysis of $\tilde{\chi}^{(5)}$. Section 4 is devoted to a floating-point diff-Pad\'e 
analysis of the 2000 terms exact series for $ \,\tilde{\chi}^{(5)}$. There is no such 
numerical work on long series in the literature and so we have had no {\em a priori} 
knowledge about what to expect. In view of this, we have been strongly motivated
to make a detailed comparison between the numerical and exact $\,\tilde{\chi}^{(5)}$ results.  
The comparison has been instructive and we now have a reasonably good basis for 
judging the efficacy and  reliability of the diff-Pad\'e method. Specifically, there 
are clear patterns showing which singularities and exponents can be obtained and 
which will be missed. We report also on the floating-point diff-Pad\'e analysis 
of the 1600 term exact series for $\tilde{\chi}^{(6)}$ where only limited exact 
results are known based on our previous~\cite{bo-ha-ma-ze-07,bo-ha-ma-ze-07b} 
toy model investigations. Section 5 is devoted to a study of the singularities that occur 
in the higher $\tilde{\chi}^{(n)}$, $n \ge 7$, by performing a diff-Pad\'e analysis on 
the full $\tilde{\chi}$ series from which the small $n$ contributions 
$\, \tilde{\chi}^{(n)},$ $n < 7$ terms have been removed. We address in Section 6 the
 ``logarithm summation'' problem discussed above, and resolve it.
A simplified model is introduced to show how the subdominant singularities in  
$\chi^{(n)}$, that individually vanish as the temperature approaches its
critical value, sum to yield the known finite amplitude at the critical point.  
In Section 7 we describe the fast Fourier transform (FFT) that was used by 
Orrick {\it et al.}~\cite{Orrick} as a diagnostic tool in the study of series singularities.  
Here we use it to confirm the absence of certain singularities on the complex 
$s$-plane unit circle and confirm the existence of a {\em natural boundary} 
for $\, \chi$.

Except for \ref{SumRule} on Fuchs' relations, the Appendices revisit and extend 
the Landau singularity analysis~\cite{bo-ha-ma-ze-07,bo-ha-ma-ze-07b,nickel-05} of  
$\, \tilde{\chi}^{(n)}$. In \ref{appendixBernie} we prove that in the absence of the 
``Fermionic factor'' $(G(n))^2$ in (\ref{chi3tild}), the singularities found 
in~\cite{ bo-ha-ma-ze-07,bo-ha-ma-ze-07b} are exhaustive.  We also prove that 
none of these singularities, beyond those found by Nickel~\cite{nickel-99}, 
can lie on the principal $s$-plane unit circle.  This dispels any hope that there 
might be singularity cancellation and that, as a result, the unit circle may not 
be a natural boundary.

The presence of the Fermionic factor in the  $\tilde{\chi}^{(n)}$ integral 
(\ref{chi3tild}) is too complicated to deal with in its entirety, and Appendices~B
and D provide a more limited perspective. We describe in \ref{app:ToyModel} what 
happens when all, but a single term of the complete Fermionic product, is dropped 
from the integrand of $ \, \tilde{\chi}^{(n)}$.  The linear ODE describing the series 
in that case has {\em additional} singularities.  On the other hand we show 
that there are no such Landau singularities~\cite{bo-ha-ma-ze-07,bo-ha-ma-ze-07b}
on various analytic continuations\footnote[5]{The  Landau singularity 
analysis~\cite{bo-ha-ma-ze-07,bo-ha-ma-ze-07b} is {\em local.} This amounts to 
{\em keeping no track} of which local square-root branch we are on for each singularity.} 
of the $\tilde{\chi}^{(n)}$. The conclusion seems to be that the singularities of 
the integral representation of $\tilde{\chi}^{(n)}$ do not identify with but are 
{\em just included} into  singularities of the linear ODE associated with a 
$\tilde{\chi}^{(n)}$. In \ref{app:Exponents} we give a power counting argument 
to determine the singularity exponents of one class of 
Landau singularities.  The close analogy of these to the principal $\, s$-plane unit 
circle singularities, and the simple formula for the exponent values, motivate us to 
call them the ``normal'' exponents. Other exponents are known to be present based on 
the exact and diff-Pad\'e analysis in Sections~\ref{sec:analysis} and \ref{sec:DPn7}, 
and, as a tentative terminology, we denote these other exponents as ``anomalous''.

\section{Extremely long series for the susceptibility \label{chilong}}

\subsection{The full susceptibility $\chi$ \label{full}}

Orrick {\it et al.}~\cite{Orrick}  used an algorithm of complexity {\rm O}$(N^6)$ to 
obtain the first $N$ coefficients of the susceptibility series. As a result, they 
generated and analysed series with more than 300 terms in both the high- and 
low-temperature regime. This remarkable {\em polynomial growth} of the algorithm is
due to the use of {\em quadratic partial difference equations} (see~\cite{Orrick})
which are {\em finite difference Painlev\'e equations}~\cite{McCoy,Perk}. 

The susceptibility $\,\chi$ is obtained via (\ref{chidef}) with this 
quadratic partial difference system of equations providing a means to calculate
the two-point correlations $\, C(m,n)$ efficiently, and simultaneously, for both 
high- and low-temperature series.  A susceptibility series of $N$ terms requires 
$\, C(m,\,n)$ on the octant $\, m+n \,\le\, 2N$, $m<n$, deduced as series from the 
quadratic difference equations, with the diagonal $\, C(n,n)$, $n \, \le \, N$, as 
initial value data.  The latter can be obtained from either a Toeplitz determinant 
expression~\cite{McCoy} or a set of non-linear recursion relations~\cite{Jimbo}. 
The calculation of $C(m,n)$ on a particular site requires the multiplication of the 
$\,C(m',\, n')$ series of length O$(N)$ for the same and/or neighbouring sites. 
If the series multiplication is done as a naive product then the complexity of the 
susceptibility determination is that of $\, {\rm O}(N^2)$ integer multiplications on 
O$(N^2)$ sites.  The Maple code used by Orrick {\it et al.}~\cite{Orrick} ran in a 
time approximately proportional to $N^6$.  This was as expected since the integer 
(digit) size grows linearly in $\, N$ and naive integer multiplication scales as 
$\, N^2$.  There was no attempt to improve on this code as the 323 term series
obtained was deemed entirely adequate.

Obvious improvements can be made. First, we know that $\, \chi$ when
expressed in the variable $\, s/2$ (high temperature) or $\, 1/(4s^2)\,$ (low temperature) 
is a series with {\em integer coefficients}. This implies that if the series generating 
program  is run using modular arithmetic~\cite{Knuth} with a number of different primes, 
then the full series coefficients can be reconstructed from the residue series via the 
Chinese remainder theorem. The number of primes necessary is easily estimated. Since 
$|s| \, =\, 1$  is the singularity boundary for both high and low temperatures, the 
high-temperature series coefficients in $\, s/2$ grow with the number of terms $\,N$ 
as $\, 2^N$, while at low temperature the series coefficients in $\, 1/(4s^2)$ grow 
as $4^N.$  The latter expressed in numbers of bits is $\, 2N$ and with primes of bit 
length 15 that we typically use, we arrive at $ \simeq \,2N/15$ as the required number 
of primes.

The modular arithmetic approach leaves the complexity of the $\, \chi$ series 
generation to $N$ terms at $\, {\rm O}(N^4)$ multiplications for a single prime. 
For the full calculation that requires $\, {\rm O}(N)$ primes the complexity is
$\, {\rm O}(N^5)$.  However, once the generating program is running in integer 
{\em mod prime} mode, it is easy to implement series multiplication via fast Fourier 
transform (FFT)~\cite{Knuth}. This changes the series multiplications from 
$\, {\rm O}(N^2)$ to $\, {\rm O}(N \log(N))$ and gives as our final 
estimate for the $\, \chi$ generation complexity $\, {\rm O}(N^4 \log(N))$. In the
 next subsection we show how similar complexity improvements
 in the generation of the $\tilde{\chi}^{(n)}$ series have been achieved.

For the extension of the $\, \chi$ series to 2000 terms as reported below, we
translated the Maple code from Orrick {\it et al.}~\cite{Orrick}
 to Fortran\footnote[5]{In preliminary studies we obtained 1600 terms of the high
and low temperature expansions of  $\, \chi$  with highly optimised C++ programs.} and 
ran it in integer mod prime mode. We also incorporated the FFT multiplication of 
series. Finally we changed the $\, C(n,\, n)$ initialization to use the 
Jimbo and Miwa recursion~\cite{Jimbo}. 

With these changes the calculation on the APAC (Australian Partnership 
for Advanced Computing) SGI Altrix cluster with 1.6GHz Itanium2
processors using 280 primes took about 240 CPU hours in total
(we note that the algorithm without FFT multiplication would
require about twice the above amount of CPU time).

The resulting series can be found on the  web-page~\cite{Iwanweb} where we give
the first 2000 coefficients of the high- and low-temperature expansions in the variables 
$u=s/2$ and $v=1/(4s^2)$, respectively. In addition we also give the expansion
for $\tilde{\chi}_H$ and $\tilde{\chi}_L$ in the self-dual variable $\, w=\, \frac{1}{2}s/(1+s^2)$.
The series in the $\, u$ or $v$ variables, or in the $\, w$ variable,
are probably the best as far as computer encoding  and modular calculations are 
concerned since all the coefficients are {\em integers} rather than rational numbers. 
For example, for any $n>2$, $\chit^{(n)}=2^nw^{n^2}[1+4n^2w^2 + {\rm O}(w^3)]$.




\subsection{The contributions  $\tilde{\chi}^{(5)}$ and  $\tilde{\chi}^{(6)}$}
\label{chi5}

The previous longest series available for $\,\chi^{(5)}$ and $\,\chi^{(6)}$ can be 
found in~\cite{Orrick} where the first 182 and 140 terms respectively of the series 
in $s$ are listed.  We have extended these series dramatically -- two extreme 
examples being the full integer series for  $\,\tilde{\chi}^{(5)}$ to 2000 terms and the 
series for $\,\tilde{\chi}^{(5)}$  modulo a single prime to 10000 terms. The latter 
extension is only possible because the complexity order for a {\em mod prime series} 
of length $N$ has been reduced to $\, {\rm O}(N^4 \log(N))$.  We outline in this
section how this reduction has been achieved.

Our method for evaluating the integral expression (\ref{chi3tild}) for $\,\tilde{\chi}^{(n)}$  
remains as described in earlier publications~\cite{ze-bo-ha-ma-05, ze-bo-ha-ma-05b}.  
We first convert (\ref{chi3tild}) back to an $n$-fold $\phi_i$ integration with the 
explicit phase constraint $\, 2 \pi \delta(\sum \phi_i)$ now in the integrand.  
This constraint is then replaced by the equivalent 
$\, \sum_k \exp(\rmi k\, \sum \phi_i)$, 
thus decoupling all $\phi_i$ integrations at the expense of a sum over the Fourier
integer\footnote[5]{ Trigonometric functions in the integrand such as 
$\, \sin((\phi_i-\phi_j)/2)$ are dealt with by expanding them into a sum of phases 
$\exp(i\sum m_i\phi_i)$ with each $\, m_i$ some small integer. These can be 
incorporated into shifts $\, k\, \phi_i \, \rightarrow \, (k\, +m_i)\cdot \phi_i$
and do not change the form of the integrand.  Since such shifts also do not change 
the complexity order of the  $\,\tilde{\chi}^{(n)}$ calculation they will not be considered 
further.} $\, k$, where the sum extends from $k=-\infty$ to $\infty$.

Next we expand all denominator factors in the integrand of $\,\tilde{\chi}^{(n)}$, thereby 
converting the integrand into a sum of $n$-fold products $\, \prod y_i\, x_i^{n_i}$.  
Each $\, \phi_i$ integration then picks out the $k^{th}$ Fourier coefficient of 
$y_ix_i^{n_i},$ namely $\, w^{|k|\;+n_i+1} \cdot a(k,n_i),$ where $\, a(k,n_i)$ is 
proportional to a hypergeometric function $\, _4F_3$ in the variable $\, 16\, w^2$.  
The net result of all these operations is that we have replaced the $\,\tilde{\chi}^{(n)}$ 
continuum integration by a nested summation of products of hypergeometric functions.

The complexity of this calculation is of some order that we can now easily determine.
If we want a series of length $\, N$, then the Fourier $k$ sum can be restricted to 
$\,{\rm O}(N)$ as can all the hypergeometric function series.  The evaluation of 
products of series of length $\,{\rm O}(N)$ is either an $\, {\rm O}(N^2)$ calculation 
if done as a naive product or $\, {\rm O}(N \log(N))$ if done by FFT.  These two
operations, namely the $k$ sum and the series multiplication, are inherent to our 
method and give an irreducible minimum complexity of either $\,O(N^3)$ or 
$\,{\rm O}(N^2 \log(N))$.  The only place where we have some freedom to reduce the 
complexity of our $\tilde{\chi}^{(n)}$ series evaluation is in the number of summations 
that are required for the expansion of the denominator factors in the original 
$\,\tilde{\chi}^{(n)}$ integrand.  The total number of products $\, \prod y_i\, x_i^{n_i}$  
must be no more than $\, {\rm O}(N^2)$ to keep the overall complexity at 
$\,{\rm O}(N^4\log(N))$ and this, in turn, implies the denominator expansion must be 
limited to two independent summations.

This limitation on the summations immediately shows that the product form (\ref{Gn}) 
for the Fermionic factor $G^{(n)}$ is not appropriate and alternatives must be used. 
The useful formulae for high-temperature series are those given in [3], in particular 
equation (5) for some low order $\, \tilde{\chi}^{(2n+1)}$ and equation (10) for the general 
case.  For low-temperature series we note that $\, G^{(2n)}$ has been shown to be a 
Pfaffian~\cite{Nappi} in the $h_{ij}$ defined in (\ref{Gn}). By a rearrangement of 
terms in the Pfaffian one finds that the analog of equation (10) in~\cite{nickel-99} is
\begin{eqnarray}
\label{Hnnew}
&&H^{(2n)} \,\, = \,\, \,{{(G^{(2n)})^2} \over {(2n)!}} \,\, \,= \,\,
{{1} \over {(2n)}} \, h_{12}\, h_{2n-1,2n} \\
&& \times  \prod_{m=1}^{n-1}\, \Bigl( h_{2m,2m+1} \, h_{2m-1,2m+2}\, 
 +{{1} \over {(2m) }} h_{2m-1,2m} \, h_{2m+1,2m+2} \Bigr).  
 \nonumber 
\end{eqnarray}
Use of label interchange symmetry, $h_{ij}=-h_{ji}$, allows one to combine 
terms in (\ref{Hnnew}) further. For the first few low order terms we have
\begin{eqnarray}
\label{H2}
&&H^{(2)} \,  =  \,  \, -{{1} \over {2}} \, (h_{12} \, h_{21}), \qquad \nonumber \\
&&  H^{(4)} \,  = \,  \, {{1} \over {8}} \, (h_{12} \, h_{21})(h_{34} \, h_{43}) \, 
\, -{{1} \over {4}} \, (h_{12} \, h_{23} \, h_{34} h_{41}), \nonumber \\
&&H^{(6)} \,  = \,  \,
 -{{1} \over {48}} \,  (h_{12} \, h_{21})(h_{34} \, h_{43})(h_{56} \, h_{65})\, 
+{{1} \over {8}} \, (h_{12}\, h_{23} \, h_{34} \, h_{41})(h_{56} \, h_{65})	
 \nonumber \\
&& \quad \quad -{{1} \over {6}} \, 
(h_{12} \, h_{23} \, h_{34} \, h_{45} \, h_{56} \, h_{61}),
\end{eqnarray}
written in an obvious cyclic form.  It is these expressions, in particular the one 
for $\, H^{(6)}$, that are the starting point of our discussion of the reduction in 
the denominator expansion summations. The most complicated term contributing to 
$\, \tilde{\chi}^{(6)}$ has an integrand that contains the last term of
$\, H^{(6)}$ in (\ref{H2}) and thus the seven denominator factors
\begin{eqnarray}
\label{x1x6}
&&(1\,-x_1\,x_2\,x_3\,x_4\,x_5\,x_6)^{-1} \,
 (1\,-x_1\,x_2)^{-1}\,(1\,-x_2\,x_3)^{-1}\,\\
&&	\times (1\,-x_3\,x_4)^{-1}\, 
(1\,-x_4\,x_5)^{-1}(1\,-x_5\,x_6)^{-1}(1-x_6\,x_1)^{-1}. \nonumber
\end{eqnarray}
Naive expansion of these denominators results in a 7-fold sum -- clearly a 5-fold 
excess that must be eliminated. The first step towards this elimination  is the use 
of partial fraction rearrangement, a trick that was already used in the evaluation of  
$\, \tilde{\chi}^{(3)}$ and $\, \tilde{\chi}^{(4)}$ in~\cite{ze-bo-ha-ma-05,ze-bo-ha-ma-05b}.
The version we use here is based on the identity
\begin{eqnarray}
\label{ABC}
\fl \qquad (1-A)^{-1}(1-B)^{-1}(1-C)^{-1}\,  = \nonumber  \\
\fl \qquad \qquad (1-ABC)^{-1}\, [1-(1-A)^{-1} \, -(1-B)^{-1}-(1-C)^{-1}+
\\
\fl \qquad \qquad  (1-A)^{-1}\,(1-B)^{-1}\, +(1-B)^{-1}\,(1-C)^{-1}\, 
                  +(1-C)^{-1}(1-A)^{-1}], \nonumber	
\end{eqnarray}
in which we first set ${A,B,C}\,=\,\, {x_1 \, x_2, \, x_3 \, x_4, \, x_5 \, x_6}$
and then ${A,B,C}\,=\, \,{x_2 \, x_3, \, x_4 \, x_5, \, x_6 \, x_1}$.  The product
of these two forms of (\ref{ABC}) enables us to replace (\ref{x1x6}) by
\begin{eqnarray}
\label{x1x2x6}
&&(1\, -x_1\,x_2\,x_3\,x_4\,x_5\,x_6)^{-3}\,
(1\,-x_1\,x_2)^{-1}\,(1\,-x_2\,x_3)^{-1}\,\nonumber	 \\
&& \qquad \times (1\,-x_3\,x_4)^{-1}\, (1\,-x_4\,x_5)^{-1}
\end{eqnarray}
plus terms with similar, or fewer, denominators. That is, (\ref{x1x2x6}) is now the 
most complicated set of denominators in the integrand contributing to $\, \tilde{\chi}^{(6)}$. 
With the replacement of the first factor by $\, (1\, -x_1\,x_2\, x_3\, x_4\, x_5)^{-1}$, 
(\ref{x1x2x6}) becomes the equivalent most complicated term in the evaluation of  
$\, \tilde{\chi}^{(5)}$ .  Thus all our subsequent remarks apply equally to both 
$\, \tilde{\chi}^{(5)}$ and $\, \tilde{\chi}^{(6)}$. 

Expansion of (\ref{x1x2x6}) results in the formal 5-fold summation
\begin{eqnarray}
\label{S}
S\,  = \, \, \sum_{m,n_1,p,q,n_5}\,  x_1^{m+n_1}\,  x_2^{m+n_1+p}\,
  x_3^{m+p+q} \, x_4^{m+q+n_5}\, x_5^{m+n_5}\,  x_6^m,
\end{eqnarray}
which is of complexity O$(N^5).$  However the $\, n_1$ and $\, n_5$ summations can be 
done independently and thus (\ref{S}) is in fact only of complexity 
$\, {\rm O}(N^3\cdot (N+N))\, =\,{\rm O}(N^4)$.  This is not yet an adequate reduction 
and we can do better by eliminating the $\, n_1$ and $\, n_5$ summations entirely via 
the use of recursion relations. For example, we define the $\, n_1$ sum of the 
$\, x_1,\,  x_2$ pair as
\begin{eqnarray}
\label{A12a}
A_{12}(m,p)\,  = \, \, \sum_{n_1=\, 0}\,  x_1^{m+n_1} x_2^{m+n_1+p} \,  = 
 \,\, \sum_{r=m} x_1^{r} \,  x_2^{r+p}
\end{eqnarray}
and note that because $m$ appears only as a limit on a dummy variable sum, 
the $\, A_{12}$ satisfies the recursion
\begin{eqnarray}
\label{A12b}
&&A_{12}(m,p)\,  =\,\, \delta A_{12}(m,p)\, + A_{12}(m+1,p), \\
&& \delta A_{12}(m,p) \,=  \,\,  \,x_1^m \, x_2^{m+p}. \nonumber 
\end{eqnarray}
Only the lower limit on the summation in (\ref{A12a}) has been given explicitly. 
There is also an upper limit that depends on the length $\, N$ of the series in  
$\, \tilde{\chi}^{(5)}$ or $\, \tilde{\chi}^{(6)}$ we want to obtain. Thus $\, A_{12}(m,p)$ 
vanishes for $m$ large enough, and (\ref{A12b}) shows that, as $m$ is decreased 
from this upper limit, each determination of $\, A_{12}(m,p)$ requires only the 
evaluation of a single product and its accumulation into a previously stored 
result\footnote[2]{Our discussion here is schematic. It is to be understood that 
the $\, \phi_i$ integrations have been carried out and the ``single product" being 
referred to is the product of the two hypergeometric function series associated 
with $x_1^m x_2^{m+p}$.  An additional implication is that $\, A_{12}(m,p)$ has 
absolutely no functional dependence on its subscripts and could equally well be 
denoted  $\, A(m,p)$.  The subscripts have only been included to indicate a 
connection to a particular factor in the  $\, \tilde{\chi}^{(5)}$ or $\, \tilde{\chi}^{(6)}$ 
integrand.}. If we now take it as given that the $\, m$ summation in (\ref{S}) 
is performed in decreasing sequence we obtain
\begin{eqnarray}
\label{A54}
S \, =\, \, \, 
\sum_{m,p,q}\, A_{12}(m,p) \cdot  x_3^{m+p+q} \cdot  A_{54}(m,q) \cdot  x_6^m,
\end{eqnarray}
which is of complexity $\, {\rm O}(N^3)$. 

It remains to be shown that the idea of recursion can be applied once more, ultimately 
reducing the complexity of the calculation of  $\,S $ to $\, {\rm O}(N^2)$. 
For this purpose define
\begin{eqnarray}
\label{B123}
B_{123}(m,q)\,  =\, \, \,  \sum_{p=0} \, A_{12}(m,p)\,  x_3^{m+p+q},
\end{eqnarray}
which, for $\, q>0$, can be put in the recursive form
\begin{eqnarray}
\label{B123A12}
B_{123}(m,q) &=&\, \,\, 
 \sum_{p=0} \, [\delta A_{12}(m,p)\, \,+A_{12}(m+1,p)] \, x_3^{m+p+q} \nonumber \\
\quad &=& \, \, \delta B_{123}(m,q)\,  + B_{123}(m+1,q-1).
\end{eqnarray}
For $\, \delta B_{123}(m,q)$, which is the first sum in (\ref{B123A12}), we have
\begin{eqnarray}
\delta B_{123}(m,q)\, =\, \, \, \sum_{p=0}\, x_1^m \,x_2^{m+p}\, x_3^{m+p+q} \, 
 = \, x_1^m \cdot  A_{23}(m,q),
\end{eqnarray}
a single product analogous to $\, \delta A_{12}$ in (\ref{A12b}). The new feature 
in the recursion (\ref{B123A12}) for $\, B_{123}$, relative to (\ref{A12b})
for $\, A_{12},$ is that we must maintain in storage an entire array of elements 
indexed by $q$.  Furthermore, for each $m$ we must supply, by a separate calculation, 
the $\, q=\, 0$ term
\begin{eqnarray}
\label{B123A12x3}
B_{123}(m,0)\,  = \,  \, \, \,  \sum_{p=0}\,  A_{12}(m,p) \cdot  x_3^{m+p}.
\end{eqnarray}
Although this does require a sum, the fact that $\, q=\, 0$ is fixed means 
the contribution of this evaluation to the complexity order 
of $\, S$ is still only $\, {\rm O}(N^2)$.  With the general $\, B_{123}$
now given either by (\ref{B123A12}) or (\ref{B123A12x3}), we obtain
\begin{eqnarray}
\label{SA54}
S \, = \, \, \, \sum_{m,q} B_{123}(m,q) \cdot  A_{54}(m,q) \cdot  x_6^m, 
\end{eqnarray}
which is the two-fold sum, and thus the O$(N^2)$ result, we were looking for. 
Note that memory requirements are also quite minimal.  We need to store the
$B$ array which has  $\, {\rm O}(N)$ elements each of which is a series of length 
$\, {\rm O}(N)$. Thus memory requirements also scale as $\, N^2$.

Our initial Fortran coding for the $\, \tilde{\chi}^{(5)}$ and $\, \tilde{\chi}^{(6)}$ series 
generation used only naive series multiplication and thus was $\, O(N^5)$ for a 
single prime. With these programs, series for  $\, \tilde{\chi}^{(5)}$ to 2000 terms
and $\, \tilde{\chi}^{(6)}$  to 3260 terms were generated in about 100000 CPU hours 
running 160 primes in parallel.  The calculations were carried out on the 
afore-mentioned APAC cluster. Series for  $\, \tilde{\chi}^{(5)}$ to 6000 terms,
modulo the single prime $p_0\,=\,\, 32749$, were obtained in about 40000 CPU hours 
using 32 processors on one of the VPAC (Victorian Partnership for Advanced Computing) 
facilities which is a Linux cluster based on Xeon 2.8Ghz CPUs.

When it was observed that the 6000 terms  were {\em not} sufficient to obtain
the linear ODE for  $\, \tilde{\chi}^{(5)}$, the Fortran codes were modified to include 
FFT series multiplication and series {\it modulo} $p_0$ for $\, \tilde{\chi}^{(5)}$ to 
10000 terms were obtained. 

The total CPU time for $\tilde{\chi}^{(5)}$ to 10000 terms was about 17000 hours
(6000 terms take around 2000 hours using the FFT algorithm on the APAC). 

The calculation was done in parallel using 128 processors. The algorithm is 
straightforward to parallelise because the calculations in the outer most loop, 
that is the sum over Fourier mode integer $k$, can be done independently for 
each value of $\, k$. The only issue is that the time required decreases with 
$\, k$. In order to use approximately the same time per processor we simply 
assign calculations with a given $\, k$ to processors in an alternating pattern 
such that processor 0 does $k=0$, $k=\, 255$,  $k=\, 256$, $k=\, 512$, $k=\, 513, \ldots$ 
while  processor 1 does $k=\, 1$, $k=254$,  $k=257$, $k=511$, $k=514, \ldots$ and so on 
up to processor 127 which does $k=\, 127$, $k=\, 128$,  $k=\, 383$, $k=\, 384$, $k=\, 639, \ldots$.
This simple assignment scheme ensures a good balance with
the total time used by various processors differing by less than $4\%$. 

It is this 10000 terms series that has enabled us to obtain the exact ODE 
{\it modulo} $p_0$ for  $\, \tilde{\chi}^{(5)}$, thus making possible the various comparisons 
found elsewhere in this paper.

We conclude with two observations on the technical aspects of the mod prime calculations.  
Firstly, a very frequent operation in our codes is the accumulation of two products, 
i.e. $\, a\, \rightarrow \, a+bc+de$.  If the variables are integers {\it modulo} $p$  
then the accumulation can be done as a standard 32 bit integer operation with the single 
(Fortran) call $\, a=\, {\rm mod}(a+bc+de,p)$ without overflow provided $p<2^{15}$.  
It is this feature that dictates our choice of primes and in particular $p_0=\, 2^{15}-19$.  
Secondly, we follow a recommended practice~\cite{Crandall2} of loading the floating point
FFT routines with mod prime variables in ``balanced" form.  That is, if any mod $\,  p $ 
variable $\, v$ is greater than $\, p/2$ it is loaded as $\, v-p$.  This eliminates most 
of the ``dc" part of the input signal and typically increases the safety margin in the 
output rounding of float to integer by several bits. We keep track of the differences in 
these rounding operations so as to guarantee our programs generate all integers correctly.

\section{Fuchsian ODEs for long series modulo a prime \label{FuchsianODEfor}}

\subsection{The linear ODE for  $\tilde{\chi}^{(5)}$ \label{linODEchi56}}

We begin with a remark that applies to all the subsequent discussion, namely
that {\it there is no single unique ODE that describes any given series.}
There is a minimum order linear ODE that is unique but this typically contains 
a very large number of apparent singularities and can only be determined from 
a corresponding larger number of series coefficients. In our quest for linear 
ODE's corresponding to given very long series
expansions, we are interested in the Fuchsian linear ODE requiring the
{\em minimum number of coefficients from the series in order to be obtained}.
In general there are any number of intermediate ODE's but unless otherwise
required by the context, we will call all of these the underlying linear ODE,
 without distinction.

The 2000 terms generated for  $\, \tilde{\chi}^{(5)}$ and
 $\, \tilde{\chi}^{(6)}$ are not sufficient 
to find the exact underlying linear ODE.  However an alternative approach is to use 
{\em mod prime series} to find the linear ODE {\it modulo} a prime.  From such a linear 
ODE we can get singularity positions {\it modulo} a prime,  indicial equations and hence 
singularity exponents {\it modulo} a prime, and indeed practically everything that could 
be obtained from the exact ODE but restricted to prime residues.  Furthermore if, for 
example, the {\it modulo} prime factorization of the head polynomial of the linear ODE 
yields factors with small coefficients, then one can, with almost perfect certainty, 
conclude that one has all the exact singularity locations.  Similar remarks would also 
apply, say, to singularity exponents\footnote[2]{Of course, if there is an ambiguity one 
can always use additional primes and resolve the ambiguity by the Chinese remainder theorem.}.  
And indeed, because we have been able to find the  {\em mod prime linear ODE} for  
$\, \tilde{\chi}^{(5)}$ and because  the singularity locations and exponents appear to be simple, 
we are confident that what we report are in fact the  {\em exact values}.

Because series generation is expensive we want to be sure our algorithm for deducing the 
underlying ODE requires the fewest number of terms.  We report in this section on a 
method~\cite{GutJoy} that is slightly different from that used for, say,  $\, \tilde{\chi}^{(3)}$ 
but that appears to have a number of appealing advantages.  We will report on some of our 
results on the analysis of  $\, \tilde{\chi}^{(5)}$, in particular those that affect the  number 
of series terms required to find the ODE.  Most of the details, such as the factorization of 
the linear ODE, will be left for a future publication.

An essential constraint on the linear ODE underlying a series $S(x)$ of the type we are 
considering here is that it must be Fuchsian.  Specifically this means that $x=0$ and 
$x=\, \infty$ are regular singular points.  A form for the linear differential operator
that automatically satisfies this $(0, \, \infty)$ regularity constraint is: 
\begin{eqnarray}
\label{28a}
	L_{MD} \,  = \,\,  
 \sum_{m=0}^{M}  \sum_{d=0}^{D}\,  a_{md} \cdot  x^d \cdot (x\, {{\rmd} \over {\rmd x}})^m,  \quad
 a_{M0}\, \ne \, 0, \,\, \, a_{MD} \,\ne \, 0.	
\end{eqnarray}
The $\, a_{M0}\, \ne \, 0$  condition is the obvious constraint to make $\, x=0$ a regular 
singular point and it is the use of the operator $x\, {{\rmd} \over {\rmd x}}$ rather than 
$\rmd/\rmd x$ that makes analysis around $x=\infty$ simple.  The change of variable $x=1/y$ 
turns (\ref{28a}) into
\begin{eqnarray}
\label{28b}
	L_{MD} \, = \,\, 
 \sum_{m=0}^{M}   \sum_{d=0}^{D} \, a_{md} \cdot  y^{D-d}
 \cdot  (-y\, {{\rmd} \over {\rmd y}})^m,
 \, \, \, \,      a_{M0} \ne \, 0,\,  \,  a_{MD} \ne \, 0,		\quad 
\end{eqnarray}
where one can see that the condition for $x=\, \infty$ ($y \, = \, 0$) to be a regular 
singular point is $\, a_{MD} \ne \, 0$.  A simple rearrangement of terms shows that 
$\, L_{MD}$ can also be written
\begin{eqnarray}
\label{28c}
	L_{MD}\,  = \, \,  
\sum_{m=0}^{M}   \sum_{d=0}^{D} \, b_{md} \cdot x^{d+m} \cdot ({{\rmd} \over {\rmd x}})^m,	
\end{eqnarray}
with the $b$ coefficients being linear combinations of the $a$.  There is no single 
$ b$  coefficient analog of $a_{MD}  \ne  0$.  We will use (\ref{28a}) exclusively, 
particularly for the purpose of determining the $ a_{md}$, but this does not preclude 
transforming to (\ref{28c}) if required.

To determine the $a_{md}$ in (\ref{28a}) we demand $L_{MD}(S(x))\,=\, \, 0$ and this 
yields a set of linear equations that we arrange in some well defined order.  There 
exists a non-trivial solution if the $\,N_{MD}\times N_{MD}$ determinant 
(with $\,\,  N_{MD}\, =\, \, (M+1)\cdot (D+1)$) corresponding to the chosen ordering 
vanishes.  We test this by standard Gaussian elimination, creating an upper triangular 
matrix $U$ in the process. If we find $ U(N,N)=\, 0$ for some $\, N,$ a non-trivial 
solution exists. If $\, N<\, N_{MD}$ we set to zero all $a_{md}$ in the ordered list 
beyond $N$. Of the remaining $a_{md}$ we set $a_{M0} = 1$, thus guaranteeing that 
$x =\, 0$ is a regular singular point, and determine the rest by back substitution.  
We guarantee that $\, x=\, \infty$ is a regular singular point by choosing the initial 
ordering of elements such that no matter what $N$ is, the first elements set to zero 
will be those from row $\,a_{0d}$, then row $\,a_{1d}$, etc. In this way the element 
$\, a_{MD}$ will never vanish unless all $a_{mD}$, $\, m\, <\, M$ vanish or the series 
$\,S(x)$ has $x=\, \infty$ as an irregular singular point.  We have not systematically 
investigated what happens when we change the element ordering within these constraints.

The $\, N $ for which $\, U(N,N)\,=\,\, 0$ is the  minimum number of coefficients needed 
to find the linear ODE within the constraint of a given $\, M$ and  $\, D$.  Obviously,
$N \, \le \, N_{MD}\, =\,\,(M+1)\cdot (D+1)$. Henceforth, $D$ will always refer to the 
minimum $\,  D$ for which a solution is found for the given $M.$  Then we can define a 
unique non-negative deviation $\, \Delta$ by $\, N\,=\,\, N_{MD}\,-\Delta$. Examples of 
such constants are given in Table~\ref{Ta:C5} based on our analysis of  $\, \tilde{\chi}^{(5)}$ and the 
combination  $\,\, 2 \cdot \tilde{\chi}^{(5)} \, -\,\tilde{\chi}^{(3)}$.

\begin{table}[htdp]
\caption{\label{Ta:C5}
$M$ is the order of the linear ODE, $D$ is the degree of each polynomial 
multiplying each derivative, $N_{MD}=(M+1)(D+1)$, $N$ is the actual number of terms
 predicted by (\ref{29}) as necessary to find an ODE of the given order $M$, 
and $\Delta$ is the difference $\, N_{MD}-N$. The first five columns gives this data
 for  $\tilde{\chi}^{(5)}$ while the next five columns gives this data for
 $\, \, 2\cdot \tilde{\chi}^{(5)}-\tilde{\chi}^{(3)}$, clearly showing the saving in 
the number of terms needed to identify the linear ODE.}
\begin{center}
\begin{tabular}{|c|c|c|c|c||c|c|c|c|c|}\hline
\multicolumn{10}{|c|}{\,\,\,\,\,\,Terms needed to find $\tilde{\chi}^{(5)}$ \,\,  Terms
 needed to find $2\cdot \tilde{\chi}^{(5)}-\tilde{\chi}^{(3)}$}\\ \hline
    $M$& $D$&  $N_{MD}$ &$N$& $\Delta$&$M$& $D$& $N_{MD}$& $N$ & $\Delta$\\ \hline
    52&  141&   7526&   7497&  29&     48&  131&   6468&   6450&  18\\
    53&  137&   7452&   7437&  15&     49&  128&   6450&   6428&  22\\
    54&  134&   7425&   7410&  15&     50&  125&   6426&   6406&  20\\
    55&  132&   7448&   7416&  32&     51&  123&   6448&   6414&  34\\
    56&  129&   7410&   7389&  21&     52&  120&   6413&   6392&  21\\
    57&  127&   7424&   7395&  29&     53&  118&   6426&   6400&  26\\
    58&  125&   7434&   7401&  33&     54&  116&   6435&   6408&  27\\
    59&  123&   7440&   7407&  33&     55&  114&   6440&   6416&  24\\
    60&  121&   7442&   7413&  29&     56&  112&   6441&   6424&  17\\
    61&  119&   7440&   7419&  21&     57&  111&   6496&   6462&  34\\ \hline
 \end{tabular}
\end{center}
\end{table}

A very striking empirical observation arises from Table~\ref{Ta:C5} and has been 
checked in many cases as summarised in  Table~\ref{Ta:ODEs}.  For reasons we do not 
understand, there exists the linear relationship
\begin{eqnarray}
\label{29}
	N\, =\, \, A \cdot M \,+ B \cdot D \,- C\, = \,(M+1) \cdot(D+1)\, -\Delta	
\end{eqnarray}
where $A,$ $B$ and $C$ are {\em constants} depending on the particular series $S(x).$  
For  $\, \tilde{\chi}^{(5)}$ they are $A=72,$ $B=33,$ $C=900$, while for the combination 
 $\, \,2 \cdot {\tilde\chi}^{(5)} \, -\, \tilde{\chi}^{(3)}$ they are $A=68,$ $B=30,$ $C=744$ 
as can be verified from Table~\ref{Ta:C5}.  Note that (\ref{29}) has no (positive) solution 
for $D$ if $M<B.$  Thus $B=\, M_0$ is the minimum order possible for the linear differential
operator that annihilates $\, S(x)$\footnote{Generically $B=M_0$. However, if the ODE is such 
that $a_{0d}=0$ for all $d$ then the constant $B=M_0-1$, that is, the minimum order minus 1.}. 
Similarly, $A=D_0$ is the minimum possible degree and thus we can rewrite (\ref{29}) in the
 more definitive form
\begin{eqnarray}
\label{30}
	N \, = \, D_0 \cdot M\,  + M_0 \cdot D\,  - C \, =\, \,
  (M+1)\cdot (D+1)\, -\Delta.	
\end{eqnarray}
The minimum order $\, M_0$ and degree $\, D_0$ can also be inferred directly from the ODE
independently of (\ref{30}).

The head polynomial $\, \sum_{d=0}^{D}\, a_{md}\, x^d$ in (\ref{28a}) can be factored 
{\it modulo} a prime and the greatest common divisor of these, from several different 
$\, L_{MD}$, is the polynomial $\,P$ whose zeros are the ``true singularities'' of the 
linear ODE.  In all cases we have tested, the degree of this head polynomial factor is 
the $\, D_0$ in (\ref{30}).  For  $\, \tilde{\chi}^{(5)}(w)$ the factor,
$\,  P(w)\, =\, \, P_{D_0}(w)\, =\,\,  P_{72}(w)$, is sufficiently simple that we are 
confident that it equals what one would obtain from the exact (not {\it modulo} a prime) 
ODE. We report it (cf. (\ref{singchi5})) in Section~\ref{sec:C5sing} in the context of a 
more general discussion of singularities and make extensive use of it in 
Section~\ref{sec:analysis} for comparison purposes with results obtained from 
floating-point diff-Pad\'e analysis of the 2000 term exact $\chit^{(5)}$ series.

The multiplicity of any zero $x=x_s$ of $\, P_{D_0}(x)$ is the number of linearly 
independent singular functions in the neighbourhood of $x_s$\footnote[8]{There is no
 difficulty in finding the roots of the associated indicial equation mod prime. In 
the case of  $\, \tilde{\chi}^{(5)}$ we are again confident that we have all the exact 
exponents and these are reported in Section~\ref{sec:analysis}.}.  
If we add to this list of multiplicities the number of independent singular functions 
at $\, x=\, 0$ and at $\, y\,=\, 1/x\, =\, 0$, then the maximum multiplicity is the 
minimum order of the ODE.  We have again verified that this agrees with $\, M_0$ in 
(\ref{30}), although there is a subtlety to determining the true multiplicities at 
$\, x=\, 0$ and $\, y=0$.  By our definition of $\, L_{MD}$ in (\ref{28a}) the indicial
 equations at these points are, $\, \sum_{m=0}^{M}\, a_{m0} \cdot p^m\,=\,\, 0$, and
$\, \sum_{m=0}^{M} \, a_{mD}\cdot (-p)^m \,=\,\, 0$,  respectively, and thus of degree 
$\, M$.  If there are positive integer roots then, even if we factor these indicial 
equations {\it modulo} a prime for several $\, L_{MD}$ and take the greatest common 
divisor, we can only be sure that the true multiplicity is less than or equal to the degree 
of the greatest common divisor.  The ambiguity at $\, x=\, \infty$ is easily resolved by 
analysing the transformed series $\, S(z)$ where $ x\,=\,\, z/(1-z/\alpha)$, thus mapping 
$x=\, \infty$ to $z\,=\, \,\alpha$.  In all our examples the situation at $\, x=0$ has 
been unambiguous but we do not see why this would be the case in general.

To interpret the constant $\, C$ in (\ref{30}) set $\, M\,=\,\, M_0$ and define 
$\Delta=\, \Delta_0$ in this case.  Then $\, D=\, (D_0-1)\, M_0\, -C +\, \Delta_0\, -1$.  
On the other hand we know that when $M\,=\,\, M_0$, the head polynomial of the  linear 
ODE factorizes into two polynomials, one of degree $\, D_0$ giving the true singularities 
of the linear ODE and one of degree $D_{app}$ whose zeros  are all apparent 
singularities.  Thus we can write $\,D\,=\,\,D_0\, +D_{app}$ which, combined with the 
solution $\,  D\,=\,\, (D_0-1)M_0\, -C\, +\Delta_0\, -1$, yields
\begin{eqnarray}
\label{(31)}
	C \,  =\, \,   (D_0-1)\cdot (M_0-1)-\,  D_{app} \, +\Delta_0\,  -2,	
\end{eqnarray}
giving a direct connection between $C$ and the apparent polynomial of the minimum order 
linear ODE.  We remark further that $D_{app}$ is related to a (true) singularity exponent
sum-rule.  In the case of  $\, \tilde{\chi}^{(5)}$ the exponents we have determined yield 
$\,  D_{app}\,=\,\, 1384$ and hence $\, C \, = \, \,(72-1)(33-1)\, -1384\, +\Delta_0\,-2\, 
= \,886\, +\Delta_0 \,\ge \,886$,  which is consistent with the observed $\,C=\, 900$. 
The details of this sum-rule calculation can be found in \ref{SumRule}.

The deviations $\,\Delta$ observed in Table~\ref{Ta:C5} are quite small and a
reasonable approximation to (\ref{30}) is obtained by setting $\Delta=\, 0$.  Then
(\ref{30}) is both a specification of $N,$ the minimum number of series terms needed 
to get a linear ODE, and a relationship between $\,M$ and $\,D$. We can use the latter 
to eliminate, say, $\,D$ and the former to find the minimum possible $N.$ The result is 
that one should be exploring the region around $\, (M+1)/(D+1)\,=\, \, M_0/D_0$ to 
obtain the minimum $N.$

In earlier work~\cite{ze-bo-ha-ma-05, ze-bo-ha-ma-05b} we observed a ``Russian-doll'' 
structure for the linear differential equations for $\tilde{\chi}^{(3)}$ and $\tilde{\chi}^{(4)}$ 
and a similar inherited Russian-doll structure on the $n$-particle contributions of the 
``diagonal susceptibility"\cite{Diag}.  We conjecture for arbitrary $\tilde{\chi}^{(n)}$ a 
``strong" Russian-doll structure for the linear differential operator for  $\tilde{\chi}^{(n)}$, 
which is to say that the linear differential operator for $\tilde{\chi}^{(n)}$ right-divides 
the linear differential operator for $\tilde{\chi}^{(n+2)}$.  We can now {\em verify this conjecture} 
on  $\, \tilde{\chi}^{(5)}$ and the results are shown in Table~\ref{Ta:ODEs}. 

\begin{table}[htdp]
\caption{\label{Ta:ODEs}
Summary of results for various series.  The equation for the size of the zero determinant 
is obtained from fits to data such as that shown in Table~\ref{Ta:C5} for 
$\tilde{\chi}^{(5)}$. The last five columns are the data for the case that the zero
determinant size is a minimum. A reduced difference in braces in column 7 appears in
those cases where a constant is a solution of the ODE and the matrix being tested  could 
have been taken as size $O \times (D+1)$ rather than $(O+1) \times (D+1).$  \\
The $\Phi_H^{(n)}$ series
are the model integrals~\cite{bo-ha-ma-ze-07b} (see (\ref{In})).
  }

\begin{center}
\begin{tabular}{|c|c|c|c|c|c|c|}\hline
    Series &  $ N=D_0\, M \, +M_0 \, D -C $&  $M$ &   $D$ &     $N_{MD}$&     $N$&   $\Delta$ \\ \hline 
\hline
    $\tilde{\chi}^{(1)}$ &$1\, M \, + \, 1\, D \, + 1$ &   1  &  1&  4&    3&  1 \\
    $\tilde{\chi}^{(2)}$& $1\, M \, + \, 2\, D \, + 2$ &  2  &  1&  6&    6&  0 \\
    $\tilde{\chi}^{(3)}$& $12\, M \, + \, 7\, D \, -40$ &  11  &  17&  216&    211&  5 \\
    $\tilde{\chi}^{(4)}$& $7\, M \, + \, 9\, D \, -36$ &  15  &  9&  160&   150&  10(0) \\
    $\tilde{\chi}^{(5)}$&  $72\, M \, + \, 33\, D \, -900$ &  56  &  129&  7410&   7389&  21 \\
    $L_7 \left(\tilde{\chi}^{(5)} \right)$ &  $60\, M \, + \, 26\, D \, +611$ &  54  &  131&  7260&   7257&  3 \\
    $6\tilde{\chi}^{(3)}-\tilde{\chi}^{(1)}$&  $12\, M \, + \, 6\, D \, -28$ &  10  &  17&  198&   194&  4 \\
    $6\tilde{\chi}^{(4)}-2\tilde{\chi}^{(2)}$&  $6\, M \, + \, 7\, D \, -17$ &  13  &  8&  126&   117&  9(0) \\
    $6\tilde{\chi}^{(5)}-3\tilde{\chi}^{(3)}$&  $68\, M \, + \, 30\, D \, -744$ &  52  &  120&  6413&  6392&  21 \\
    $L_2$&   $65\, M \, + \, 28\, D \, -526$ &  50  &  117&  6018&  6000&  18 \\
    $L_3$&   $64\, M \, + \, 27\, D \, -409$ &  49  &  117&  5900&  5886&  14 \\
    $\Phi_H^{(3)}$&  $10\, M \, + \, 5\, D \, -21$ &  8  &  13&  126&   124&  2 \\
    $\Phi_H^{(4)}$&  $5\, M \, + \, 6\, D \, -12$ &  9  &  6&  70&   69&  1 \\
    $\Phi_H^{(5)}$&  $45\, M \, + \, 17\, D \, -277$ &  28  &  80&  2349&   2343&  6 \\
    $\Phi_H^{(6)}$&  $26\, M \, + \, 27\, D \, -342$ &  48  &  39&  1960&   1959&  1 \\
    $\Phi_H^{(7)}$&  $145\, M \, + \, 49\, D \, -1943$ &  92  &  257&  23994&  23990&  4 \\ \hline
 \end{tabular}
\end{center}
\end{table}

Further, a stronger property amounts to saying that we actually have, in the
decomposition of the linear differential operator for $\tilde{\chi}^{(n+2)}$,
the linear differential operator for $\,\tilde{\chi}^{(n)}$ occuring as part of
a {\em direct sum}. Such a reduction was 
found~\cite{ze-bo-ha-ma-04,ze-bo-ha-ma-05,ze-bo-ha-ma-05b} for the combinations 
$ \,\, 6\cdot  \tilde{\chi}^{(n+2)}-n \cdot \tilde{\chi}^{(n)}$, $\, n=1$ or $2,$ 
and we now {\em verify this conjecture for the case} $n=3$.  Detailed results that 
we referred to earlier are in Table~\ref{Ta:C5}.  A summary of all the observed operator 
reductions appears in Table~\ref{Ta:ODEs}. 

Some operator reduction data in Table~\ref{Ta:ODEs} calls for explanation. Since the 
differential operator $\,L_7$ for $\tilde{\chi}^{(3)}$~\cite{ze-bo-ha-ma-05}, acting 
on the $\tilde{\chi}^{(5)}$ series gives an order 26 ODE (see sixth line in Table~\ref{Ta:ODEs}) 
and the minimal ODE for  $\,6\, \tilde{\chi}^{(5)}-3\, \tilde{\chi}^{(3)}$ is of order 30, 
i.e. less than 33 (see ninth line in Table~\ref{Ta:ODEs}), one can conclude that this order 30 
differential operator contains an order 4 differential operator occurring in the known $\,L_7$.
From this order 4 differential operator, we focus here on the differential operators of order 1, 
which because they have no apparent singularities, are most effective in reducing the number 
of series coefficients that need to be generated. There are two such order 1 
operators\footnote[8]{They already occurred~\cite{ze-bo-ha-ma-04} as solutions of $\, L_7$,
the differential operator for $\tilde{\chi}^{(3)}$.}; their solutions are
\begin{eqnarray}
\label{(32)}
S_1\, =\, \, w/(1-4\, w), \quad \quad 
S_2\, =  \,\,  w^2/((1-4\, w)\sqrt{1-16\, w^2}), 
\end{eqnarray}
We have also found an order 1 operator whose solution
\begin{eqnarray}
\label{(32b)}
  S_3\, =\,\,  w^2/(1-4\, w)^2 
\end{eqnarray}
is a solution of the order 30 differential operator for 
$\, 6 \,\tilde{\chi}^{(5)}-3 \,\tilde{\chi}^{(3)}$, but not a 
solution of the differential operator for $\tilde{\chi}^{(3)}$. 
Let us introduce the second order differential operator\footnote[2]{The differential 
operator $L_2$ was given as $O_1 \cdot N_1= T_1 \cdot L_1$ in eq.(7) of~\cite{ze-bo-ha-ma-05c}. 
It is thus the direct sum of the order 1 operators associated with $\, S_1$ and $\, S_2$.}
$L_2$ which simultaneously annihilates $S_1$ and $S_2$ and the order 3 differential
operator $L_3$ which annilates $S_3$ as well.
When these act on the series $\,6 \, \tilde{\chi}^{(5)}\, -3\tilde{\chi}^{(3)}$, the
reductions in $N$ are shown in Table~\ref{Ta:ODEs}
(tenth and eleventh lines, labelled respectively as $L_2$ and $L_3$).
The essential observation is that, although we 
found in our first computer runs that 6000 terms were not enough
to get the linear ODE mod $\, p_0$ for  $\, \tilde{\chi}^{(5)}$, this is now more than
adequate to get the linear ODE {\it modulo} $p$ for  $\, \tilde{\chi}^{(5)}$
for as many new primes $\, p$ as we wish to investigate.

\subsection{On the linear ODE for  $\tilde{\chi}^{(6)}$}

We do not have corresponding results for  $\, \tilde{\chi}^{(6)}$ and it seems likely that
obtaining long enough series in this case is beyond our presently available computing 
resources.  We base this on the following very crude correspondence: determining the 
exact linear ODE for  $\, \tilde{\chi}^{(3)}$ required a minimal series of about 220 terms. 
The exact ODE for $\, \tilde{\chi}^{(4)}$ required about 170 terms (in $\, x\, = \, w^2$).  
The ratio  $N_{\tilde{\chi}^{(4)}}/N_{\tilde{\chi}^{(3)}}\, \simeq \, 170/220 \, \simeq \, 0.77$.  
There is a similar ratio for  $\, \Phi_H^{(n)}$, integrals without the Fermionic factor 
introduced in~\cite{bo-ha-ma-ze-07b}. From  Table~\ref{Ta:ODEs}, we have 
$\, N_{\Phi^{(4)}}/N_{\Phi^{(3)}}\, \simeq \,  80/130  \, \simeq \, 0.62$ and
$\, N_{\Phi^{(6)}}/N_{\Phi^{(5)}}\, \, \simeq \, 2000/2400 \, \simeq \, 0.83$. 
A reasonable guess then might be  $N_{\tilde{\chi}^{(6)}}/N_{\tilde{\chi}^{(5)}}\, \simeq \, $ in
the range 0.9 to 1.1.  Although we generated 10000 terms for  $\, \tilde{\chi}^{(5)}$ only 
about 7500 were actually required.  Our guess is that about the same number will be 
required for  $\, \tilde{\chi}^{(6)}$.  Now our codes for  $\, \tilde{\chi}^{(5)}$ and  
$\, \tilde{\chi}^{(6)}$ are such that 7500 terms for  $\, \tilde{\chi}^{(6)}$ is roughly the 
equivalent in time to 15000 terms for  $\, \tilde{\chi}^{(5)}$ and this means a running time 
for  $\, \tilde{\chi}^{(6)}$ about $\, 1.5^4 \, \simeq \,  5$ times in excess of that taken 
for the 10000 terms of $\, \tilde{\chi}^{(5)}$.  This might be reduced somewhat if we rely 
on some direct sum assumptions about the linear ODE for  $\, \tilde{\chi}^{(6)}$ similar to 
those we have found work for  $\, \tilde{\chi}^{(5)}$.  On the other hand, a guess for a series 
length $N$ that is too small might leave us with no results whatsoever and thus we would 
probably want to err on the conservative side and require a calculation of  
$\, \tilde{\chi}^{(6)}$ with a run time cost as much as 10 times that for $\, \tilde{\chi}^{(5)}$.

\subsection{Singularities of the linear ODE for $\tilde{\chi}^{(5)}$ 
and Landau singularities \label{sec:C5sing} }

From the linear ODE for $\tilde{\chi}^{(5)}$ obtained modulo a prime, one can
easily reconstruct the singularity polynomials of the ODE as they appear
at the highest derivative. These polynomials read
\begin{eqnarray}
\label{singchi5}
&& w^{33} \cdot (1-4w)^{22} (1+4w)^{16} (1-w)^4 (1+2w)^4 (1+3w+4w^2)^4 \nonumber \\
&& (1+w)(1-3w+w^2)(1+2w-4w^2)(1 -w -3w^2 +4w^3) \nonumber \\
&& (1 +8w +20w^2 +15w^3 +4w^4)(1-7w+5w^2-4w^3) \nonumber \\
&& (1+4w+8w^2) (1-2w).
\end{eqnarray}
All these singularities, except $(1-2w)$, have been predicted by the model integrals
we introduced in~\cite{bo-ha-ma-ze-07b}. The models considered in 
\cite{bo-ha-ma-ze-07,bo-ha-ma-ze-07b} are  integral representations (one-dimensional 
and multidimensional) which belong to the ``Ising class''~\cite{Crandall}. These 
integrals are  holonomic and we obtained the linear ODE's of these sets of integrals 
through series expansions~\cite{bo-ha-ma-ze-07,bo-ha-ma-ze-07b}. In~\cite{bo-ha-ma-ze-07b} 
a detailed analysis of the multiple integrals $\Phi_H^{(n)}$ was performed. These 
$\, n$-fold integrals correspond to removing the Fermionic factor $(G^{(n)})^2$ in 
(\ref{chi3tild}), so that
\begin{eqnarray}
\label{In}
\Phi_H^{(n)}(w) \, \,= \,\,\, {\frac{1}{n!}}  \cdot 
\Bigl( \prod_{j=1}^{n-1}\int_0^{2\pi} {\frac{d\phi_j}{2\pi}} \Bigr)  
\Bigl( \prod_{j=1}^{n} y_j \Bigr)  \cdot  
 {\frac{1\,+\prod_{i=1}^{n}\, x_i}{1\,-\prod_{i=1}^{n}\, x_i}} .
\end{eqnarray}

We obtained (after eliminating the apparent singularities) the following polynomial factors 
for the head polynomial of the linear ODE's satisfied by the $\Phi_H^{(n)}$, expressed in 
terms of Chebyshev polynomials of the first and second kind~\cite{bo-ha-ma-ze-07b}: 
\begin{eqnarray}
\label{famil1corps}
&& T_{2p_1} \left( 1/2w +1 \right)\,  \,=\,\,\,\, \,\, 
 T_{n-2p_1-2p_2} \left( 1/2w-1 \right), \\
&& 0\, \le\, p_1 \,\le \,[n/2], \quad \quad 
 0 \,\,\le\,\, p_2\, \,\le\, \,[n/2]\,\,-p_1, \nonumber
\end{eqnarray}
and the polynomial arising from the elimination of $z$ in:
\begin{eqnarray}
\label{famil2corps}
&& T_{n_1}( z)\,\,  
- T_{n_2} \Bigl(  {\frac{4w-z}{1-4w\,z }} \Bigr) \,\, = \,\,\,  0, \nonumber  \\
&& T_{n_1} \left( {1 \over 2w}- z \right)\,\, \,
 - T_{n_2} \Bigl( {1 \over 2w}- {\frac{4w-z}{1-4w\,z }} \Bigr) 
\,\, = \,\, \, 0, \nonumber \\
&& U_{n_2-1} (z) \cdot  
U_{n_1-1}\Bigl( {1 \over 2w}- {\frac{4w-z}{1-4w\,z }} \Bigr)\nonumber \\
&& \qquad  \qquad 
- U_{n_2-1}\left( {1 \over 2w}- z \right)\cdot 
U_{n_1-1} \Bigl(  {\frac{4w-z}{1-4w\,z }} \Bigr) \,\,=\,\,\, 0,  \\
&&  n_1 \,=\, p_1, \qquad \qquad \qquad  n_2 \,=\,\, n\,-p_1\,-2p_2,  \nonumber   \\
&&  0 \,\,\le \,\,p_1\, \le\, n, \,\, \qquad \qquad 
0\,\, \le\,\, p_2 \,\, \le\, \,[(n-p_1)/2].  \nonumber 
\end{eqnarray}

Our motivation was to obtain ``good candidates'' for the factors in the head polynomial 
of the linear ODE of $\, \tilde{\chi}^{(n)}$, expecting that the Fermionic factor 
$(G^{(n)})^2$ may not introduce ``too many additional singularities''.

There is also the possibility of singularity cancellation. Indeed, in the even 
simpler integral in which the $R^{(n)}$ factor (\ref{Rn}) is dropped from (\ref{In}), 
all the singularities predicted in \cite{nickel-05} are seen; these are the 
singularities (\ref{famil1corps}) but with the even $2p_1$ replaced by an integer 
that can also be odd. Thus the inclusion of $R^{(n)}$ has eliminated a whole class of 
singularities.  The presence of $\left(G^{(n)}\right)^2$, at least in $\chit^{(5)}$, has 
been much less dramatic and led only to the one extra factor $(1-2w)$ in (\ref{singchi5}).

To get a better understanding of the effect of  $\left(G^{(n)}\right)^2$ we revisit the 
Landau singularity approach in Appendices B through E but performing the calculations 
on the $(n-1)$-fold integrals (\ref{chi3tild}) rather than the original 
$(2n-2)$-fold integrals \cite{wu-mc-tr-ba-76} that was the basis of calculations in 
\cite{bo-ha-ma-ze-07b}.  The new approach detailed in \ref{appendixBernie} confirms our 
previous Landau singularity calculations \cite{bo-ha-ma-ze-07,bo-ha-ma-ze-07b},
the difference being in the number of integration variables considered. The original 
representation for $\tilde{\chi}^{(n)}$  given in~\cite{wu-mc-tr-ba-76} is an integral 
over two sets of $(n-1)$ independent phases $\phi_i$, $\zeta_j$ with an integrand that 
is symmetric under the interchange of these sets. Integrating out one set to arrive 
at (\ref{chi3tild}) has obviously broken this symmetry but a vestige of it remains, 
such that for every combination of the $ \, \zeta_j$, $\phi_i$ leading to a singularity 
there is another set $ \, \phi_j$, $ \, \zeta_i$ obtained by 
$ \, \zeta \,  \leftrightarrow \,  \phi$ interchange that leads to the same singularity.
In addition another symmetry arises such that, for a given $\phi_i$, $\, \zeta_i$ 
combination, the Landau conditions allow the reversed combination $-\phi_i$, $-\zeta_i$
{\em explaining the $\, n \, \, \rightarrow \,\,  n-2m$ replacement singularities}
seen to occur in the analysis of~\cite{bo-ha-ma-ze-07b}. Our analysis proves
that for the $\Phi_H^{(n)}$ there are no singularities other than those given by 
(\ref{famil1corps}) and (\ref{famil2corps}). We also show that $\left(G^{(n)}\right)^2$ neither 
reintroduces the singularities cancelled by the $R^{(n)}$ nor leads to further cancellation 
and thus the singularities of $\Phi_H^{(n)}$ are included in the singularities of 
$\chit^{(n)}$.  The $\left(G^{(n)}\right)^2$ factor has a dramatic effect on the 
exponents of the singularities and a power counting argument is given in \ref{app:Exponents} 
to predict the exponents at all singularities.  Those calculations are not intended to cover 
all possible contingencies and while many predicted exponents are observed there remain 
a number of ``anomalous'' cases. 

Our analysis confirms the fact that the Landau singularities of $\chit^{(n)}$, at least those 
given by $\Phi_H^{(n)}$, are included in the set of singularities of $\chit^{(n+2k)}$, $k\geq 1$.  
This is exactly what we have found for the linear ODE for $\chit^{(5)}$. The first line in 
(\ref{singchi5}) corresponds to the singularities occurring in the linear ODE for $\chit^{(3)}$.  
This is a necessary condition for $\chit^{(3)}$ to be embedded in $\chit^{(5)}$.  However this 
comes with an important caveat.  We prove in \ref{appendixBernie} that none of the embedded 
singularities can lie on the $|s|=1$ boundary of the principal disc of the $\chit^{(n)}$ function 
defined by the integral (\ref{chi3tild}).  We also show that of all the different classes of 
singularities, only those we call Case 2 irreducible singularities in \ref{appendixBernie} lie on 
$|s|=1$ of the principal disc.  These correspond to the singularities derived in 
\cite{nickel-99,nickel-00} and are elsewhere called the Nickelian or circle singularities.  
The importance of this lies in the fact that we cannot expect cancellation between different 
singularities on the principal disc and thus the elimination of $|s|=1$ as a natural boundary.

While some aspects of the effect of $\left(G^{(n)}\right)^2$ have been determined, a complete Landau 
singularity analysis of $\chit^{(n)}$ has not been done; in particular we cannot definitively 
state whether the $(1-2w)$ factor in (\ref{singchi5}) does or does not identify with a Landau 
singularity of $\chit^{(5)}$.  However, a toy integral intermediate between $\Phi_H^{(n)}$ and 
$\chit^{(n)}$ is discussed in \ref{app:ToyModel} and provides an example in which it can be shown 
that the ODE describing an integral has more singularities than the integral.  The analogy to 
$\chit^{(n)}$ is sufficiently close that we believe it is likely that $(1-2w)=0$ is not a singularity 
of  $\chit^{(5)}$.  Clearly it is of interest to know whether there are such additional singularities 
in the ODE for  $\chit^{(n)}$ for larger $n$ and whether they are also on $|s|=1$.

\ref{Singularities} lists, for the first few $n$ of $\Phi_H^{(n)}$, the singularities corresponding 
to (\ref{famil1corps},\ref{famil2corps}) and also derived in \ref{appendixBernie}.  
In the next section we will see whether 
these singularities appear in the analysis of our long series with exact coefficients.

\section{Diff-Pad\'e analysis of the long series for $\tilde{\chi}^{(5)}$
and $\tilde{\chi}^{(6)}$ \label{sec:analysis} }

We present, in this section, a diff-Pad\'e analysis\footnote[5]{We call diff-Pad\'e 
the type of analysis detailed in~\cite{ze-bo-ha-ma-05}. For a given number of terms 
$N$ of the series, {\it there is} a linear ODE of order $q$ that reproduces the 
first $\, N-q$ terms but may fail for subsequent coefficients. The same analysis, called 
``method of differential approximants'' was described in \cite{Rehr}.} of our long 2000 
coefficient series (these are the actual series coefficients, not the coefficients
{\it modulo} a prime, for which we have longer series) in floating point form, in
order to obtain the singularities (together with their associated exponents) that 
should occur in the corresponding linear ODE's.

\subsection{Singularities and  indicial exponents 
of the linear ODE of $\tilde{\chi}^{(5)}$}
\label{diff-Pade5}

We begin with the series for $\tilde{\chi}^{(5)}$, for which we have obtained
the linear ODE (modulo a prime) and have recognized all the singularities as 
given in (\ref{singchi5}). Our calculations provides a check on whether or not a
diff-Pad\'e analysis on a series too short to find the exact ODE can nevertheless
yield enough information to locate the singularities precisely and
determine the associated local exponents accurately.

A diff-Pad\'e analysis with just 400 coefficients (using approximating linear 
ODE's of order ten or eleven) already confirms, with $\, 36$ digit accuracy, 
the occurrence of the singularities given by the roots of the factors
\begin{eqnarray}
\label{11}
\left( 1+w \right)   \left( 1-3\,w+{w}^{2} \right)
 \left( 1+2\,w-4\,{w}^{2} \right),
\end{eqnarray}
which are the Nickelian singularities labelled as $P(^{2} \, 5)$,
and the roots of the factors
\begin{eqnarray}
\label{12}
\left( 1-7\,w+5\,{w}^{2}-4\,{w}^{3} \right) 
 \left( 1+8\,w+20\,{w}^{2}+15\,{w}^{3}+4\,{w}^{4} \right),
\end{eqnarray}
with twelve digit accuracy.
The roots of the factors
\begin{eqnarray}
\label{13}
\left( 1-w \right)  (1\, +2\,w) 
(1-w-3\,{w}^{2}+4\,{w}^{3}) 
\end{eqnarray}
are confirmed with three or four digit accuracy, and, finally,
the roots of the factor $\, 1+4\,w+8\,{w}^{2}$
are obtained with just one digit accuracy.
The roots of $\, 1+3\,w+4\,{w}^{2}$ do not yet appear in the analysis using
400 coefficients.

A generalized diff-Pad\'e analysis is conducted by steadily increasing the order of 
the linear ODE and the degree of the polynomials, while looking for the roots which 
stabilise with increasing accuracy. For instance, using 1250 coefficients, the roots 
(\ref{12}) now appear with fifteen digit accuracy, and similarly for the other roots, 
the accuracy increases. While the roots of the factor $\, 1+3\,w+4\,{w}^{2}$ are not 
yet observed, we see the emergence of a {\em new polynomial} not found among the set
of singularities of $\, \Phi^{(5)}_H$: the singularity $\, w\, = \, +1/2$ 
{\em is actually observed with twelve digit accuracy}.

 Further increasing the degrees and the order of the linear ODE to {\em fully utilise 
the 1980 coefficients at our disposal}, the accuracy  is dramatically improved.
The roots  (\ref{12}) {\em are now confirmed up to $\, 67$ digits}, the  two 
complex roots of polynomial $\, 1-w-3\,{w}^{2}+4\,{w}^{3}$ {\em are observed with
more than $\, 51$ correct digits}, and the real root is  observed with more than 7
correct digits. The roots of the factor $\, 1+4\,w+8\,{w}^{2}$ {\em are now 
observed with 17 digit accuracy}. The roots of $\left( 1-w \right) \left( 1+2\,w \right)$
are seen with 5 correct digits for $\, w=1$ and 7 correct digits for  $\, w=\, -1/2$.
Finally, the roots of  $\, 1+3\,w+4\,{w}^{2}$ {\em are seen with 4 digit accuracy},
and the ``new'' $( 1\, -2\, w)$ factor {\em is observed with $\, 27$ correct digits}. 
We summarise this discussion in Table \ref{Ta:C5sing}.

\begin{table}[htdp]
\caption{This table shows the number of significant digits found, in the case
of each singularity of $\tilde{\chi}^{(5)}$, from a differential approximant (diff-Pad\'e) 
analysis of series of length, respectively, 400, 1250 and 1980 terms. \label{Ta:C5sing} }
\begin{center}
\begin{tabular}{|c|c|c|c|c|}\hline
Label of singularity & Associated polynomial & 400 &  1250 &  1980 \\
 \hline
 $P(^{2} \, 5)$      & $ (1+w)$                   &36    &     & \\
$P(^{2} \, 5) $      & $(1+2w-4w^2)$              &36    &     & \\
$P(^{2} \, 5) $      & $(1-3w+w^2)$               &36    &     &  \\
$P(^{3}\, 5_{1,4})$   & $(1+8w+20w^2+15w^3+4w^4)$ &12    &15   &67 \\
$P(^{3}\, 5_{3,2})$   & $(1-7w+5w^2-4w^3)$        &12    &15   &67\\
$P(^{4}\, 5_{4,1})$    & $(1-w-3w^2+4w^3)$        &4     &15   &51\\
$P(^{4}\, 5_{3,2})$    & $(1+4w+8w^2)$            &-     &8    &17\\
$P(^{5} \, 5/\, ^{2} \, 3)$  & $(1+2w)$              &3     &5    &7\\
$P(^{5} \, 5/\, ^{2} \, 3)$  & $(1-w)$              &3     &3    &5\\
$P(^{5} \, 5/\, ^{3}\, 3_{1,2})$  & $(1+3w+4w^2)$   &-     &-    &4\\
Unknown & $(1-2w)$                                &-     &12   &27\\
\hline
 \end{tabular}
\end{center}
\end{table}

These diff-Pad\'e calculations are, in fact, sufficiently robust {\em to allow us 
to predict the minimum multiplicity of some singularities}. When a given singularity 
(say $w=1$) is observed, it is put in the head polynomial for a second run 
using more coefficients. If this singularity appears again it must be a double root 
(one exact root $w=1$ and a second root $w \simeq 1$ observed numerically with 
sufficient accuracy). Next the factor $(1-w)^2$ is included in the head polynomial 
for another run and so on until no further occurrences of the given root are found.

With the number of series coefficients at hand we find for the head polynomial
of the ODE the following factors occurring {\em with the indicated multiplicity}:
\begin{eqnarray}
(1\, +2\, w)^4 \cdot (1\, -w)^4 \cdot (1\, +3w\, +4w^2)^2, \nonumber 
\end{eqnarray}
while all other roots (except $1\, -16\, w^2$) occur
{\em with multiplicity one.} We are very close to the exact multiplicities
of the exact linear ODE (see (\ref{singchi5})).

This knowledge can then be used, in a kind of converging procedure, to improve the 
accuracy of our diff-Pad\'e calculations. Having a totally unambiguous location 
of the singularities and {\em a minimum value} for the multiplicities, we revisit 
the diff-Pad\'e calculations using  the following Ansatz for the polynomial in front 
of the $i^{th}$  derivative
\begin{eqnarray}
\label{ansatz}
&& w^{i-1}\cdot \left( 1-16{w}^{2} \right)^{i-2}
 \cdot \Bigl( \left( 1+2w \right)
\cdot \left( 1-w \right) \Bigr)^{i-q+4} \nonumber \\
&&\qquad  \times \left( 1+3w+4{w}^{2} \right)^{i-q+2}\cdot 
 P(w)^{i-q+1} \cdot Q_i(w) 
\end{eqnarray}
with $P(w)$ containing the other singularities with multiplicity one:
\begin{eqnarray}
&& P(w)\,= \, \, \left( 1+w \right) 
\left( 1-3w+{w}^{2} \right)  \left( 1+2w-4{w}^{2} \right)
   \left( 1+4w+8{w}^{2} \right)
    \nonumber \\
&& \qquad \times \left( 1-2w \right) \left( 1-7w+5{w}^{2}-4{w}^{3} \right) 
 \left( 1-w-3{w}^{2}+4{w}^{3} \right) \,
    \nonumber \\
&& \qquad \times \left( 1+8w+20{w}^{2}+15{w}^{3}+4{w}^{4} \right)  
 \nonumber 
\end{eqnarray}
Here $q$ is the order of the linear ODE and the index  $\,i\,=\,0\,,\,  \cdots, \, q$ 
denotes the successive derivatives in the linear ODE (the actual exponents in the 
Ansatz are zero when a negative value is encountered). The $Q_i(w)$ are unknown 
polynomials, and $Q_q(w)$ at the highest derivative is included in order to handle 
the expected apparent singularities.

The Ansatz (\ref{ansatz}) is then used in a diff-Pad\'e analysis to obtain in floating 
point form, at each singularity, the associated critical exponents as roots of 
the indicial equation.
Our findings are listed in Table~\ref{Ta:C5expo}.
Let us explain how we display our results. Consider for instance the singularity 
$\, 1\, -w\, =\, 0$. The successive roots of the indicial equation appear as the
integers $0, \, 1, \, \cdots, \, q-5$ (which is by construction) together
with  the integers $\, 2,\,  3,\,  3,\,  4$. The results for this case will be 
displayed as $2, 3^2, 4$, but note that the roots 2 and 4 both appear twice, and 
the root 3 appears three times. 
%
Recall that the roots of the indicial equations appear in floating point form
and {\em we recognize these roots as being integers or half integers}. The accuracy 
of the indicial exponents can be as low as three correct
digits as is the case at the singularities given by $1+3\,w+4\,{w}^{2} =\, \,  0$.

\begin{table}[htdp]
\caption{Singularity exponent list for ${\tilde{\chi}}^{(5)},$ based on the exact
({\it modulo} a prime) linear ODE compared to those found by the (floating-point)
diff-Pad\'e analysis. The mod prime exponent list is followed by a number in
braces, which is the sum  in (\ref{(B.7)}) of Appendix A.
A final column gives the expected ``normal"
exponents using (E.1).  \label{Ta:C5expo}}

\begin{center}\small
\begin{tabular}{|c|c|c|c|c|}\hline
 Singularity& Exponents (from &Exponents&Exponents \\
 Polynomial& mod prime analysis)& (diff-Pad\'e)& (E.1)\\
\hline \hline
            $w^{33}$& $1^5,2^4,3^4,4^3,5^3,6^3,7^2,8^2,9^2,$& &--\\
   &$10,12^2,15,25,(192)$&& \\
           $(1-4w)^{22}$& $-2,-7/4,-3/2,-5/4,-1^3,-1/2,0^4,$& $-3/2,-1,0^4$& -1\\
&$1/2,1^2,2^2,3,4,5,6,7,(549/2)$&&\\
           $(1+4w)^{16}$&$ -1,-1/2,0^4,1/2,1^2,3/2,2^2,3^2,$&$ -1/2,0^4,1^2$& 0\\
&$4,5,(315/2)$&&\\
   $1/w^{19}$&$ 0^3,1^4,2^2,3^3,4^2,5^2,6,7,8,(56+91)$& $0^3,1^4,2$& --\\
$ (1+2w)^4$&$ 2,5/2,3^2,(41/2)$& $2,5/2,3^2$&$ 3$\\
$ (1-w)^4$&$ 2,3^2,4,(22)$&$ 2,3^2,4$& 3\\
$ (1+3w+4w^2)^4$&$ 0,1^2,2,(14)\times2$&$ 0$& 1\\
$ (1+w)$& $11,(12)$& 11& 11\\
 $(1+2w-4w^2) $&$11,(12)\times 2$ &11& 11\\
$ (1-3w+w^2)$&$ 11,(12)\times2$& 11 &11\\
$ (1+8w+20w^2+15w^3+4w^4)$&$ 7,(8)\time4$& 7& 7\\
$ (1-7w+5w^2-4w^3)$&$ 5,(6)\times3$&5& 5\\
$ (1-w-3w^2+4w^3)$&$ 7,(8)\times 3$& 7& 7\\
$ (1+4w+8w^2)$& $5,(6)\times2$& 5& 5\\
           $(1-2w)$& $7/2,(9/2)$&$ 7/2$&---\\ \hline
 \end{tabular}
\end{center}
\end{table}

Since  we have obtained the exact linear ODE  for $\tilde{\chi}^{(5)}$ this case
provides a valuable test of our diff-Pad\'e analysis. Furthermore, the Landau 
singularity analysis has been extended, in \ref{app:Exponents}, to include a power 
counting argument for the exponents at each singularity.
All these results, the exponents from a diff-Pad\'e analysis on 2000 terms,
the exponents from the (modulo a prime) linear ODE, and those derived
in \ref{app:Exponents} are displayed together in Table \ref{Ta:C5expo}.

Even if we did not know the exact (modulo a prime) linear ODE of $\tilde{\chi}^{(5)}$, 
our diff-Pad\'e analysis can provide accurate information about the exact ODE.
We note that the indicial exponents found above are accurate enough that we can be 
confident in their exact values, but the set of exponents may be incomplete. 
Let us consider the case $1-2w=\, 0$ to show what we mean. This factor was taken with 
a multiplicity of one in the head polynomial and the roots of the indicial 
equation show up as $ 0,\,  1, \, \cdots,\,  q-2,\,  7/2$. Assume that in the exact 
linear ODE, the multiplicity is two. Then, the roots of the indicial equation will be
$ 0, \, 1, \, \cdots, \, q-3,\,  7/2$ plus an unknown exponent. If we had more series 
coefficients, with a further increase of the order and the degrees, this unknown indicial 
exponent may be obtained as another half integer, the exponent 7/2 or an integer.
The dominant singular behavior at $w=\, 1/2$ would change accordingly.

The comparison with the exact ODE results shows that the diff-Pad\'e analysis 
on only 2000 terms (which are insufficient to encode the linear ODE for 
$\tilde{\chi}^{(5)}$) is able to correctly give all the local exponents for the 
singularities together with the correct multiplicity. The singularity polynomial
$(1+3w+4w^2)$ was used in the head polynomial of the linear ODE
with a multiplicity of two instead of the correct multiplicity four.
One should then obtain two local exponents.
The missing local exponent was obtained as 0.87 instead of 1.

These results give us considerable confidence that our numerical
analysis of $\tilde{\chi}^{(6)}$ in the following subsection,
and of higher order susceptibility components subsequently,
are completely correct.

\subsection{Singularities and indicial exponents of the 
linear ODE of $\tilde{\chi}^{(6)}$}
\label{diff-Pade6}

Similar calculations to those detailed in the previous section for $\, \tilde{\chi}^{(5)}$
have been performed for $\tilde{\chi}^{(6)}$ in the variable $x=\, w^2$.

In a diff-Pad\'e analysis, increasing the order of the linear ODE and the
degree of the polynomials, the singularities predicted by the $\Phi_H^{(6)}$
model (labelled as $P(^{2} \, 6) P(^{3}\, 6_{4,2}) P(^{4}\, 6_{5,1})$
and $P(^{2} \, 4)$) are obtained with increasing accuracy.

We should note that the additional singularity $\, 1\, -2w=\, 0$, occurring for
$\tilde{\chi}^{(5)}$, was seen in our diff-Pad\'e analysis {\em before} we
obtained the exact ({\it modulo} a prime) linear ODE. Let us detail for $\tilde{\chi}^{(6)}$ 
how our numerical procedure proceeds. First we check whether the ``candidate" 
singularities appear as roots of the head polynomials,  and we also check whether 
some of the other roots stabilize as the number of terms and the order of 
the linear ODE increases. If so this root is a true singularity.
Next the well confirmed ``candidate" singularities are put into
the head polynomial and another run is carried out with more terms to
confirm the ``new singularity".

Table \ref{Ta:C6sing} shows the results of three specific runs. The third column 
shows results from a run using 387 terms with order 12 ODE.
An unknown (with respect to our candidates) singularity is seen to stabilize
around a value consistent with $\, 1-8x=0$ to an accuracy of 3 digits.
Keeping the same number of terms but increasing the order to 16 this root
would change if it were a root of the apparent polynomial.
The results in the fourth column show that the new unknown singularity
remains and is seen with 4 digit accuracy with respect to its exact value.
Finally all the other singularities are put into the head polynomial and
the fifth column shows the results for an order 31 analysis using 997 terms.
These calculations show the {\em  existence of a new singularity at $\, x\, = \, 1/8$}
and thus arising from the factor $1-8w^2=\, 0$ in the head polynomial of the true ODE.

\begin{table}[htdp]
\caption{This table shows the number of significant digits found, in the case
of each singularity of $\tilde{\chi}^{(6)}$, from a diff-Pad\'e  analysis ($x\, = \, \, w^2$). 
\label{Ta:C6sing}}
\begin{center}
\begin{tabular}{|c|c|c|c|c|}\hline
Singularity & Singularity & 387 terms, & 387 terms, & 997 terms,\\
Label       & Polynomial  & order 12   & order 16   & order 31 \\ \hline
$P(^26)$        &  $1-x$      & 28 & 29 & \\
$P(^56/^24)$    &  $1-4x$     & 30 & 30 &\\
$P(^26)$        &  $1-9x$     & 30 & 30 & \\
$P(^36_{4,2})$  &  $1-25x$    & 13 & 14 &\\
$P(^36_{4,2})$  &  $1-x+16x^2$   & 8  & 10 & \\
$P(^46_{5,1})$  &  $1-10x+29x^2$ & 10  & 12 & \\
Unknown         &  $1-8x$        & 3  & 4 & 26\\ \hline
 \end{tabular}
\end{center}
\end{table}

Various runs (with linear ODE's ranging in order from $\, 26$ to $\, 31$) show that 
the factor $(1\, -4\, x)$ occurs with a power five which leads us to the following Ansatz:
\begin{eqnarray}
&&{x}^{i-1} \cdot \left( 1-16\,x \right)^{i-2} \cdot 
\left( 1-4\,x \right)^{i-q+5} \cdot  P(x)^{i-q+1}\cdot  Q_i(x)
\end{eqnarray}
for the polynomials in front of the $\, i^{th}$ derivative  in the linear 
ODE where $\, P(x)$ reads: 
\begin{eqnarray}
&&P(x) \, = \, \,\, \, (1-x)  \, (1-9\,x)
  \, (1-25\,x) 
 \left( 1-x-16\,{x}^{2} \right)  \nonumber \\
&& \qquad \times \, (1-10\,x+29\,{x}^{2})  \, (1-8\,x).
\nonumber 
\end{eqnarray}
The singularities and corresponding exponents for $\tilde{\chi}^{(6)}$
are summarised in Table \ref{Ta:C6expo}, and we also show the exponents
predicted by the local analysis given in \ref{app:Exponents}.

\begin{table}[htdp]
\caption{Singularity exponent list for $\tilde{\chi}^{(6)}$ found by the
(floating-point) diff-Pad\'e analysis together with the expected ``normal''
exponents (E.1). \label{Ta:C6expo}}
\begin{center}
\begin{tabular}{|c|c|c|c|}\hline

Singularity & Singularity& Exponents &Exponents\\
Label& Polynomial& (diff-Pad\'e)& (E.1)\\ \hline
&  $x$  &$ 0, -1^2, -1/2$&  ---\\ 
&  $1-16x$&$ -3/2, -1, 0^5, 1$ & -1\\
&  $1/x$ &$ -1^2, 0^2, -1/2^2, 1/2^6$ &--\\
$P(^26)$& $1-x$ & 33/2 & 33/2 \\
$P(^56/^24)$& $ 1-4x$ & $11/2, 13/2^2, 15/2, 33/2 $& 13/2\\
$P(^26)$& $ 1-9x$ & 33/2 & 33/2 \\
$P(^36_{4,2})$& $ 1-25x$ & 17/2 & 17/2\\
$P(^36_{4,2})$&  $ 1-x+16x^2$ & 17/2 & 17/2 \\
$P(^46_{5,1})$&  $ 1-10x+29x^2$ & 23/2 & 23/2\\
Unknown & $1-8x$ & 7 &--\\ \hline
 \end{tabular}
\end{center}
\end{table}

Again we see that with some 1600 terms (which are insufficient to
encode the linear ODE for $\tilde{\chi}^{(6)}$) the diff-Pad\'e analysis
is able to give the correct exponents in agreement with \ref{app:Exponents}.
Depending on the multiplicity of the singularity polynomial put into the
linear ODE, the exponents may be incomplete. Again we find rational exponents 
for {\em all} the singularities of the linear ODE for $\, \tilde{\chi}^{(6)}$.

From our diff-Pad\'e calculations we thus have confirmation that the singularities 
encoded in the head polynomial of the as yet unknown linear ODE for $\tilde{\chi}^{(6)}$ 
are (at least) the ones of $\Phi^{(6)}_H$ {\em together with the roots of a new polynomial} 
$\, 1\, -8\,w^2$. Note that these $\, 1\, -8\,w^2=0$ {\em additional 
singularities lie on the unit circle} $| s | =\, 1$.
At this stage, and since all the series coefficients of $\tilde{\chi}^{(6)}$
have been used, one may ask if there are other singularities like
$w^2=\, 1/8$ ``still to be discovered''?
Without the exact linear ODE of $\tilde{\chi}^{(6)}$ we cannot give a definitive
answer to this question.

\section{Diff-Pad\'e analysis for $\tilde{\chi}^{(n)}$, $n \ge 7$ \label{sec:DPn7}}

It is known~\cite{Orrick} that the first non-zero coefficients in $\tilde{\chi}^{(n)}$ 
appear at order $n^2$ in $w$. The high- or low-temperature  series for $\tilde{\chi}$ 
up to $N=2000$ then contains contributions from all the odd, respectively even,
$\tilde{\chi}^{(n)}$ up to $n \,= \, \sqrt{N}$. 

Since we know the first 2000 series coefficients for $\tilde{\chi}^{(5)}$ 
and the first 1630 coefficients for $\tilde{\chi}^{(6)}$ as well as the series for 
$\tilde{\chi}^{(n)}$, $n \le 4$,  up to an arbitrary number of coefficients,
one may ask whether the total $\tilde{\chi}$ with these lower $\tilde{\chi}^{(n)}$ terms 
removed, can yield any information about the singularities that should occur
in the linear ODE of $\tilde{\chi}^{(n)}$, $n \,\ge\, 7$? As far as $\tilde{\chi}^{(5)}$ and 
$\tilde{\chi}^{(6)}$ are concerned and in view of the ``limited" analysis done here for 
$\tilde{\chi}^{(n)},\, n \geq 7$, our conclusion is that there are no new singularities that 
are not in the ``known" set (i.e. irrespective of the index $n$).

\subsection{High temperature analysis}
\label{diff-Pade2}

The diff-Pad\'e analysis of the long high-temperature series
\begin{eqnarray}
\label{chi7plus}
\tilde{\chi}_H\, - \tilde{\chi}^{(1)}\,-\tilde{\chi}^{(3)} \,
-\tilde{\chi}^{(5)}\, =\, \,\,
128\,w^{49} \,+ 25088 \,w^{51}\, + \cdots 
\end{eqnarray}
amounts to looking for the singularities of a ``linear ODE" approximation to the
(infinite non-holonomic)  sum,
$\,\tilde{\chi}^{(7)}\,+\tilde{\chi}^{(9)}\,+\tilde{\chi}^{(11)}\, + \cdots \,$,
as roots of the head polynomial of a given diff-Pad\'e approximant. These singularities 
occur grouped together (that is to say we find several singularities simultaneously) and 
each singularity can be attributed to a given $\tilde{\chi}^{(n)}$ according to the 
polynomials given in \ref{Singularities}, i.e. arising from the $\Phi_H^{(n)}$ model.

With an order eight linear ODE and 312 terms, the roots of the following polynomials
(besides $\, 1\, -16 w^2\, =\, 0$) are recognized:
\begin{eqnarray}
&&P_1:\,  1 + 2w - 8w^2 - 8w^3\, =\, \,0, \nonumber \\
&&P_2:\,  1 + 2w - w^2 - w^3\, =\,\, 0, \nonumber \\
&&P_3:\,  1 - 5w + 6w^2 - w^3\, =\,\, 0. \nonumber 
\end{eqnarray}
These polynomials can be identified with all the Nickelian singularities corresponding 
to $\tilde{\chi}^{(7)}$. The roots of the polynomial $P_1$ are obtained with 5, 10 and 12
correct digits, while two roots of the polynomial $\, P_2$ are obtained with 3 and 9 correct 
digits.

A summary of various runs using  500, 900 and 1956 terms is given in Table \ref{Ta:CH}, 
where we display the number of significant digits found, in the case of each singularity 
of the partial high-temperature susceptibility (\ref{chi7plus}). The Ansatz for the linear 
ODE search is to include only the factor $(1 - 16w^2)^i$,  where $i$ is a positive integer 
discussed in the previous section. The first block of singularities correspond to
Nickelian singularities, given as Case 2 in \ref{Singularities}.

\begin{table}[htdp]
\caption{\label{Ta:CH} 
This table shows the number of significant digits found for each singularity of the 
partial high-temperature susceptibility (\ref{chi7plus}) using a diff-Pad\'e analysis.
The singularities equal those predicted from  $\Phi_H^{(n)}$ as occurring in 
$\tilde\chi^{(n)}$. The second column gives the polynomial factor, the zeros of which 
corresponds to the singularity locations. Subsequent pairs of columns give, firstly,
the number of significant digits found for that singularity, and secondly the order 
of the linear ODE for which this singularity was found.}

\begin{center}
\begin{tabular}{|c|c|c|c|c|c|c|c|}\hline
 & &\multicolumn{2}{c|}{500 terms}&\multicolumn{2}{c|}{900 terms}&\multicolumn{2}{c|}{1956 terms}\\ \hline
$n$ & polynomial & Digits & Order & Digits & Order & Digits & Order  \\ \hline
7& $1-5w+6w^2-w^3$&18&12&33&19&50&18\\
7& $1+2w-w^2-w^3$&16&11 &28&15&58&18\\
7& $1+2w-8w^2-8w^3$&18&11&34&15&58&18\\ \hline
9&  $1-w$ &-& &-& &15&14\\
9&  $1+2w$ &-& &6&13&30&16\\  
9& $1+3w-w^2$&5&11&16&15&45&15\\
9& $1-6w+9w^2-w^3$&7&12&20&13&45&18\\
9&$1-3w^2-w^3$&-& &8&15&30&18\\ 
9& $1-12w^2+8w^3$&-& &14&14&40&20\\  \hline
11& $1-9w+28w^2-35w^3+15w^4-w^5$&-& &7&13&30&18\\
11& $1+2w-5w^2-2w^3+4w^4-w^5$&- & &-& &23&17\\
11& $1+2w-16w^2-24w^3+48w^4+32w^5$&-& &-& &20&20\\ \hline
13& $1-11w+ \cdots + w^6$&-& &-& &16&16\\ 
13& $1+2w-20w^2+ \cdots -64 w^6$&-& &-& &7&18\\ \hline 
15& $1+2w-4w^2$&-&&-& &5&14\\
15& $1-9w + \cdots +w^4$&-& &-& &5&14\\ \hline \hline \hline
7&  $1+12w+54w^2+\cdots +4w^6$ &-& &6&13&12&17 \\
7& $1-3w-10w^2 +\cdots -16w^8$ &-& &-& &4& 12\\ \hline
9&  $1+16w+104w^2 +\cdots +4w^8$ &-& &-&&7&17 \\  \hline \hline \hline

 \end{tabular}
\end{center}
\end{table}

Again we have confirmation that some of the singularities of the linear
ODE of $\Phi_H^{(n)}$ are {\em actually singularities of the linear ODE} of the 
$\tilde{\chi}^{(n)}$\footnote[2]{Note however that without an ``exact" series 
for each $\, \tilde{\chi}^{(n)}$, one cannot safely attribute, for instance, 
the non-Nickelian singularities polynomial 
$$(1+16w+104w^2+352w^3+660w^4+672w^5+336w^6+63w^7+4w^8)$$ to $\, \tilde{\chi}^{(9)}$.
These singularities may well come from $\tilde{\chi}^{(7)}$, but not be predicted
by $\Phi_H^{(7)}$.}.
From Table~\ref{Ta:CH} we note that the first singularity polynomials
to appear are the Nickelian ones. All the Nickelian singularities for
$n=7$ and $n=9$ are confirmed. Of the four singularity polynomials for $n=11$,
one is missing and among the five singularity polynomials for $n=13,$
three are missing. We also note that once the Nickelian singularities for
a given $n$ have appeared the non-Nickelian ones begin to show up.

\subsection{Low temperature analysis}

Similarly, the diff-Pad\'e analysis of the long low-temperature series ($x=w^2$)
\begin{eqnarray}
\label{chi8plus}
\tilde{\chi}_L \, - \tilde{\chi}^{(2)}\, -\tilde{\chi}^{(4)} \, 
-\tilde{\chi}^{(6)}\,  =\, \,  \,  \, 
 256x^{32} + 65536 x^{33}  \, +  \, \cdots 
\end{eqnarray}
amounts to looking for the singularities of the ``linear ODE" of the (infinite) sum,
$\,\tilde{\chi}^{(8)}\, +\tilde{\chi}^{(10)}\, +\tilde{\chi}^{(12)}\, + \cdots$, 
as roots of the head polynomial of the diff-Pad\'e approximant.

The Ansatz for the ODE search is to include only the factor $(1 - 16x)^i$.
Here we show the results in Table \ref{Ta:CL} for two orders fully utilising
the series coefficients at our disposal. The first block of singularities 
correspond to Nickelian singularities given as Case 2 in \ref{Singularities}.

\begin{table}[htdp]
\caption{\label{Ta:CL}
The number of significant digits found for each singularity of the partial 
low-temperature susceptibility (\ref{chi8plus}) using a diff-Pad\'e analysis.
The singularities  are those predicted from $\Phi_H^{(n)}$ to occur in $\tilde\chi^{(n)}$.
The second column gives the polynomial factor whose roots give the singularity locations. 
Subsequent  columns give the number of significant digits found for that singularity
at orders 14 and 20.
}

\begin{center}
\begin{tabular}{|c|c|c|c|}\hline
$n$ & polynomial & Order 14 & Order 20   \\ \hline
8, 12, 16&    $1-4x$    &  69      &    72         \\    
8, 16&  $1-2x$      &  51      &    52          \\
8, 16&  $1-8x$      &  74      &    75          \\ 
8, 16&  $1-12x+4x^2$&  73      &    79          \\ \hline
10, 12&    $1-x$     &  21      &     24        \\  
10& $1-5x$      &  43      &    46           \\
10& $1-7x+x^2$  &  42      &    44          \\ 
10& $1-12x+16x^2$ &  49        &  53            \\
10& $1-15x+25x^2$ &  55        &  59            \\ \hline
12&    $1-9x$    &  12      &    16          \\  
12& $1-3x$        &  10        &  12            \\
12& $1-12x$       &   9       &   13           \\ 
12& $1-14x+x^2$   &  35        &  40            \\ 
12& $1-8x+4x^2$   &   6       &    6          \\ \hline
14& $1-21x+98x^2-49x^3$  &  21        & 24             \\   \hline
16& $1-24x+148x^2-176x^3+4x^4$   &  10        &  12            \\ \hline \hline \hline \hline
8&  $1-20x+16x^2-16x^3$          &   8       &    7           \\   
8& $1-26x+242x^2-960x^3+1685x^4-1138x^5$    &   6   &      6     \\   \hline 
10& $1-24x+128x^2-289x^3$          &  -      &    4           \\   
10& $1-46x+866x^2+\cdots-56642x^9$      &   -       &     6          \\   \hline \hline \hline

 \end{tabular}
\end{center}
\end{table}

As for the high temperature analysis we have confirmation that some of the singularities 
of the linear ODE of $\Phi_H^{(n)}$ are {\em actually singularities 
of the linear ODE} of the $\tilde{\chi}^{(n)}$.
All the Nickelian singularities for $n=8, 10, 12$ are confirmed.
Of the five (nine) Nickelian singularity polynomials for $n=14$
($n=16$) four (eight) are missing.

As for the new singularities, $1-2w=0$ for the ODE of $\tilde{\chi}^{(5)}$ and
$1-8w^2=0$ for the ODE of $\tilde{\chi}^{(6)}$, we should say that these
singularities occur for higher index $n$. Thus, as far as $\tilde{\chi}^{(5)}$ and 
$\tilde{\chi}^{(6)}$ are concerned, and in view of the ``limited" analysis carried 
out for $\tilde{\chi}^{(n)}$, $n >6$, there is no new singularity discovered that is 
not in the ``known" set (given by the Nickelian singularities and the singularities 
of the $\Phi_H^{(n)}$ integrals, irrespective of the index $n>6$).

\subsection{Local exponents}

We turn now to the indicial exponents at some of the singularities
found in the analysis of (\ref{chi7plus}) and (\ref{chi8plus}).

Here, the situation may seem different from the equivalent analysis made
for $\tilde{\chi}^{(5)}$ and  $\tilde{\chi}^{(6)}$. For instance, when we
consider the indicial exponents for the singularities that should occur
in the still unknown linear ODE for, e.g. $\tilde{\chi}^{(8)}$, the series 
that we analyse contain contributions from all the even  $\tilde{\chi}^{(n)}$ 
up to $\sqrt{N}$. For instance, for $(1-4x)=0$, this singularity is also a 
singularity of the linear ODE of $\tilde{\chi}^{(10)}$, $\tilde{\chi}^{(12)}$, 
etc. One may thus expect to obtain for the Nickelian singularity $(1-4x)=0$ the local 
exponent~\cite{nickel-00} $(n^2-3)/2$, for $n=8, 10$ and $n=12$. The differences 
between all three exponents being integers, it is the value $\, 61/2$ corresponding
to $n=\, 8$ which should appear. What we have obtained is indeed that
for the Nickelian singularities, when the exponents have stabilized, they
agree with $(n^2-3)/2$, where $n$ is the lowest index.

Here we give some other examples of local exponents. For the non-Nickelian singularity polynomial
$(1-26x+242x^2-960x^3+1685x^4 -1138x^5)$ corresponding to $n=8$, the local exponent 
is $23.5$ (with 4 digits for $q=14$, 5 digits for $q=17$ and 3 digits for $q=20$).
For the non-Nickelian singularity polynomial $(1-20x+16x^2-16x^3)$ corresponding to $n=8$,
the local exponent is $18.5$ (with 2 digits for $q=14$, 3 digits for $q=17$ and 2 digits 
for $q=20$). Both exponents agree with (\ref{eqE1}) with $n=8$ and respectively $m=1$
and $m=2$.

Similarly for the high temperature analysis, and for instance, for the Nickelian 
singularities given by roots of the polynomials $P_1, \, P_2$ and $P_3$ corresponding 
to $\tilde{\chi}^{(7)}$, the indicial exponent 23 appears with 10 correct digits using 
480 terms at  order 25. The exponents for the non-Nickelian singularity
polynomial $(1+12w+54w^2+112w^3+105w^4+35w^5+4w^6)$, which should correspond to 
$\tilde{\chi}^{(7)}$, shows up as $17$ with 2 correct digits. 

Let us close this analysis with the following remark: the diff-Pad\'e analysis of 
the sum (\ref{chi7plus}) has shown the singularity $w=-1/2$ which was attributed 
to $\tilde{\chi}^{(9)}$. Actually $w=\, -1/2$ is known to be a Nickelian singularity 
for $\tilde{\chi}^{(9)}$. From the analysis of the $\Phi_H^{(n)}$ we found that $w=-1/2$ 
is also a singularity of $\Phi_H^{(7)}$. So this singularity may well be attributable 
to $\tilde{\chi}^{(7)}$. The diff-Pad\'e analysis gives for this singularity $39$ as 
the local exponent, which is the value predicted by $(n^2-3)/2$ for  
$\tilde{\chi}^{(9)}$\footnote{We may thus conclude that either the linear ODE of 
$\, \tilde{\chi}^{(7)}$ does not have 
this singularity or it has this singularity, but the exponent is compatible with $39$, 
i.e. it may be $(\rho)^p$ with $\rho \,\ge \, 39, \,\,p \,\ge\, 1$.}.

\subsection{``Indicial'' exponents of non-holonomic sums}
\label{indicialnonholo}

We showed in  previous subsections that even if the series of $\tilde{\chi}^{(7)}$ 
is ``polluted" by the terms of the other $\tilde{\chi}^{(n)}$ the diff-Pad\'e analysis
is {\em efficient enough to give the correct singularities and the corresponding 
indicial exponents}.

Recall that the sums (\ref{chi7plus}, \ref{chi8plus}) are believed to be {\em non-holonomic}, 
thus our linear ODE search (diff-Pad\'e analysis) is just an approximation and another way 
to encode, via  the singularities and their local exponents, the information contained 
in the series coefficients.

For this series with infinitely many singularities, generically not suitable
for a diff-Pad\'e analysis, it is interesting to see how the local exponents
actually appear for those singularities occurring in all $\tilde{\chi}^{(n)}$, 
i.e. $w=\pm 1/4$.

To more fully appreciate the results of this section, consider the "holonomic"
sum $\tilde{\chi}^{(1)}+\tilde{\chi}^{(3)}+\tilde{\chi}^{(5)}$. One may ask what
are the local exponents that appear for $w=\, 1/4$, knowing the exponents for
each term? Recall that at $w=\, 1/4$, the local exponent for the linear ODE of
$\tilde{\chi}^{(1)}$ is $-1$, for $\tilde{\chi}^{(3)}$ they are $-3/2$, $(-1)^2$ 
and $(0)^2$ and for the linear ODE of $\, \tilde{\chi}^{(5)}$, (from diff-Pad\'e)
the observed exponents are $-3/2$, $-1$ and $(0)^4$.

A diff-Pad\'e analysis on the holonomic sum
$\tilde{\chi}^{(1)}+\tilde{\chi}^{(3)}+\tilde{\chi}^{(5)}$ should show
all these exponents (with possibly a change in the multiplicity due
to auto-cancellation) and possibly other exponents that differ by integer values 
from the exponents in each individual $\tilde{\chi}^{(n)}$  term. This last possibility
comes from a cancellation between the initial terms of the series.

Putting the known singularities into the head polynomial of an order fifteen
linear ODE and using 860 terms we obtain:
\begin{eqnarray}
 w=\,1/4, \,\,\qquad \quad -3/2,\,\,\, -1,\,\,\, 0^3,\,\, \,1.
\end{eqnarray}
Thus we see that here is {\em no new exponent} not differing by integer values from
the exponents of each individual $\tilde{\chi}^{(n)}$ term.

Let us now return to considering the {\em non-holonomic} infinite sum,
$\tilde{\chi}_H=\, \, \tilde{\chi}^{(1)}\, +\tilde{\chi}^{(3)}\, +\cdots$, for
which a diff-Pad\'e analysis (with some 2000 terms and orders 15
and 18) gives, at $w=\, 1/4$, the local exponents:
\begin{eqnarray}
w\,=\, 1/4, \,\,\qquad \quad -1,\,\, \,-1/8,\,\, 
 \, (3/8)^2,\,\,\,   (15/8)^2, \,\,\,(35/8)^2 
\end{eqnarray}
Here we see the {\em generation of new exponents} that can only be seen as a 
consequence of the {\em  non-holonomic character of the sum}.
Note that the new exponents are still {\em rational numbers}. 
Note also that the new exponents appear not only in a diff-Pad\'e
analysis of $\tilde{\chi}_H$,
but also in $\tilde{\chi}_H\,-\tilde{\chi}^{(1)}\,-\tilde{\chi}^{(3)}$ or in
$\tilde{\chi}_H\,-\tilde{\chi}^{(1)}\,-\tilde{\chi}^{(3)}\,-\tilde{\chi}^{(5)}$. 
One has some kind of ``self-similarity''
in $\tilde{\chi}$ as far as the indicial exponents are concerned.
Note that the multiplicities above may increase with more terms.

The local exponents are those of the full $\tilde{\chi}_H$ around $w=\, 1/4$. They 
can be checked in 
\begin{eqnarray}
{\frac{(1-s^4)^{1/4}}{s}} \,\,\tilde{\chi}_H \,= \, \, \,
 const \,  \cdot \tau^{-7/4}\,F_{+}\,  +\,\,   B_f 
\end{eqnarray}
given by Orrick {\it et al.}, using $F_+$ and $B_f$ in the Appendix in
\cite{Orrick},  by switching from the variable $\tau$
to the variable $(w-1/4)$. 

The following section presents a simple model showing the mechanism by which a 
{\em resummation of the  infinite number of logarithmic singularities proliferating
in the holonomic $\, n$-fold integrals $\, \tilde{\chi}^{(n)}$ can lead to 
the known power-law singularities in $\chi$}.

\section{Convergence of the $n$-particle sequence}
\label{natural}

It was already observed by Wu {\it et al.}~\cite{wu-mc-tr-ba-76} that for
the leading divergence of the susceptibility  proportional to $|\tau|^{-7/4}$ 
(where $\, \tau=(1/s-s)/2$), the sequence of partial sums of either $\chi^{(2n)}$ 
or $\chi^{(2n+1)}$ appears to converge exceptionally rapidly to the corresponding 
susceptibility below or above the critical temperature.  This has been confirmed  
numerically to higher order by Bailey {\it et al.}~\cite{Crandall}, who estimated 
that, asymptotically, the amplitudes $I_n$ of the leading divergence of $\chi^{(n)}$ 
are in the ratio $I_{n+2}/I_n < 1/1000$.  On the other hand, the leading correction 
terms in $\chi^{(1)}$ and $\chi^{(2)}$ are easily seen to be of order $|\tau|^{1/4}$
and $|\tau|^{1/4}\log|\tau|$ respectively whereas the leading correction in $\chi$ 
approaches a constant in the limit $\tau \rightarrow\,  0$. In fact the exact 
solutions for $ \chi^{(3)}$ and $\chi^{(4)}$, found by 
Zenine {\it et al.}~\cite{ze-bo-ha-ma-04, ze-bo-ha-ma-05b} make it plausible that 
every partial sum of $\chi^{(2n)}$ or $\chi^{(2n+1)}$  will, asymptotically in the 
limit $\tau \rightarrow 0,$  give a vanishingly small contribution to the leading 
correction term. This raises the issue of the nature of the convergence
of the $n$-particle sequence to the susceptibility. 
 
With the exceptionally long series now available for $\chi^{(5)}$  and $\chi^{(6)}$ 
we can, even in the absence of an exact linear ODE solution, make numerically 
precise estimates of the correction terms in these functions. A scheme that works 
well is a combination of unwanted singularity suppression as described 
in~\cite{nickel-99} and function fitting. The latter is done by assuming various 
combinations of powers and powers of logarithms with unknown coefficients and 
generating the corresponding series in $s.$  The unknown coefficients can then 
be estimated by least squares fitting to the highest order terms of the exact
 (unwanted singularity suppressed) series.  This fitting is typically done 
iteratively with the leading order exact coefficient values substituted for
 the numerically estimated ones from an earlier iteration.

Then, by combining this information with what is already known for the lower orders 
$\chi^{(n)}$, we can build a plausible  ``toy model'' $\, \chi_{Toy}^{(n)}$ that can 
be easily extrapolated to $\, n \rightarrow \infty.$  This gives us, at least 
qualitatively and even semi-quantitatively, a picture of the convergence of the 
partial sums of $\chi^{(n)}$ to the full $\chi.$  The details of this extrapolation 
procedure via $\chi_{Toy}^{(n)}$ is described below.
 
\subsection{Behaviour of $\chi^{(n)}$}
 \label{behavior}
We begin by providing a summary of the behaviour of $\chi^{(n+1)}$
near the ferromagnetic critical point.  

The normalization factor included for convenience on the left-hand side of the 
equations below is $4\pi^n n! $ while the factor $\sqrt{s}$ is crucial to give 
series increasing in even powers of $\tau$ only. An empirical observation is that 
$\tau=(1/s-s)/2$ appears almost universally in all formulae in the combination 
$|\tau|/4$ so we have defined $\tau_4 = |\tau|/4$. The known $ \chi^{(n)}$ terms 
are\footnote{The formula given here for $\chi^{(4)}$ corrects misprints in 
\cite{ze-bo-ha-ma-05c}.}
 \begin{eqnarray}
\label{con1}
\fl \qquad 4\, \sqrt{s}\cdot \chi^{(1)}\,  = \, \, 
 4 \cdot \left(\frac{1}{8}\tau_4^{-7/4}\, + \tau_4^{1/4}\right) \,
 - \tau_4^{1/4}\,  + {\rm O}(\tau_4^{9/4}), \nonumber \\ 
\fl \qquad  4\pi\,\sqrt{s}\cdot  \chi^{(2)} \, = \, \,
 \frac{1}{3} \cdot \left(\frac{1}{8}\tau_4^{-7/4}\, + \tau_4^{1/4}\right) + 
\left(\log\tau_4+\frac{11}{12}\right)\cdot  \tau_4^{1/4} \, 
+ {\rm O}(\tau_4^{9/4}), \nonumber  \\
\fl \qquad  8\, \pi^2 \, \sqrt{s}\cdot \chi^{(3)}\, =\, \, 
  8\, I_3^+\pi^2  \cdot \left(\frac{1}{8} \tau_4^{-7/4}\, + \tau_4^{1/4}\right) \nonumber \\
\fl \qquad \qquad \qquad 
 - \left(\log^{2}\tau_4 
+\frac{23}{6}\log\tau_4+\frac{\pi^2}{3}+\frac{41}{36}\right) \cdot  \tau_4^{1/4}\,
 + {\rm O}(\tau_4^{9/4-\epsilon}), \nonumber \\
\fl \qquad  24\pi^3\,\sqrt{s} \cdot  \chi^{(4)} \, =\, \,  
 24I_4^-\pi^3 \cdot  \left(\frac{1}{8} \tau_4^{-7/4}\, + \tau_4^{1/4}\right) + 
\left(\log^3\tau_4+\frac{35}{4}\log^2\tau_4 \right. \nonumber \\
\fl \qquad \qquad \qquad  + \left. \left(2\pi^2+\frac{107}{12}\right) \log\tau_4 
\, +34.3411462895\ldots \right) \cdot  \tau_4^{1/4}\, 
 + {\rm O}(\tau_4^{9/4-\epsilon}),  \nonumber \\
\fl \qquad  96\, \pi^4\, \sqrt{s}\cdot \chi^{(5)}\, =\, \, 
 96I_5^+\pi^4 \cdot \left(\frac{1}{8} \tau_4^{-7/4}\, + \tau_4^{1/4}\right) \, 
- \left( \log^4\tau_4+\frac{47}{3}\log^3\tau_4\, \right. \nonumber \\
\fl \qquad \qquad \qquad  +\left(6\pi^2+\frac{245}{6}\right) \cdot \log^2\tau_4 \,
+  305.6550541085\ldots \, \, \log\tau_4 \nonumber \\
\fl \qquad  \qquad  \qquad \left. \phantom{\frac12} 
+ 375.271992213596336733341995793\right) \tau_4^{1/4}\,
 + {\rm O}(\tau_4^{9/4-\epsilon}), \nonumber \\
\fl \qquad 480 \, \pi^5\, \sqrt{s}\cdot \chi^{(6)}\, =
\, \,  480I_6^-\pi^5 \cdot  \left(\frac{1}{8} \tau_4^{-7/4}\,+ \tau_4^{1/4}\right) \, + 
\left (\log^5\tau_4 \,  +\frac{295}{12}\log^4\tau_4\, \right. \nonumber \\
\fl \qquad  \qquad \qquad
+\left(\frac{40\pi^2}{3}+\frac{2275}{18}\right) \cdot \log^3\tau_4 \, 
 + 1437.2558956691\ldots\ln^2\tau_4\,\nonumber \\
\fl \qquad  \qquad \qquad +4238.8509988858869410798745 \log \tau_4 \nonumber \\
\fl \qquad  \qquad \qquad  \left. \phantom{\frac12} 
 +  5284.244417602341195112209\right) \cdot  \tau_4^{1/4} \, 
 + {\rm O}(\tau_4^{9/4-\epsilon}) 
\end{eqnarray} 
while the complete summation of all $\, \chi^{(n)}$ yields the exact  
 \begin{eqnarray}
\label{chisqr}
\fl \qquad \sqrt{s} \cdot \chi \, =\, \, \, 
1.00081526044021 \, \ldots \,\,  \left(\frac{1}{8} \tau_4^{-7/4}\,\, 
 + \tau_4^{1/4}\right)  \, \, -0.1041332450938 \ldots \nonumber \\
\fl \qquad \qquad +(0.0323522684773 \ldots \, \log(\tau)\,
-0.074368869753\ldots)\cdot \tau 
\, +{\rm O}(\tau^2) 
\end{eqnarray}
for $\tau>\, 0$ and 
\begin{eqnarray}
\label{con2}
\fl \qquad \sqrt{s} \cdot \chi \, =\, \,  \,   
\frac{ 1.00096032872526...}{12\pi} \cdot \left(\frac{1}{8} \, \, \tau_4^{-7/4} \,
 + \tau_4^{1/4}\right) \, \,  -0.1041332450938 \ldots \nonumber  \\
\fl \qquad \qquad +(0.0323522684773\ldots \, \, \log(-\tau)\,\, 
 -0.074368869753\ldots) \cdot \tau\,  +{\rm O}(\tau^2) 
\end{eqnarray}
for $\tau < \, 0$.
We have used O$(\tau_4^{9/4-\epsilon})$ to indicate O$(\tau_4^{9/4})$ with 
logarithmic corrections. We also distinguish between numerical constants. 
Those with trailing $\ldots$ are known to much higher accuracy but have not 
yet been recognized in terms of elementary constants\footnote[1]{The exact 
values for $I_3^+$ and $I_4^-$ have been given by Tracy~\cite{Tracy} (see also note 
added in proof in~\cite{ze-bo-ha-ma-05c}). Highly accurate numerical values for 
$I_5^+$ and $I_6^-$ are given by Bailey {\it et al.}~\cite{Crandall}.
For the constants in $\chi$ see Orrick {\it et al.}~\cite{Orrick}.
The constant of $\chi^{(4)}$ in (\ref{con1}), to higher accuracy is  
$34.3411462895318287823760337015134958201194891883303683732355064700755161069888189217774\\$
$8221462488591136755297844305262613713337765 \, \cdots $}.
The others are believed, but not guaranteed, to be accurate to the number of digits given.

An important remark about the form of equations (\ref{con1}) is that we have split the
$\tau_4^{1/4}$ contribution to create the combination 
$(\frac{1}{8} \tau_4^{-7/4}+ \tau_4^{1/4})$ which is the scaling part of the full
susceptibility in (\ref{con2}). The original motivation for this was the observation 
by Orrick {\it et al.}~\cite{Orrick} that {\em the terms in the expansion 
of the scaling function in the susceptibility are $ \tau_4^{-7/4+2n}$ without 
logarithms, while in the expansion of the background part of the susceptibility 
there are only integer powers of $\, \tau_4$ but now with logarithmic corrections}.
A simple way to reproduce the scaling part of the susceptibility from the $\chi^{(n)}$
sum is to replace every $\tau_4^{-7/4}$ by the scaling function, as we have done. 
Then the sum rule on the $I_n$ that ensures the correct $\tau_4^{-7/4}$ amplitude 
in the susceptibility will also automatically yield the correct scaling function. 
Such separation into ``scaling'' and ``background'' in individual $\chi^{(n)}$ would 
appear to be completely arbitrary {\em except that we now find that the ``background'' 
remainder in $\chi^{(n)}$ has a simple dependence on $n$ in which the same 
formulae apply simultaneously to both odd and even $n$}.

This was quite unexpected and 
the separation may well only apply to the leading correction term we are concerned with here.
As a practical matter, it leads us directly to the conjecture
 \begin{eqnarray}
\label{con3}
\fl 
\quad  4n!\, \pi^n\,  \sqrt{s}\cdot  \chi^{(n+1)} \, =  \, \,  \, \,   
4n! \, \pi^n \, I_{n+1} \cdot  \left(\frac{1}{8} \tau_4^{-7/4}\,\,+\tau_4^{1/4}\right) \nonumber \\
\fl \qquad  \quad   -(-1)^n \left[\log^n\tau_4\, +n\left(n-\frac{1}{12}\right)
\cdot  \log^{n-1}\tau_4 \right. \nonumber \\
\fl \qquad  \quad   +{n \choose 2}\left(n^2+\frac{(2\pi^2-19)n}{6}\, 
+\frac{125-12\pi^2}{36}\right)\cdot \log^{n-2}\tau_4 \nonumber \\
\fl \qquad  \quad   +  {n \choose 3}\left(n^3+\frac{(4\pi^2-37)n^2}{4}\,
+\frac{(374-37\pi^2)n}{12}\, -0.441452610\ldots\right)\cdot \log^{n-3}\tau_4\nonumber \\
\fl \quad \quad  \, \ldots  +    \left. {n \choose k} \left( n^k+k\left(k(\pi^2-9)\, +\frac{17}{2}\,
-\pi^2\right) \frac{n^{k-1}}{6}\, 
+ \ldots\right) \cdot \log^{n-k}\tau_4 \, +\ldots \right] \cdot \tau_4^{1/4} 
\nonumber \\
\fl \qquad  \quad + \quad {\rm O}(\tau^{9/4-\epsilon})  
\end{eqnarray}
based on the known low order results and applicable equally to the logarithmic 
and non-logarithmic terms. Note that the numerical constant in (\ref{con3}) is 
simply related to the constant in $\chi^{(4)}$ deduced from its linear ODE solution.
Specifically, one has $-0.44145 \, \ldots \, = \, \, 34.341 \, \ldots\, +\pi^2/4\, -149/4$.
 
\subsection{Resummation of the toy model}
\label{resumtoy}
The conjectured general term (\ref{con3}) forms the basis for our $\,\chi_{Toy}^{(n+1)}.$  
The leading divergence in $\, \chi^{(n+1)},$ now combined with the scaling function, is 
not under consideration here since we have nothing to add to what is already known.  
Of the correction terms in (\ref{con3}) we will capture exactly the leading $\log^n\tau_4$ 
into our toy model. The appearance of binomial coefficients in the next three terms is 
suggestive of a formula like $(\log \tau_4 + n)^n$ which would capture the leading $n$ 
dependence  correctly. We do not have enough information to be confident about the 
behaviour of any lower order logarithmic terms and so at this point our model becomes 
dictated by the criteria of simplicity.  The numerical values of the lower order terms 
suggest the simple formula $(\log\tau_4+\lambda n)^n$ as reasonable, where 
$\lambda \approx 1$ is some as yet undetermined constant. If $\lambda$ is very close to 
$1.0$ then our constant term $(\lambda n)^n$ is probably an underestimate but it is 
worth remarking that in the $\tau\rightarrow 0$ limit, $ \log\tau_4$ is negative.  
This implies successive terms in the expansion of $\, (\log \tau_4+\lambda n)^n$ 
alternate in sign and the final value, just as in the exact $\, \chi^{(n+1)}$,
is {\em the result of large cancellations between terms}. Thus it is more important 
that we capture correctly the ``smoothness" by which successive terms vary and this 
is hard to estimate from the limited data available. Let these caveats be understood. 
Let us denote by $\, \Delta$ the  correction to scaling. Then our toy model 
for the correction to scaling terms is 
\begin{eqnarray}
\label{con4} 
 &&\sqrt{s}\cdot \Delta\chi_{Toy}^{(n+1)} \,\, =\, \, \, \,
 -\frac{ \tau_4^{1/4}}{4\pi^n n!} \cdot (\log(1/\tau_4) \,-\lambda n)^n,
 \nonumber \\
&& \sqrt{s}\cdot \Delta\chi_{Toy}  \,\, =\, \,\,  \,
 \sum   \sqrt{s} \cdot  {\Delta}\chi_{Toy}^{(n)} 
 \end{eqnarray}
with the sum understood to be over even $n$ for $T\,<\, T_c$ and odd $n$ for 
$\,T>\, T_c$. As $\tau\rightarrow 0,$ the sum in (\ref{con4}) is dominated by 
large $n$ and can be replaced by an integral that is easily treated by steepest 
descent methods.  The value of the integral will depend on the undetermined $\lambda$ 
and we will choose $\lambda$ such that, in the limit $\,\tau\, \rightarrow\, 0$,
$\sqrt{s} \cdot \Delta\chi_{Toy} \, \rightarrow\,  A_{Toy}$ is  a constant.  
That this is possible is verified by explicit calculation below. 
{\em The reasonableness of our toy model can then be judged by how close $\, A_{Toy}$ 
is to the exact} $A = \, - 0.1041\ldots $ from (\ref{con2}). 
 
The dominant $n$ dependence of the integrand in the integral approximation for 
$ \,\sqrt{s}\cdot  \Delta\chi_{Toy}$ is the factor 
$\, \exp[n\log(\log(1/\tau_4)-\lambda n)\, -n\, \log(n\pi)+n]\, $
and this has a maximum at $n_{p}$ where: 
 \begin{eqnarray}
\label{con5} 
 n_{p} \, =\, \, \,\, \frac{p}{1\, +p\lambda} \cdot  \log(1/\tau_4),  
\,\,\, \qquad 
  p\cdot \lambda \,+ \log(\pi p)\, =\, \,\,  0. 
 \end{eqnarray}
The exponential at its maximum evaluates to $\, \exp(p\log(1/\tau_4))\, =\, 1/\tau_4^p$
which establishes the dominant  $\tau$  dependence of the integral as a function of 
$\lambda$ since $\, p \,  = \,  \, p(\lambda)$ is the solution of the transcendental 
equation in (\ref{con5}).  Expanding the exponential about its maximum then gives the
approximate $\, \sqrt{s}\cdot \Delta\chi_{Toy}^{(n+1)}$ and the required 
sum in (\ref{con4}), namely 
\begin{eqnarray}\label{con6} 
\fl \quad \sqrt{s}\cdot \Delta\chi_{Toy} \, \approx\, 
 \frac{-\tau_4^{1/4-p}}{8\sqrt{2\pi n_p}} \int\! dn  
\exp[-(1+p\lambda)^2(n-n_p)^2/(2n_p)] \,=\,  
  -\frac{\tau_4^{1/4-p}}{8\, (1+p\lambda)},
\end{eqnarray}
valid in the limit $ \tau\rightarrow 0$ both above and below $T_c$. The choice 
$p \, =\, \,  1/4$ is now seen as necessary and we get from (\ref{con5}) and (\ref{con6}) 
 \begin{eqnarray}
\label{con7}
 &&\lambda \, =\, \, \,  \,   \, 
4\,\log(4/\pi)\, \,  \,  \approx\, \, \,   0.966,  \nonumber \\
 &&A_{Toy} \, =\, \,  \, 
\lim_{\tau\rightarrow  0} \, \sqrt{s}\cdot  \Delta\chi_{Toy} 
\, = \,\, -\frac{1}{(8+8\log(4/\pi))}\,  \approx\, \,  \,  -0.1007.  
 \end{eqnarray}
{\em The close agreement of the asymptotic amplitude with the exact 
$A\, \approx\,  -0.1041$ gives us confidence that (\ref{con4}), with the
specific choice of $\lambda$ from (\ref{con7}), will be usefully predictive for 
finite $\tau$ and $n$}. 
 
\begin{figure}
\begin{center}
\includegraphics[scale=0.9]{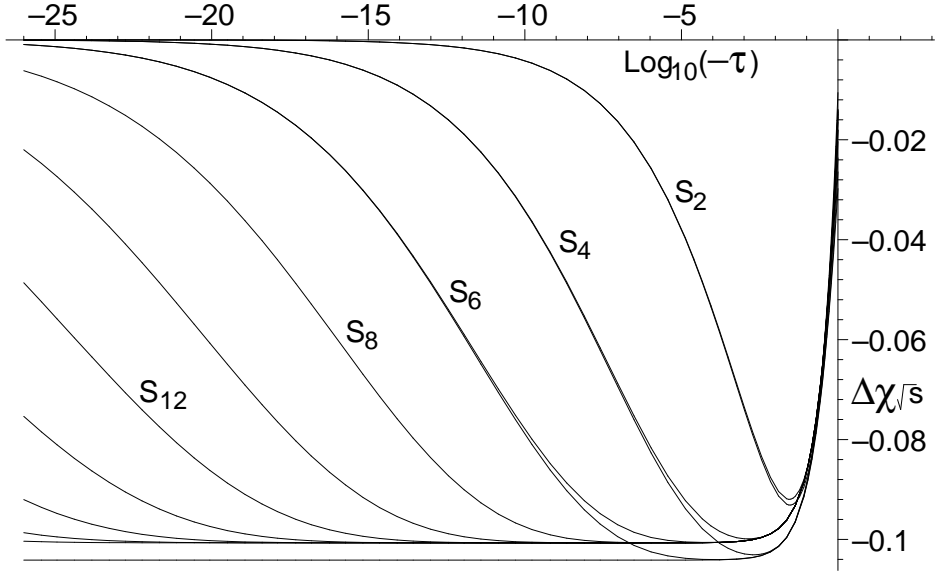}
\end{center}
\caption{\label{fig:conv}  The contribution to the ``background'' part
 of the susceptibility from the partial sums $S_2\, =\, \chi^{(2)}\sqrt s$,
 $ S_4=\, S_2+\chi^{(4)}\sqrt{s},$  
$ S_6=\, S_4+\chi^{(6)}\sqrt{s}$ and the exact $\chi\sqrt{s}$ based 
on the formulae (1) and (2) exclusive of the
scaling $\frac{1}{8}\tau_4^{-7/4}+\tau_4^{1/4}$ combination.
Also shown are the corresponding toy model $S_n$ based
 on (\ref{con4}) now extending to larger $n.$}
\end{figure}

The result of numerical computation for finite $\tau$ and $n$ is shown for 
$ T\, <\, T_c$ in Fig. \ref{fig:conv}. A similar plot could be made for 
$T\, >\, T_c$.  The agreement between exact and toy partial sums at low order 
is of course by design and the agreement in the limit $\tau\rightarrow 0$ has 
already been remarked on in connection with (\ref{con7}).  The striking feature 
of the shifts in  $\log_{10}(-\tau)$ with order can be deduced from (\ref{con5}).  
The $\, n_p$ in (\ref{con5}) which corresponds to the $n$ of maximal contribution 
is also the $n$ characterising the transition region of the partial sum 
approximations to $\, \sqrt{s}\cdot \Delta\chi_{Toy}$ between the asymptotes
$\, 0$ and $\, A_{Toy}\, =\, -0.1007$. Then, since each additional order is a 
change in $n$ of $2,$ we deduce a transition region shift of 
$\, \Delta\log(1/\tau_4)\, = \, 8\cdot  (1+\log(4/\pi)) \,\approx \,10$,
or about four decades in temperature. 

\section{Natural boundary: power spectrum analysis}
\label{power}
In this section we are using the power spectrum analysis introduced
in Sec 6.3 of Orrick {\it et al.}~\cite{Orrick}, though not fully described there. The model of $n$-fold 
integrals~\cite{bo-ha-ma-ze-07b} has shown that the singularities of the 
linear ODE of $\Phi_H^{(n)}$ occur in the linear ODE of the higher order 
$\Phi_H^{(n+2m)}$. These model integrals are assumed to ``mimic'' correctly 
the $\chi^{(n)}$ in terms of the locus of the singularities. With the singularities 
obtained here from diff-Pad\'e analysis of $\tilde{\chi}^{(5)}$ and $\tilde{\chi}^{(6)}$ 
we have confirmation of this feature. If we focus on the singularities which 
are on the unit circle $| s |\, =\, 1$, the Nickelian singularities 
of $\tilde{\chi}^{(3)}$, (i.e. $(1-w)(1+2w)\, =\, 0$) are also singularities of the 
linear ODE of $\tilde{\chi}^{(5)}$.

However, the detailed Landau conditions analysis (see \ref{appendixBernie}) 
proves that the above singularities lying on $\, |s|=\, 1$ do not occur on the
principal disc of $\tilde{\chi}^{(5)}.$  Indeed, this is just a special case of the general
theorem that applies to all the Case 3-5 singularities.  Furthermore, the
analysis in \ref{app:ToyModel} suggests that the singularities associated with the
ODE head polynomial factors $(1-2w)$ for $\, \tilde{\chi}^{(5)}$ and $\, (1\, -8 \, w^2)$ 
for $\, \tilde{\chi}^{(6)}$ are not even Landau singularities of the integrals 
$\, \tilde{\chi}^{(5)}$ and $\, \tilde{\chi}^{(6)}$, at least in a Landau singularities 
analysis that does not fully take into account the Fermionic term.

Here we confirm the limited result of the absence of these singularities on 
$\, |s|=\, 1$ on the principal disc using a technique~\cite{Orrick} based on
the fast-Fourier transform of the series of $\chi^{(n)}$. The series is first 
appropriately modified by removing the dominant singularities and smoothing.

The Nickelian singularities $s_0$  have very small amplitude, roughly 
$\, (1-s/s_0)^{n^2/2}$  for $\tilde{\chi}^{(n)}$. For $\tilde{\chi}^{(5)}$ the effect on the 
series coefficients (for series in the $s$ variable) at $N=2000$ is roughly 
$1/(2000)^{13} = 10^{-43}$ times smaller than that from the  ferromagnetic 
singularity. The amplitude of the dominant ferromagnetic divergence being known 
this contribution can be subtracted. As the smoothing continues  (see details 
in~\cite{nickel-99}), the series coefficients will start to decay rapidly with 
order $N$ and this must be corrected for by multiplying, at each step,  by some 
power of $N$ to again make the series coefficients roughly constant in $N$. Once 
the series have been smoothed and multiplied up by about $N^{13}$ the FFT is performed.

The FFT is obtained by using say 512 coefficients $a_{n+N}$ starting from some
large $N$ to utilise the highest order coefficients available. We get
\begin{eqnarray}
b_m\,\, = \,\, \sum a_{n+N} \cdot \exp \left(2\,\pi \,\rmi \,n\,{m \over 512} \right), \qquad
m  \, =  \, 1,\, \,  2,\, \,  ...\, \,  , 512.
\end{eqnarray}
and the power spectrum $P_m=b_m\cdot b_{512-m}$ which runs over 256 points.
The ``frequency", $m$,  is directly interpretable as an angle $\theta=\pi m/256$ giving 
the locations of the singularities $\exp(\pm \, \rmi\, \theta)$ in the complex $s$ plane. 
Windowing $\, a_n \, \rightarrow \,  a_n\, \sin^p(n\, \pi/512)$
can be used to reduce background (this broadens the spectral lines).

\begin{figure} 
\begin{center}
\includegraphics[scale=0.5,angle=-90]{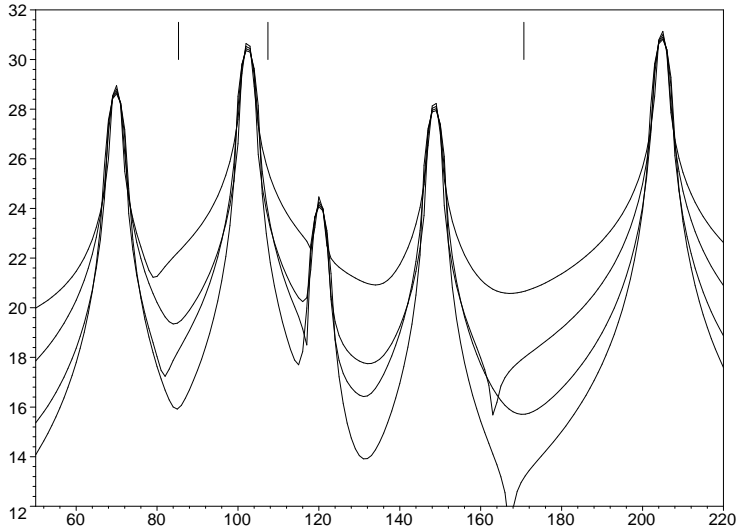}
\end{center}
\caption{\label{fig:xxx} 
Power spectrum of $\chi^{(5)}$ series on $\log_{10}$ scale vs ``frequency'' $m$} 
\end{figure}

The power spectrum of $\tilde{\chi}^{(5)}$ series is shown on Fig.~\ref{fig:xxx}.
The curves correspond to various values of the integers $p=\, 2,\,  \cdots, \, 5$.
From left to right, the first and third spikes correspond to the singularities of 
$1-3w+w^2=\, 0$. The second and fifth spikes correspond
to $\,1+2w-4w^2\,=\, 0$ and the singularity $w=\, -1$ appears as the fourth spike.

{\em The singularities $w=1/2$, $w=1$ and $w=-1/2$ are not seen}. Their positions
are indicated, respectively, by the vertical lines at the top of the spectrum.

Similar analysis of the series of $\tilde{\chi}^{(6)}$ has shown the non-occurrence
of the spike corresponding to the non-Nickelian singularity $w^2=1/8$
which is on $| s | =\,1$.

The FFT analysis does show, in all examples so far, that there is no evidence
for any singularities on the $|s|=1$ boundary of the physical sheet other
than the Nickelian singularities. {\em Thus the possibility that there will be
``destruction" of the natural boundary by cancellation of the singularities 
accumulating on the unit circle is becoming more and more remote}.

\section{Conclusion}
\label{conc}

We have calculated series of some $2000$ or more terms for the magnetic
susceptibility (high- and low-temperature regime of the isotropic Ising model)
as well as for the individual five- and six-particle contributions
$\tilde{\chi}^{(5)}$ and $\tilde{\chi}^{(6)}$. This was achieved by use of modular
arithmetic which amounts to generating the coefficients {\it modulo}
a set of primes, then using the Chinese remainder to obtain the exact
coefficients.

By extending the series for $\tilde{\chi}^{(5)}$ modulo a single prime to 10000 terms
we have discovered the exact linear ODE for $\tilde{\chi}^{(5)}$ {\it modulo} that prime.
The Russian-doll structure previously found to occur for $\, \tilde{\chi}^{(3)}$
and $\, \tilde{\chi}^{(4)}$, and  conjectured for the linear differential operators
of the  $\, \tilde{\chi}^{(n)}$, is actually confirmed for 
$\, \tilde{\chi}^{(5)}$, {\em as well as a stronger direct sum structure}.

We have described our present algorithm for finding the linear ODE satisfied by a series 
$S(x)$ in some detail because it is robust and particularly efficient.  It is based on an 
ansatz of a rectangular array of coefficients of size fixed by degree $D$ in $x$ and order 
$M$ in $x\rmd/\rmd x$. The use of the operator $x\rmd/\rmd x$ rather than $\rmd/\rmd x$ 
guarantees that the resulting ODE is Fuchsian. Now suppose we have found the ODE with minimum 
$D$ for given $M$.  If $N$ is the number of series terms required for this $D$, $M$ combination 
then an empirical observation is that there exists a linear relationship between $N$, $M$ and $D$ 
that enables us to infer the order $M_0$ of the minimum order ODE required by $S(x)$ without 
having to obtain the minimum order ODE itself or even having to obtain any singularity information 
whatsoever.  We have no explanation for this linear relationship and leave it as a challenge for 
the reader to provide an analytic basis for our ``magic formula'' (\ref{30}).

The exact long series have been used in diff-Pad\'e analysis to obtain the singularities 
that should occur in the linear ODE of $\tilde{\chi}^{(5)}$ and $\tilde{\chi}^{(6)}$. We have 
confirmation that the singularities of the ODEs of the $n$-fold integrals $\Phi_H^{(5)}$ and 
$\Phi_H^{(6)}$ (which differ from $\tilde{\chi}^{(5)}$ and $\tilde{\chi}^{(6)}$ by the absence 
of the Fermionic factor) are all singularities of the ODEs of $\tilde{\chi}^{(5)}$ and 
$\tilde{\chi}^{(6)}$. However, {\em our diff-Pad\'e analysis showed that these linear ODEs
have additional singularities, namely $w = 1/2$ for the linear ODE of $\tilde{\chi}^{(5)}$ and
$w^2 = 1/8$ for the ODE of $\, \tilde{\chi}^{(6)}$, not predicted by the corresponding 
$\Phi_H^{(n)}$}.  We see that the Fermionic factor does not affect the singularities 
corresponding to $\, \Phi_H^{(n)}$, but 
it does bring additional singularities to the linear ODE. From a Landau singularity
analysis viewpoint this is not a surprise: A new factor cannot move a singularity, 
but it can either give rise to cancellations or introduce new singularities, and we have 
seen that cancellation does not occur. Our exact mod prime ODE for $\chit^{(5)}$ 
confirms the accuracy of the diff-Pad\'e results and in particular proves that the ODE 
for  $\chit^{(5)}$ carries the extra singularity at $w=1/2$.

We do not know whether or not these extra singularities of the $\chit^{(n)}$ ODEs are 
singularities of the integrals (\ref{chi3tild}). The most common case is that in which 
all the singularities of the ODE and the integral  are the same; this is the situation 
for $\chit^{(3)}$ if one takes into account its various analytical continuations 
(cf. first footnote in \ref{appendixBernie}). As a specific example, both the Landau 
analysis and the ODE for $\chit^{(3)}$ predict a singularity at $s=(-1+\rmi\, \sqrt{7})/4$ 
for which it is to be noted $|s|<1$.Ê While $\chit^{(3)}$ on the principal disc is not 
singular at this point there exists an analytic continuation of $\chit^{(3)}$ that 
is\footnote{This can be inferred from the connection matrices provided in \cite{ze-bo-ha-ma-05c}.}. 
Our analysis of toy integrals in \ref{app:ToyModel} provides an example of a difference 
between the singularities of the ODE and those of the integral. The ODE for the toy analog 
of $\chit^{(5)}$ has singularities at $w=1/2$ and $w^2=1/8$ but our Landau analysis of 
the toy integral fails to find singularities at these points. This may be a genuine distinction, 
or it may be that we missed something in the Landau analysis, or that the Landau analysis can't 
be guaranteed to give all singularities. It would be extremely useful to have other (simpler) 
examples where it can be shown that the ODE defined by an integral has more singularities than 
the integral.

A further check on the occurrence of the singularities (of the linear ODE's of 
$\, \Phi_H^{(n)}$) has been made for some $\, \tilde{\chi}^{(n)}$, $n \ge \, 7$, by 
subtracting, from the full $\, \tilde{\chi}$, the known 
$\tilde{\chi}^{(1)}$, $\tilde{\chi}^{(3)}$ and 
the now long $\, \tilde{\chi}^{(5)}$ series. {\em The diff-Pad\'e analysis again confirms
a large number of singularities occurring in the linear ODE's of the $\Phi_H^{(n)}$
that can, thus, be attributed to the corresponding linear ODE of $\, \tilde{\chi}^{(n)}$}.
A similar analysis has been performed for the low-temperature regime.

While these diff-Pad\'e analyses only yield approximate linear ODE's, they are 
efficient enough to give the indicial  exponents at each singularity with good 
accuracy in most cases.  An example of failure is that in the ODE for $\chit^{(5)}$ 
at $w=1/4$ where we detected $-3/2$ but not $-2$,$-7/4$ and $-5/4$, exponents also 
known to be present from our exact mod prime analysis.  Thus while in all cases we 
have observed that indicial exponents are rational numbers, in the absence of this 
example we might have erroneously concluded that indicial exponents are either 
integer or half-integer.  As a particularly striking success we obtained for 
$\chit^{(6)}$ at $w=\pm\, 1/4$ the indicial exponents 0 with a multiplicity of five ($0^5$).  
While the diff-Pad\'e analysis can only confirm this multiplicity of five as a lower 
bound it is the correct one at $w=1/4$ as shown by our exact (conjectured) equation (\ref{con3}).

From the linear combination of the long series for $\, \tilde{\chi}^{(5)}$, 
$\, \tilde{\chi}^{(6)}$ and $\, \tilde{\chi}$, we were able to make some serious progress on 
two important questions for a deeper physical and mathematical
understanding of the full susceptibility $\, \tilde{\chi}$.

Firstly, we have, finally, resolved the issue of the power/log behaviour
of each $\chi^{(n)}$ at the singular points versus the behaviour of the full $\chi$.
The diff-Pad\'e analysis performed for the full $\tilde{\chi}$ shows the non-occurrence, 
at $w=\, \pm 1/4$, of the logarithmic singularities corresponding to the individual 
$\tilde{\chi}^{(n)}$, but shows with good accuracy the indicial exponents known to occur 
in $\tilde{\chi}$. We presented a model showing the mechanism of the resummation of the
infinite number of logarithmic singularities that proliferate in the
holonomic $\, n$-fold integrals $\, \tilde{\chi}^{(n)}$ {\em building the known
divergence of $\tilde{\chi}$ at scaling}.

Secondly, we have proved by a Landau analysis that no non-Nickelian singularities of 
$\Phi_H^{(n)}$ can lie on the $|s|=1$ boundary of the principal disc. Since we do not
 understand the origin of the additional ODE singularities at $w=1/2$ for $\chit^{(5)}$ 
and $w^2=1/8$ for $\chit^{(6)}$ we cannot make the same analytic claim.  Instead we have 
verified this numerically by fast-Fourier transform on appropriately modified series.  
The FFT results are {\em surprisingly clear-cut, and display
peaks at the precise points of Nickelian singularities on the unit circle}.
The possibility of cancellation becomes more and more remote, up to a point where 
we are able to {\em confirm the existence of a natural boundary for the full 
susceptibility}.

In conclusion, we can say that, with this kind of experimental mathematics
based on ``extreme'' massive computer calculations, we are exploring a new
kind of ``modular'' lattice statistical mechanics, getting results that
were unthinkable  before. Indeed, we are now far along the road to a complete 
synthesis and understanding of the Ising model susceptibility.
However, these experimental mathematical ideas for studying Ising model
integrals  are much more widely applicable, as they can be applied to
any $n$-fold  integral corresponding to a Feynman diagram.
In such cases we are likely to be restricted only to the experimental
aspects, leaving open the challenge for a subsequent theoretical
understanding, such as that  which we have been able to provide in the case
of the Ising model.

\ack
We would like to thank B. McCoy for many illuminating comments.
We would like to thank A. Bostan for helping to find the $\tilde{\chi}^{(5)}$
ODE, {\it modulo} a prime, with his MAGMA program.
We thank  J. Dethridge for some optimization of earlier C++ programs.
The calculations presented in this paper would not have been possible
without a generous grant of computer time on the server cluster of the
Australian Partnership for Advanced Computing (APAC). We also gratefully acknowledge use
of the computational resources of the Victorian Partnership for Advanced 
Computing (VPAC). One of us JMM would like to thank MASCOS, Melbourne
for hospitality where part of this work was initiated and completed.
This work is partially supported by a PICS/CRNS grant.
This work has been performed without any ANR or ERC support.
IJ and AJG gratefully acknowledge financial support from the 
Australian Research Council, and the hospitality of LPTMC,
Universit\'e de Paris 6 where much of this work was carried out.

\section*{E-mail or WWW retrieval of series}

The series for the various generating functions studied in this paper
can be obtained via e-mail by sending a request to 
I.Jensen@ms.unimelb.edu.au or via the world wide web on the URL
http://www.ms.unimelb.edu.au/\~{ }iwan/ by following the instructions.

\appendix

\section{Singularity exponent sum-rules (Fuchs' relations) \label{SumRule}}

Suppose the order $\,M$ and degree $\, D$ linear ODE, $\, L_{MD}(S(x))\,  = \, 0$, 
with $\, L_{MD}$ given by (26), has singularities at $\, x = \, x_i$ with 
multiplicities $\,q_i$.  Besides the true singularities of the generating 
function $\, S(x)$, the $\, x= \, x_i$ may be apparent singularities when 
$M=\, M_0$, the minimum order, or they may be $\, M$ dependent spurious 
singularities when $\,  M >\, M_0$. In any case the ODE is necessarily of the form
\begin{eqnarray}
\label{(B.1)}
&&	P_D(x) \cdot L_x 
\cdot (x \cdot {{\rmd} \over {\rmd x}})^{M \, -1} (S(x))	
	+ \cdots \,   = \,  \,  0,	\\
&& L_x \, = \,  x \cdot {{\rmd} \over {\rmd x}} -s_0 \,
    +\sum_i(q_i \cdot M \,  -{{1} \over {2}} \, q_i \, (q_i+1)\, -s_i)x/(x-x_i),
 \nonumber 
\end{eqnarray} 
with $\, x_i$ the roots of the degree $\, D$ head polynomial $\, P_D(x)$.  One 
can explicitly verify that $\, s_0$ is the coefficient of $p^{M-1}$
in the indicial equation of degree $\,M$ at $\,x=\, 0$, 
\begin{eqnarray} 
\label{(B.2)}
        p^M \, - s_0 p^{M-1}\,  + \cdots\,  + constant\, \, =\, \,  0,
\end{eqnarray} 
and thus is the sum of $\,M$ exponents at $ \, x=\, 0$.  Similarly, $s_i$ is the
 coefficient of $p^{ q_{i}-1}$ in the  indicial equation of degree $\, q_i$ at $\, x=\, x_i$,
\begin{eqnarray} 
\label{(B.3)}
	p^{q_i}\,  - s_i \cdot p^{ q_{i}-1} \, \,  +\, \cdots  \, + 
\, constant\,  = \, \, 0,	
\end{eqnarray} 
and is the sum of the $\, q_i$ singularity exponents at $\, x= \, x_i$. 

If we set $\, x\,=\, 1/y$, the transformed ODE (\ref{(B.1)}) is
\begin{eqnarray} 
\label{(B.4)}
&& P_D(1/y) \cdot L_y \cdot (y \cdot {{\rmd} \over {\rmd y}})^{M -1} (S(1/y))
			+ \cdots  \, =  \, \, 0,	 \\
&& L_y \, = \, y \cdot {{\rmd} \over {\rmd y}} \,
 +s_0 \, -\sum_i(q_i \cdot  M \, -{{1} \over {2}} q_i \, (q_i+1)\,
 -s_i)/(1\, -y\, x_i), \nonumber 
\end{eqnarray} 
and we can identify
 $-s_0 \, +\sum_i \, (q_i \cdot M  \, 
-{{1} \over {2}} \, q_i \, (q_i+1) \, -s_i)$ with 
$\, s_{\infty}$, the
 coefficient in of $p^{M-1}$ in the  indicial equation of degree $\, M$ at $\, y=\, 0$,
\begin{eqnarray} 
\label{(B.5)}
        p^M - s_{\infty}\cdot p^{M-1} \, +\, \cdots\, + \,constant \,  = \, \, 0
\end{eqnarray} 
and this coefficient is thus the sum of the $\, M$ exponents at $\,y=\, 0$. 
 This exponent equivalence is the sum-rule
\begin{eqnarray} 
\label{(B.6)}
	s_0 \, + s_{\infty}\, 
+\sum_{i, \, all}\,(s_i\, + {{1} \over {2}}\,q_i \,(q_i+1))
\, = \, \,M \cdot \sum_i \, q_i
 \,=\,\, M \cdot D,	
\end{eqnarray} 
and is the starting point for our specialisations below. It is easy to show that 
(\ref{(B.6)})  is equivalent to the usual Fuchs' relations~\cite{Waall,Kodai}
which are sum-rules on all exponents.  The advantage of (\ref{(B.6)}) is that
it makes explicit the role of those singularity exponents that arise as solutions 
of the indicial equations of (typically) much smaller degree dictated by the 
multiplicity of the head polynomial zeros of the linear ODE.

We now assume that the true singularities and their associated exponents have been 
determined.  If $\, M>\, M_0$, then (\ref{(B.6)}) is a constraint on the spurious
singularity exponents and is of little interest.  On the other hand if $\, M\,=\,\, M_0$,
then there are no spurious singularities and the $\, x= \, x_i$ are either true 
singularities or apparent singularities. Furthermore, in many cases of interest 
as we found in our analysis of $\, \tilde{\chi}^{(3)}$ and $\, \tilde{\chi}^{(4)}$, the multiplicities 
of the apparent singularities at $\, x=\, x_i$ are all $\, q_i =\, 1$ and
the exponents are all $s_i\, =\,\, M_0$.  Then the contribution to
$\, \sum_i (s_i\, +q_i \, (q_i+1)/2)$ coming from the apparent  singularities is simply
$\, \sum_i \, (s_i+1) \, = \, \sum_i (M_0+1)\,=\,\, D_{app} \cdot (M_0+1)$ with 
$\, D_{app}\, =\,\,  D\, -D_0$ and $\, D_0$ the minimum possible degree of the ODE.  The
sum-rule (\ref{(B.6)}) can now be rewritten as
\begin{eqnarray} 
\label{(B.7)}
&&s_0 \, + s_{\infty} \,
 +\sum_{i, \, true}(s_i\,+ \, {{1} \over {2}} \, q_i \cdot (q_i+1))
 \, =\,\,  \\
&& \qquad \, \,\,\,\,  M_0 \cdot D\,\, -D_{app} \cdot (M_0+1)\, 
 =\,\,   M_0 \cdot D_0 \, - D_{app},	\nonumber 
\end{eqnarray} 
which is an explicit formula for $\, D_{app}$ in terms of the true singularities only. 
To emphasize this point we note that for any physical problem the true singularities 
are determined and encoded in the generating function $\, S(x)$.  By deciding to 
represent this information as a linear ODE of minimum order of the form (\ref{28a}) we 
are forced to specify a total of $\, (M_0+1)(D_0\,+D_{app}\, +1)$ coefficients with 
$\, D_{app}$ given by (\ref{(B.7)}).  For the $\, S= \, \tilde{\chi}^{(5)}$ example treated 
in the text, $\, s_0\,=\,\, 192$ from Table~\ref{Ta:C5expo} and $\, s_{\infty}= \, 147$,
which is 56 from Table~\ref{Ta:C5expo} plus the 14 term sum $\, 0+1\, +\cdots \,+13\,=\, 91$ 
from the additional regular solution exponents inferred to be in the degree $\, M_0=\, 33$ 
indicial equation.  The remaining true singularity exponent sum in (\ref{(B.7)}) is 653 
from Table~\ref{Ta:C5expo}. On solving (\ref{(B.7)}) for $D_{app}$ we get the value 
$\, D_{app} \, = \, \, 1384$ given in the text.  The number of ODE coefficients
is $\, (M_0+1)(D_0\, +D_{app} \, +1)\,  = \,(33+1)(72+1384+1)\, =\,\, 49538$.

There are a number of related observations that are significant.  Firstly, the presence of 
apparent singularities in the minimum order ODE implies that there are constraint 
conditions~\cite{Ince}, namely $\, M_0-1$ conditions for each singularity in addition 
to the observed exponent value $\, s_i =\, M_0$. Thus $\, D_{app} \cdot M_0$ coefficients
in total are fixed by the constraints and this means that in principle we need only 
$\, (M_0+1)(D_0+1) \, +D_{app}$ series terms to determine the minimum order ODE. 
In the $\, \tilde{\chi}^{(5)}$ example this is $ \,(33+1)\,(72+1) \, +1384 \, =  \, \,3866$
which is much smaller than the $>\, 7000$ terms for any $\, \tilde{\chi}^{(5)}$ ODE given 
in Table~\ref{Ta:C5}.  Unfortunately, we know of no practical way to implement the 
constraints as they are in general non-linear.  Specifically, let us write the minimum
order ODE operator as $\, \sum_m \, f_m (x \cdot {{\rmd } \over {\rmd x}})^{M_0 \, -m}$  
with $\, f_m\,=\,\, 0$ for $\, m\,>\, M_0$ and the head polynomial $\,  f_0$ factored as
$\,f_0\,  =\, \, P_{true} \cdot P_{app}$, thus clearly separating the true and 
apparent singularities.  The apparent singularity constraint conditions are then the 
statement that each
\begin{eqnarray} 
\label{(B.8)}
&&F_m \,  = \,\, \,  [f_m \cdot (f_m\, + x \cdot {{\rmd f_{m-1}} \over {\rmd x}}) \, 
-f_{m-1} \cdot (f_{m+1} \, +x \cdot {{\rmd f_m} \over {\rmd x}})]/P_{app}, \nonumber 	\\
&& \qquad  \qquad  m=\, 1,\,2,\, \cdots , \, M_0, 	
\end{eqnarray} 
is a polynomial of degree $\, 2\cdot D_0 \, +D_{app}$.  It is the exact 
division and reduction in degree from $\, 2 \, (D_0 \, +D_{app})$ that implies the 
existence of $\, D_{app}$ conditions in the numerator of the right hand side of 
(\ref{(B.8)}) for each individual $\, m$.

Secondly, although the number of coefficients needed in an ODE can be dramatically reduced 
by moving away from minimum order\footnote[8]{In the $\, \tilde{\chi}^{(5)}$ example we can 
move from the $ \,49538$ coefficients given above for $M\,=\, M_0 =\, 33$ to 
$\, (M+1)(D+1)\, =\, \, 7410$  given in Table~\ref{Ta:C5} for $\, M= \, 56$.}, this is offset 
by a dramatic increase in the size of the integer coefficients specifying the ODE. This 
observation is based on our experience with a   $\, 6 \,\tilde{\chi}^{(3)}\,-\tilde{\chi}^{(1)}$ 
analysis.  The minimum order ODE has $\, M_0\,=\,\, 6$ and degree$\, D= \, 40$.  
With the normalization choice $\, a_{M0} \, =\, 1$ in (\ref{28a}), the remaining coefficients 
are integers and can be found by a mod prime and Chinese remainder theorem analysis 
using five primes.  In contrast, with the same normalization, the ``best" choice ODE 
from Table~\ref{Ta:ODEs} with $M =\, 10$, $\, D=\, 17$ has as its remaining coefficients 
rational fractions with numerator and denominator integers each typically 160 digits in 
length.  To find these requires a supplementary continued fraction calculation starting
from Chinese remainder theorem residues of 320 digits and requires about 70 primes.  
Similar results hold for other non-minimum order ODE's for  
$\,\,  6\tilde{\chi}^{(3)}\,-\tilde{\chi}^{(1)}$ and we believe analogous disparities are likely 
in any  $\, \, 2\tilde{\chi}^{(5)}\,-\tilde{\chi}^{(3)}$ analysis.  But we need to emphasize 
that the utility of the ``best'' mod prime ODE as a recursion relation device \cite{GutJoy} 
remains. It enables us, for any given prime, to extend the shortest possible generating 
function series to the length necessary to find the minimum order ODE.

\section{On the Landau singularities \label{appendixBernie}}

The discussion of the Landau singularities of Ising like integrals in~\cite{bo-ha-ma-ze-07b} 
is based on analogues of the $(2n-2)$-dimensional integrals for $\tilde{\chi}^{(n)}$ given by 
Wu {\it et al.}~\cite{wu-mc-tr-ba-76}.  Here we derive and extend those results based on 
analogs of the $(n-1)$-dimensional integrals (\ref{chi3tild}).  The calculations are 
complementary, each having certain advantages and disadvantages.

In the forthcoming technical discussion we will often use the following
definitions supplementing those in (\ref{chi3tild}-\ref{they}).
The square root factor appearing in $x_i$ and $y_i$ we denote by:
\begin{eqnarray} 
\label{seven}
 f_i\, =\, \,\frac{1}{2 w}\cdot \sqrt{(1-2w \cos(\phi_i))^2 -4 w^2 }
 \,= \,\,1/y_i.
\end{eqnarray} 
We also define 
\begin{equation} 
\label{eight}
\cos( \zeta_i)\, =\,\, 1/(2w )-\cos(\phi_i), \qquad \quad  \, 
 \sin(\zeta_i) = \,\rmi \cdot f_i  
  \end{equation}
so that we can write: 
 \begin{equation} 
\label{nine}
 x_i\, =\, \,\exp(\rmi \zeta_i).
\end{equation} 
 
The formulae (\ref{seven}-\ref{nine}) are understood to apply for $s$ or $w$ small 
and elsewhere by analytic continuation. Furthermore we will take it to be understood 
that by the integral $\tilde{\chi}^{(n)}$ we mean (\ref{chi3tild}) and various analytic
continuations of the series analytic at $\, w\, = \, \, 0$. This means we do not 
distinguish between ``integral" and that ``particular solution" of the linear ODE 
that agrees with (\ref{chi3tild}) for small $w$ (or $s$) but whose domain is not restricted 
in any way\footnote[1]{Thus completing the generalizations already in print.  
In~\cite{ze-bo-ha-ma-04} ``integral" was understood to mean a single-valued 
function in the cut $w$ plane.  The definition of ``integral" was extended 
in~\cite{ze-bo-ha-ma-05c} to be a single valued function in a cut $s$ plane 
and hence double-valued in $w$.  Finally, in a Landau singularity analysis 
as in~\cite{bo-ha-ma-ze-07b} one no longer attempts to specify on which {\em local}
Riemann sheet any particular singularity occurs.}. This is a great simplification 
for our discussion below since the value of $\tilde{\chi}^{(n)}(w_c)$ reached by analytic 
continuation in general depends on the path chosen for $w$ between 0 and $w_c$ and 
this can lead to very complicated topological considerations. We do not address any 
of that here but note that some of the differences in $\tilde{\chi}^{(n)}$ on different 
branches might arise because analytic continuation requires the displacement of 
branch cuts which result in changes in the signs of $f_i$ and $\zeta_i$
in (\ref{seven}-\ref{nine}).   The results we describe below {\em allow for all 
possible sign changes} and thus the singularity list is the complete list 
covering all these {\em local} signs.

This must be borne in mind when, for example, we say $\tilde{\chi}^{(n)}$ is singular at 
certain points $| s | < 1$ since it is certainly the case that the radius 
of convergence of $\tilde{\chi}^{(n)}$  is $| s | =\, 1$. The latter observation 
follows trivially from the fact that if the phases in (\ref{chi3tild}) are
real then for $| s |\, < \,  1$ one finds that $f_i$ cannot vanish 
and also $| x_i | < | s |$. When we want 
to restrict the domain on which the integral is defined\footnote[5]{And is univalued!}
to the region $| s | \le 1$  (and on which for $| s | \, < \,1$ the 
series applies), we will explicitly indicate this by making reference to the
``principal disc" or the ``integral on the principal disc".

The points $s=\pm 1, \pm \rmi$ or equivalently $w=\pm 1/4, \infty$ are fairly obvious
singularities of $\tilde{\chi}^{(n)}$ and to simplify the analysis we
explicitly exclude these points.  A consequence of this exclusion is that (\ref{eight}) implies that if $\sin(\zeta_i)=\, \sin(\phi_i)=0$  then 
$w=\pm 1/4$ or $\infty$.  Thus when we exclude these singular points 
{\it we are preventing $\sin(\zeta_i)$ and $\sin(\phi_i)$ from vanishing simultaneously, a result we will use on a number of occasions below}.

The discussion of the Landau singularities of an integral like (\ref{chi3tild}) is 
somewhat simplified by the fact that the integration is over a unit cell of a periodic
function and thus there are no end-point singularities. In fact it is best to think 
of the $\phi_i$ integrations as closed contour integrations in $z_i=\, \exp(\rmi\phi_i)$ 
that can be arbitrarily deformed away from the unit circle provided no integrand 
singularities are crossed while the constraint $\prod z_i =\, 1$ is maintained.  
These singularities in (\ref{chi3tild}) are at $f_i=\, 0$ for all 
$i$, $\, x_i\,  x_j\, =\, 1$ for all $i \ne j$ and $\prod x_i=\, 1$. They are not 
all independent since, for example, $f_i=\, 0$ implies $\zeta_i=0$ or 
$\pi$ and $x_i=\pm 1$.  Thus the vanishing of $f_i$ can overlap with the $x_i$ 
product singularities. However it is important to note that $f_i=0,$ which is a 
condition on a single $\phi_i,$ is very different from $\prod x_i=\, 1,$ which is a
relationship between all  $\phi_i$. If  $f_i=\, 0$ cannot be avoided by  the 
$z_i$ contour for some particular $w,$ then the integral $\tilde{\chi}^{(n)}$ is singular at that $w.$  
On the other hand $\prod x_i=1$ can lead to a singularity of the integral only if 
$\prod x_i$ is also stationary with respect to variation in all $\phi_i$ subject 
to the phase constraint in (\ref{they}). Both $f_i\, =\, 0$ and $\prod x_i=\, 1$ 
are examples of pinch singularities, but to distinguish the simpler $f_i=\, 0$ case 
we will refer to it exclusively as a pinch singularity in the discussion below.
The $\prod x_i=1$  case or any similar situation in which a non-trivial
stationary condition must also be satisfied we will call a van Hove singularity 
in recognition of his analysis~\cite{vanHove} that predates that of 
Landau~\cite{Landau} by several years. Another singular integral situation arises 
when, say, $\, x_1 \, x_2\, = \, 1$ and $\prod x_i=1$ are simultaneously satisfied 
but rather than $x_1x_2$ and $\prod x_i$ being separately stationary the normals 
to these two hypersurfaces are parallel (\cite{eden}, p.48). In this case the 
integration variables are trapped between two distinct hypersurfaces that touch 
tangentially.

Both to simplify the discussion and because this includes the most important 
situations, we start with the Landau problem of only $\prod x_i=1$  and $f_i=\, 0$
as singularities in the integrand of $\tilde{\chi}^{(n)}$. Since $x_i=\, \exp(\rmi\,\zeta_i)$  
this product constraint together with the phase constraint in (\ref{they}) results 
in the symmetric pair
\begin{eqnarray}
\label{ten}
 \sum \phi_i = 0 \,\, \quad  {\rm mod} \, \, 2\pi, \qquad \,\quad 
  \sum \zeta_i = 0 \, \,\quad   {\rm mod} \,\, 2\pi.
\end{eqnarray}
The phase constraint we handle directly by taking $\phi_i$, $i=\, 1, \cdots, n-1$, as 
independent so that $\zeta_i=\zeta_i(\phi_i)$, $i=\, 1,\,  \cdots,\,  n-1$, and 
$\zeta_n=\, \zeta_n(\phi_n)=\, \zeta_n(2\pi k\, -\phi_1- \cdots - \phi_{n-1})$. The 
requirement that $\sum \zeta_i$ be stationary with respect to phase variation is that 
the derivative combinations $\zeta_i'-\zeta_n'$ vanish where $i=\, 1,\, \cdots,\, n-1$ 
and $\zeta'_i=\, \partial{\zeta_i}/\partial{\phi_i}=-\sin(\phi_i)/\sin(\zeta_i)$. The last 
equality follows from the definition (\ref{eight}). The derivative conditions can be 
rewritten as
\begin{eqnarray}
\label{eleven}
\sin(\zeta_i)\cdot \sin(\phi_j) \,= \,\,\sin(\zeta_j)\cdot \sin(\phi_i), 
\quad \quad \quad   \,\,\,i \ne j
\end{eqnarray}
and our derivation requires that (\ref{eleven}) be subject to the restriction that no
$\sin(\zeta_i)$ vanishes. However this restriction can be dropped because we are looking 
only for solutions for which $\sin(\phi_i)$ and $\sin(\zeta_i)$ are not simultaneously 
zero and the case that (\ref{eleven}) yields all $\sin(\zeta_i)=0$ is nothing but the 
pinch singularity condition we must investigate also.  We can rewrite (\ref{eleven})
for each $i,j$ combination as the pair
\begin{eqnarray} 
\label{twelve}
\fl \qquad (\cos(\phi_i)-\cos(\phi_j))\cdot (4w\,-\cos(\phi_i)\,-\cos(\phi_j)\,
+4w \cdot  \cos(\phi_i)\cos(\phi_j))
\, =\, \,  0,\nonumber \\
\fl \qquad (\cos(\zeta_i)-\cos(\zeta_j))\cdot (4w\,-\cos(\zeta_i)\,-\cos(\zeta_j)\,
 +4w \cdot \cos(\zeta_i)\cos(\zeta_j))
\,  =\,\,   0
 \end{eqnarray}
by squaring, rewriting the sine functions in terms of cosines, and utilizing the 
definition (\ref{eight}) which we reproduce here in symmetric form
  \begin{equation} 
  \label{thirteen}
\cos(\phi_i)\,+\cos(\zeta_i) \,= \, \, 1/(2 w),
 \,\quad \quad \quad   \, \, i\,=\,\, 1 \,\ldots \, n.
 \end{equation}
Note that (\ref{twelve}) and (\ref{thirteen}) allow all possible sign combinations 
$\pm \phi_j$, $\pm \zeta_j$ for a given $\phi_i$, $\zeta_i$ pair and one must in 
all cases check that (\ref{eleven}) is also satisfied.  That there is a remaining 
sign degeneracy allowed by (\ref{eleven}) is a consequence of our decision to 
define our integral to include all possible sign combinations in the local Riemann 
sheet. If the domain of $w$ were to be restricted to a particular Riemann sheet 
then additional analysis would be required to determine the uniquely signed solution.

The equations (\ref{ten}-\ref{thirteen}) are the Landau conditions for our
reduced problem of $\prod x_i =1$ and $f_i=0$ as the only singularities.  
We now consider specific situations, Case 1 to Case 5.

\subsection{Case 1: the phases $\phi_i$ and $\zeta_i$  equal 0 or $\pi$}
\label{case-i}

The simplest case is that of phases $\phi_i$ and $\zeta_i$  equal 0 or $\pi$.
The stationary constraint (\ref{eleven}) is trivially satisfied and  (\ref{thirteen}) implies singularities at $w=\pm 1/4$ or $\infty$.  
These are the points we excluded from the analysis and we again note that $\sin(\phi_i)$ and 
$\sin(\zeta_i)$ cannot simultaneously vanish.

\subsection{Case 2: all $\phi_i$  equal and all $\zeta_i$ equal}
\label{case-ii}

Equally simple is the case of all $\phi_i$  equal and all $\zeta_i$ equal.  
Again the stationary constraint (\ref{eleven}) or (\ref{twelve}) is satisfied
automatically and the phase constraints (\ref{ten}) are satisfied with
\begin{eqnarray}
\label{fourteen}
\phi_i\,= \,\,2 \pi k/n, \quad   \quad 	\zeta_i\, =\,\,  2 \pi m/n, 
\quad \quad \quad   0 \,\le\, k,\,\,m\, <\, n	
\end{eqnarray}
and with $\phi_i$ and $\zeta_i$ not both 0 or $\pi$ so as to exclude Case 1.  
The associated singularities are at
\begin{eqnarray}
\label{fifteen}
1/(2 w_{km})\, = \,\,s_{km}\, + 1/s_{km}\, =
\, \,\cos(2 \pi k/n)\,+\cos(2 \pi m/n).
\end{eqnarray}
Because the stationary constraint for these singularities is automatic the
full Landau formalism is not necessary and  (\ref{fourteen}), (\ref{fifteen})
could have been guessed just as they were in~\cite{nickel-99}.  Note that 
$| s_{km} | =1$ and the calculation in~\cite{nickel-99} shows them to 
be principal disc singularities. In the following we will designate them 
as Nickelian singularities to distinguish them from other van Hove singularities. The 
singularities $\, w_{km}$  can be given as the roots of polynomials in $\, w$ 
with integer coefficients.

{\bf Remark 1}
A new situation not considered
in~\cite{nickel-99} arises out of the Landau formalism.
For a given $\phi_i$, $\zeta_i$ combination, the constraints
(\ref{eleven}-\ref{thirteen}) allow $\phi_j$, $\zeta_j$ to be  $-\phi_i$, $-\zeta_i$  
in addition to the $\phi_i$, $\zeta_i$  we have considered. {\em For every such sign 
reversal there is one pairwise cancellation in the constraint sums (\ref{ten}) so 
that we should add to (\ref{fourteen}), (\ref{fifteen}) new singularity conditions 
obtained by the replacements $n \rightarrow n-2$, $n \rightarrow n-4$, etc.}  
We will however leave (\ref{fourteen}), (\ref{fifteen}) unchanged and if there
is a possibility of confusion, refer explicitly to (\ref{fourteen}), (\ref{fifteen}), 
for which there has been no pairwise cancellation, as the ``irreducible" conditions.  
The $n \rightarrow n-2m$ replacement singularities will be treated separately
as Case 5 below as this situation arises numerous times.

{\bf Remark 2}
{\em A very important observation concerning the $n \rightarrow n-2m$
replacement singularities is that none of them are principal disc singularities}.
To see this note that (\ref{seven}), (\ref{eight}) defines $\zeta_j$  as an even 
function of $\phi_j$ for $| s | < 1$ on the principal disc and by 
continuity to the limiting case $| s | = 1$  as well. On the other hand, 
for fixed $\phi_i$, $\zeta_i$,  (\ref{eleven}) requires $\zeta_j$ to be an odd 
function of $\phi_j$. Thus while (\ref{eleven}) allows $\phi_j$, $\zeta_j$ to 
be $-\phi_i$, $-\zeta_i$ in addition to the $\phi_i$, $\zeta_i$, this singularity 
cannot be on the principal disc but {\em rather must be on those other 
``Riemann sheets'' on which the $f_j$ square root function has the opposite sign}.  
The numerical evidence is consistent with this result. It was already shown
in~\cite{nickel-99} that $\tilde{\chi}^{(3)}$ principal disc singularities were not 
present on the principal disc of $\tilde{\chi}^{(5)}$.

\subsection{Case 3: all $\sin(\zeta_i)=0$, i.e. all $f_i=0$, 
so that $\zeta_i=0$ or $\pi$}
\label{case-iii}  

At the next level of complexity consider the possibility that all $\sin(\zeta_i)=0$, 
i.e. all $f_i=0$, so that $\zeta_i=0$ or $\pi$. The constraint (\ref{eleven}), which 
in this case specifies a pinch singularity, is trivially satisfied. We find from
 (\ref{thirteen}) that there are only two possible values for $\cos(\phi_i)$, 
namely $\cos(\phi^{(+)})=\, 1/(2w)+1$ and $\cos(\phi^{(-)})=\, 1/(2w)-1$. The former 
is associated with $\zeta_i=\pi$ and if we demand that the singularity condition in 
(\ref{ten}) also be satisfied then we must have an even number of these terms. We implement 
the phase 
constraint (\ref{ten}) on $\phi_i$  as follows. Since there are only 
two possible values\footnote[2]{Exactly as in Case 2 we consider only the ``irreducible" 
case in which a given value of $\cos(\phi_i)$ defines a uniquely signed $\phi_i$ and 
hence there are no pairwise cancellations in the constraint $\sum \phi_i$. We will 
remark further on the general situation, considered as Case 5, later.} we can write 
$k\phi^{(+)} +(n-k) \phi^{(-)} =0$  mod $2\pi$ and from the preceding remarks, $k$ 
is even.  Equivalently, $\exp(\rmi k\phi^{(+)}+\rm i(n-k)\phi^{(-)})=\, 1$ or  
$\exp(\rmi k\phi^{(+)})=\exp(-\rmi(n-k)\phi^{(-)})$. Now add to this last equation the 
reciprocal and obtain $\cos(k\phi^{(+)})=\cos((n-k)\phi^{(-)})$. This form is 
convenient because each $\cos(m \phi)$ can be expressed simply in terms of 
$\cos(\phi)$. On using the definition of Chebyshev polynomials we get
\begin{eqnarray} 
\label{sixteen} 
&& T_k(1/(2 w_k)+1) \, = \, \,  T_{n-k}((1/(2 w_k)-1), \,\,  \\
&& \quad \quad \quad \quad  0 < k < n, \,\, 
\quad \quad  {\rm and}\,\,\quad\quad   k \,\, {\rm even},  \nonumber 
\end{eqnarray}
as the defining equation(s) for the singularities. Note that we have excluded 
$k=0$ or $n$ since all $\phi_i$ are equal and (\ref{sixteen})
becomes a special case of (\ref{fourteen}), (\ref{fifteen}).

{\bf Remark 3}
The singularity conditions (\ref{sixteen}) {\em do not allow for singularities on} 
$| s | =1,$  as we now show. The condition $| s | =\, 1$ is the 
condition $w$ real and $-2 < 1/(2w) <2$. Consider first the interval $0 < 1/(2w) <2$.  
Then the argument of $T_k$ in (\ref{sixteen}) lies between 
1 and 3 and $\vert T_k \vert > 1,$ while the argument of $T_{n-k}$ 
lies between $-1$ and 1 and $\vert T_{n-k} \vert \le 1$. Thus (\ref{sixteen}) cannot 
be satisfied. A similar argument applies on the interval $-2 <1/(2w) <0$.  
Since we excluded $w=\pm 1/4$ and $\infty$ at the outset, the proof is complete.

{\bf Remark 4}
The same integral singularity conditions (\ref{sixteen}), {\em but with odd index 
$k$ allowed as well}, were given in~\cite{nickel-05} based on the integrand 
singularity condition $f_i=0$ for all $i$ irrespective of $\prod x_i$. We see here 
the partial overlap between the all $f_i=0$  and $\prod x_i=\, 1$ conditions, but 
because $f_i=\,0$ is the more general condition it would appear that odd $k$ should 
be included in (\ref{sixteen}). In fact (\ref{sixteen}) {\em is correct because of 
the very special structure of the Ising integrals} (\ref{chi3tild}). This will be 
proved below. But first note that the integrals evaluated in~\cite{bo-ha-ma-ze-07},
where the integrand in (\ref{chi3tild}) was replaced by the simpler pole product 
$\left( \prod y_i \right)^2$, all show that both odd and even $k$ are required in 
that case. That is to say, for these integrals the pinch singularity conditions are 
both necessary and sufficient and the same should apply also to the Ising $\, \tilde{\chi}^{(n)}$.
The only way contours that are trapped in the process of analytic continuation in the 
toy integrals could escape being trapped in  $\tilde{\chi}^{(n)}$ is for at least one $f_i$ 
factor to become non-singular. {\em The mechanism for this to happen is that in the 
vicinity of a potential pinch an $f_i$ appears not in a Laurent series but rather in 
a Taylor series in powers of $(f_i)^2$. Given the complexity of the Ising $\tilde{\chi}^{(n)}$  
integrand this would seem rather miraculous but in fact happens as we now show.}  
Not too surprisingly, this point was entirely missed in~\cite{nickel-05}.

The argument below is for a pinch singularity defined by $f_i=0$ for all $i$.  
We begin by noting that at a point where $f_i$ is small, $\zeta_i$ is either small 
or near $\pi$. For purposes of the present argument only, we replace those $\zeta_i$
near $\pi$ by $\zeta_i-\pi$ and incorporate this change consistently by the simultaneous 
replacement $x_i \rightarrow -x_i$. The factor 
$\left( 1+\prod x_i \right)/\left( 1-\prod x_i \right)$ in the $\tilde{\chi}^{(n)}$
integrand can then be written as $\cot\left( \sum \zeta_i /2 \right)$ if an even number 
of such replacements were made or $\tan\left( \sum \zeta_i /2 \right)$ if an odd 
number of such replacements were made. In the first case, only a Laurent expansion is 
possible and the pinch singularity is qualitatively like that in the toy integral with 
integrand $\left( \prod y_i \right)^2$. Since that integral was observed to have the 
singularities (\ref{sixteen}) the same must be true here and we conclude that the 
necessary conditions (\ref{sixteen}) are also sufficient for an integral singularity.

However, in the odd replacement case a Taylor expansion in odd powers of $\sum \zeta_i$ 
exists.  This will yield a sum of terms of the form $\prod \left( \zeta_i \right)^{p_i}$ 
with with $\sum p_i$ odd. Since $f_i\, =\, \rmi\cdot \sin(\zeta_i)$ implies $\zeta_i$ is 
an odd function of $f_i$ the expansion could equally well be written as a sum of 
$\prod \left( f_i \right)^{p_i}$ with again $\sum p_i$ odd. Because $\sum p_i$ is odd 
there is at least one $p_i$ which is odd and of course this $p_i >0$.  Now recall that 
the $\tilde{\chi}^{(n)}$ integrand also contains the factor $\prod y_i = \, \prod (1/f_i)$ and 
this shifts every $p_i$ down by one. In particular the original odd $p_i$ has now become 
even and is still non-negative so that the integration over the analytically continued 
$\phi_i$ encounters no singularity, i.e. no pinch.  This completes the argument for 
those proxy Ising integrals that do not contain the $\left( G^{(n)} \right)^2$ factor.

To show that the presence of $\left( G^{(n)} \right)^2$ in the integrand does not 
change the argument for an absence of singularities we first note that $G^{(n)}$
is a product of terms each of which is of the form
\begin{equation} 
\label{seventeen}
 2\, \sin((\phi_i-\phi_j)/2) \cdot \frac{\sqrt{x_i x_j}}{(1-x_i x_j)}. 
\end{equation} 
The $x_i x_j$ dependent factor is proportional to either $\sin((\zeta_i + \zeta_j)/2)$ 
or  $1/\sin((\zeta_i + \zeta_j)/2)$, after making the replacements described in the 
penultimate paragraph, so that now $\zeta_i$ and $\zeta_j$ are understood to always 
be near zero. The sine function in the numerator of (\ref{seventeen}) is crucial in 
canceling a potential zero in the denominator, but has a complicated representation in 
terms of $\zeta_i$ and $\zeta_j$. For purposes of a singularity existence analysis it 
is legitimate to replace it by any more convenient analytic and asymptotically 
linear\footnote[3]{Linear in the deviation of $\phi_i$ from the singular point.  
This means a linear function in the quadratic $f_i^2$.} function in the neighbourhood 
of $\zeta_i =\zeta_j =0$ and we choose instead of $\sin((\phi_i-\phi_j)/2)$ the 
difference $(f_j)^2-(f_i)^2 = \sin^2(\zeta_i)-\sin^2(\zeta_j)$.  Then (\ref{seventeen}) 
is proportional to one of
\begin{eqnarray} 
\label{eighteen}
&&(\sin^2\zeta_i-\sin^2\zeta_j)\sin((\zeta_i+\zeta_j)/2),  \nonumber \\
&&(\sin^2\zeta_i-\sin^2\zeta_j)/\sin((\zeta_i+\zeta_j)/2) \, =\,  \\
&& \qquad \qquad \qquad 
 2\, (\sin\zeta_i - \sin\zeta_j)\cdot \cos((\zeta_i-\zeta_j)/2).\nonumber 
\end{eqnarray}

Both of the expressions in (\ref{eighteen}) have a Taylor expansion with terms of the 
form  $\left( \zeta_i \right)^{p_i} \left( \zeta_j \right)^{p_j}$ with $p_i+p_j$ odd.
An equivalent expansion is in terms 
$\left( f_i \right)^{p_i} \left( f_j \right)^{p_j}$  with $p_i+p_j$ again odd.  
The square of this expansion will be similar but with $p_i+p_j$ now even.  
Finally the expansion of $\left( G^{(n)}\right)^2$ must be a sum of terms of the 
form $\prod \left( f_i \right)^{p_i}$ with $\sum p_i$ even and multiplying the 
expansion obtained in the absence $\left( G^{(n)}\right)^2$  by terms
$\prod \left( f_i \right)^{p_i}$ for which $\sum p_i$ is even cannot change the 
argument for the absence of a pinch.
 
To summarize this case, the pinch conditions give (\ref{sixteen}) as necessary. 
The evidence from the toy model study~\cite{bo-ha-ma-ze-07} is almost
certainly a proof that (\ref{sixteen}) is also
sufficient for these singularities of $\tilde{\chi}^{(n)}$.

{\bf Remark 5}
A situation closely related to $\sin(\zeta_i)=\, 0$, for all $i$, as treated above is 
$\sin(\phi_i)=0$, for all $i$.  This leads to van Hove singularities in which the 
stationary conditions (\ref{eleven}) and both conditions (\ref{ten}) must be 
satisfied. Each $\phi_i$ is either 0 or $\pi$.  The phase constraint then requires 
an even number of $\phi_i=\, \pi$ values.  On imposing the singularity constraint 
in (\ref{ten}) the same solution (\ref{sixteen}) is obtained.

{\bf Remark 6}
This equality of solutions for all $\sin(\zeta_i)=0$ and all $\sin(\phi_i)=\, 0$ 
is expected and illustrates a very important aspect of the Landau analysis of the 
integral representation (\ref{chi3tild}) for $\tilde{\chi}^{(n)}$. The original representation 
for $\tilde{\chi}^{(n)}$  given in~\cite{wu-mc-tr-ba-76} is an integral over two sets of 
$n-1$ independent phases with an integrand that is symmetric under the interchange of 
these sets\footnote[4]{The invariance under phase variable interchange is not always 
obvious in a particular formula. See the remarks in Wu {\it et al.}~\cite{wu-mc-tr-ba-76}
following their equation (4.87) on this point.}. Integrating out one set to arrive 
at (\ref{chi3tild}) {\em has obviously broken this symmetry, but a vestige of it remains 
in the two conditions} (\ref{ten}). The $\phi_i$ are the remaining dummy phase variables 
of integration while the $\zeta_i=\, \zeta_i(\phi_i)$ are the specific values that result 
from evaluating the other set of phases at the Wu {\it et al.}~\cite{wu-mc-tr-ba-76}
integral singular points. Finding singularities of the integral such as (\ref{sixteen}) 
amounts to completing the integration process and while the steps associated with
dealing with the ``integrated" $\zeta_i$ versus the ``unintegrated" $\phi_i$ are 
obviously different, once the $\phi_i$ values have been fixed by the Landau conditions 
{\em the symmetry is restored} and the final answers one obtains must be the same.
That is, {\it for every combination} of the $\zeta_i$, $\phi_j$ leading to
an integral singularity there is another set $\phi_i$, $\zeta_j$ obtained by 
$\zeta \leftrightarrow \phi$ {\it interchange that leads to the same singularity}
and we can always choose the combination that requires the least computational effort.

{\bf Remark 7}
The above comparison of alternatives also illustrates that {\em there is no fundamental 
distinction between the pinch and van Hove singularities}. {\em Rather it is just a 
distinction in the ordering of the steps of the calculation}.  This is particularly 
relevant when comparing the calculations here to those in~\cite{bo-ha-ma-ze-07}.
There the Landau analysis is done at the $(2n-2)$ integral stage and there is no analogue
of the pinch singularity in the restricted sense we have defined here.
{\em Instead, every singularity evaluation requires a supplementary stationary
condition evaluation and thus all singularities are of van Hove type.}

This concludes our treatment of the singularities associated with the pinch condition 
$\sin(\zeta_i)\,=\,\,0$ and the corresponding special van Hove case $\sin(\phi_i)=0$. 
This list is exhaustive as there can be no integral singularities for the mixed case 
that $\sin(\zeta_k)=\,0$, $k=\,1,\, \cdots, \,m$, and 
$\sin(\zeta_j) \,\ne\, 0$, $j\,=\, m+1, \cdots, n$, with $1 \,\le\, m\, <\, n$.  
To see this note that while the $\zeta_k$ is a singular, i.e. square root, function of 
$\phi_k$ in the neighbourhood of $\sin(\zeta_k)=0$, the $\phi_k$ is an analytic 
function of $\zeta_k$. Specifically, $\phi_k$ is asymptotically a constant with an
added quadratic dependence on the deviation $\delta \zeta_k$ of $\zeta_k$ from 0 or $\pi$.
If, for each $k$, $1 \le k \le m$, we replace 
$\int d\phi_k /\sin(\zeta_k) \simeq \int d\phi_k/f_k$ by $-\int d\zeta_k/\sin(\phi_k)$
then the only integrand singularity condition remaining is $\sum \zeta_i =0$ mod $2 \pi$. 
Furthermore, because we are explicitly demanding $\sin(\zeta_n) \ne 0$, $\zeta_n(\phi_n)$ 
is an analytic function in the neighbourhood of the integrand singularity. As before we 
eliminate $\phi_n$ using the constraint $\sum \phi_i=\,0$ mod $2\pi$ except that here 
we also express $\phi_k$ as $\phi_k(\zeta_k)$  for $\,1 \,\le\, k \,\le\, m$.  
The leading dependence of $\zeta_n$ on $\zeta_k$ is the quadratic dependence 
$\delta \zeta_k^2$ and this implies that no variation such as
$\partial \sum \zeta_i/\partial \zeta_k =\,$
$\partial \zeta_k/\partial \zeta_k-\partial \zeta_n/\partial \zeta_k$
$=\,1-{\rm constant\,} \delta \zeta_k \zeta_n'$    
can ever vanish\footnote[1]{That $\zeta_n'$ is finite is crucial.
This fails in the previous pinch situation  where all $\sin(\zeta_i)=0$.}.
That is, no integral singularity is possible in the mixed case.

\subsection{Case 4: neither $\sin(\phi_i)$ nor $\sin(\zeta_i)$ vanish}
\label{case-iv}  

In all of the above cases {\em the Landau stationary condition was either automatic 
because of symmetry or did not need to be invoked because the same singularities arose 
from the $f_i=0$ for all $i$ condition.}  References~\cite{bo-ha-ma-ze-07,bo-ha-ma-ze-07b} 
showed how other singularities can arise out of the Landau rules for some toy integrals 
and we give in this section the first and simplest example of a case where {\em the 
Landau stationary conditions are non-trivial for the Ising} $\, \tilde{\chi}^{(n)}$.
Consider the situation\footnote[2]{As in Cases 2 and 3 we restrict our attention to the 
``irreducible" situation in which there is no cancellation between pairs of phases 
differing only in sign. The general situation is treated as Case 5.} that in 
(\ref{ten}-\ref{thirteen}) $k$ of the $\phi_i$, $\zeta_i$ take on the values $\phi_a$, 
$\zeta_a$ and the remaining $(n-k)$ are $\phi_b$, $\zeta_b$. The phase constraints can be 
written as polynomial relations in cosines as in (\ref{sixteen}) but with the difference 
that the individual terms like $\cos(\phi_a)$ are not explicitly given as functions of 
$w$ but are to be determined self-consistently together with $w$. In summary we must 
solve the analogue of (\ref{sixteen}) which is the pair
\begin{eqnarray} 
\label{nineteen}
&&T_k(\cos\phi_a) \, =\, \,\,  T_{n-k}(\cos\phi_b),  \nonumber \\
&&T_k(\cos\zeta_a) \, =\, \,\,  T_{n-k}(\cos\zeta_b), 
\,\, \qquad 0\, <\, k\, <\, n 
\end{eqnarray}
together with the stationary constraints (\ref{twelve})
\begin{eqnarray} 
\label{twenty}
&&4 w\,  - \cos\phi_a\, -\cos\phi_b\, 
+\, 4 w \cdot \cos\phi_b\cos\phi_a \, =\, \, 0, \nonumber \\
&&4 w\,  - \cos\zeta_a\, -\cos\zeta_b\, 
+4 w \cdot \cos\zeta_b\cos\zeta_a \, =\,\,  0,
\end{eqnarray} 
and the  definitions (\ref{thirteen}) 
\begin{eqnarray} \label{twentyone}
\cos\phi_a\, +\cos\zeta_a \, = \, \,1/(2 w), \quad \quad \quad 
\cos\phi_b\, +\cos\zeta_b\,  = \, \,1/(2 w).
\end{eqnarray}

We can reduce these equations to somewhat simpler form by using (\ref{twenty}) 
to eliminate $\phi_b$ and $\zeta_b$. The result is a triplet of equations
\begin{eqnarray} 
\label{twentytwo}
&&T_k(\cos\phi_a) \,=\, \,  T_{n-k}((4 w-\cos\phi_a)/(1-4 w \cos\phi_a)),
 \nonumber \\
&&T_k(\cos \zeta_a) \,=\, \,  T_{n-k}((4 w-\cos\zeta_a)/(1-4 w \cos\zeta_a)),
 \qquad  0<k<n, \nonumber \\
&&\cos\phi_a+\cos\zeta_a \,= \, \, 1/(2 w) 
\end{eqnarray} 
defining the van Hove singularities when neither $\sin(\phi_i)$ nor 
$\sin(\zeta_i)$ vanish.

To produce the singularity polynomials, we use the last condition in (\ref{twentytwo}) 
to eliminate $\cos(\zeta_a)$. Then, the elimination of $\cos(\phi_a)$ from the first 
two polynomial conditions in (\ref{twentytwo}) gives the roots that are inserted into 
(\ref{twentytwo}) to determine which are actually solutions.  We must also go back and verify
that (\ref{eleven}) is satisfied as our candidate polynomials are all based on the 
squared form of (\ref{eleven}) as expressed in (\ref{twelve}), (\ref{thirteen}).

Our explicit procedure for eliminating the phase variables in (\ref{twentytwo}) is as 
follows. The first condition in (\ref{twentytwo}) can be expressed as a polynomial 
equation $\, p_1(\cos(\phi_a))=0$. The second condition can be reduced to another 
$\, p_2(\cos(\phi_a))=\, 0$ if we use the last condition to eliminate $\, \cos(\zeta_a)$.  
We are not interested in the solutions $ \, \cos(\phi_a)=\, \pm 1$ as these are a subset 
of those already found.  We therefore take it as given that $\, p_1$ and $\, p_2$ are
reduced polynomials which do not contain the factors $ \, \cos(\phi_a)=\, \pm 1$. They 
are both of degree $\, m=\, n-2, \,n-1,\, n,$  depending on whether 
$\, k$ and $\, n-k$ are even-even, odd-even or odd-odd combinations.  We now eliminate 
$ \, \cos(\phi_a)$ by the following iterative process. We first reduce the algebraic 
complexity of the polynomials $\, p_1$ and $\, p_2$ by dividing them at each stage by 
their greatest common divisor.  Assume this has been done and they are now written as 
$\, p_1\,=\, a_1+\, \cdots \,  +b_1 \cos^m( \phi_a),$ 
$\,  p_2=\, a_2+ \, \cdots \,  +b_2  \cos^m(\phi_a)$.  We then generate two new 
polynomials $\, q_1=\, (A_2 \, p_1\, -A_1\, p_2)/\cos(\phi_a)$ and 
$\, q_2=\, B_2\, p_1\, -B_1\, p_2 $ where $A_1$ and $A_2$ are just  
$a_1$ and $a_2$ divided by their greatest common divisor and similarly for 
$B_1$ and $B_2.$  By construction $q_1$ and $q_2$ are of degree one less than 
$\, p_1$ and $\, p_2$ and after division by their greatest common divisor are the 
$\, p_1$ and $\, p_2$ of the next iteration. The process stops when we reach degree 0. 
The greatest common divisors are polynomials in $\, w$ whose zeros are candidates
for solutions of the original equations.

{\bf Remark 8}
We find empirically that for given $n$ and $k$ in (\ref{twentytwo}) the degree of 
the $w$ polynomials giving the Landau singularities is bounded by $n(n-2)/4-m(m-1)/2$, 
$m=\min(k,n-k)$.  There is evidence from~\cite{bo-ha-ma-ze-07b} based on toy integrals 
that the predicted singularities are seen in the ODE so that just as in Case 3 we conjecture that 
our polynomial solutions for Case 4 are both necessary and sufficient. For $n=2N+1$, 
the number of $s$ plane singularities is $N(N+3)$, $N(3N+1)$ and $5N(N^2-1)/3$ for 
Cases 2-4 respectively. The corresponding number of $s$ plane singularities for $n=2N$ 
is $2[(N-1)(N+3)/2]$, $6[(N-1)^2/2]$ and $2[(N-2)(2N-3)(5N+4)/12]$ barring accidental 
degeneracies\footnote[3]{$N=6$ affords examples. Here the $[k,m]$ combinations $[0,3]$ 
and $[2,2]$ in $\cos(k \pi /N)+\cos(m \pi /N)$ yield the same $1/(2w)=\,1$. Other 
degenerate combinations are $[0,4]$ and $[2,3]$ and those obtained by the replacements 
$k \rightarrow 6-k$, $m \rightarrow 6-m$.}.

{\bf Remark 9}
As continuation of our empirical observations we note that of the Case 4 singularities, 
$N(N-1)$ are on $\vert s \vert =1$ for $n=2N+1$ and $2[(N-2)^2/2]$ for $n=2N$. However, 
{\em in none of these cases is the singularity on the principal disc of} $\tilde{\chi}^{(n)}$
and the general proof that no Case 4, $\vert s \vert =\,1$, singularity lies on the 
principal disc is as follows. Observe first that a singularity with $\vert s \vert =1$ 
on the principal disc requires that the integrand of $\tilde{\chi}^{(n)}$ be singular for 
real phases $\phi_i$. That is, both $\phi_a$ and $\phi_b$ must be real. Equivalently 
$\cos(\phi_a)$ and $\cos(\phi_b)$ are real and of magnitude less than one.  Furthermore, 
the cosines are related by (\ref{twenty}) or 
$\cos(\phi_b)=(4 w -\cos(\phi_a))/(1-4 w \cos(\phi_a))$. If $4 w$ is real and 
$\vert 4 w \vert > 1,$ which is the condition for $\vert s \vert \,=\,1$, then 
$\vert \cos(\phi_a) \vert\, <\,  1$ yields $\vert \cos(\phi_b) \vert > 1$ and 
vice versa. An important consequence of this proof is that {\em there are no 
Case 4 singularities to cancel any of the Case 2 Nickelian singularities.}

Although we started the discussion of Case 4 with the remark that our calculation 
was just the first and simplest example of a non-trivial van Hove singularity, one 
observes that the singularity conditions (\ref{twelve}), for example, require 
$\cos(\phi_j)=\cos(\phi_i)$ or $\cos(\phi_j)=(4 w -\cos(\phi_i))/(1-4 w \cos(\phi_i))$ 
and the latter is an example of a one to one M\"obius' mapping. Thus it is not possible, 
for any given $w$, to have more than two distinct values in  the set 
$\cos(\phi_i)$, $i=1, \cdots, n$. This in turn implies that as long as we consider 
$\prod x_i =1$ as the only singularity we have exhausted all possibilities for the 
values of $\cos(\phi_i)$ in Cases 2-4 above. Of course this applies only to the 
``irreducible" situations and we turn now to the implications of the $\pm \phi_i$ 
and $\pm \zeta_i$ sign degeneracy. 

\subsection{Case 5: $\pm \phi_i$ and $\pm \zeta_i$ sign degeneracy}
\label{case-v}

Once the ``irreducible" solutions Cases 2-4 are known, all other solutions can 
be obtained by recursion which we now describe. Suppose the combinations $S^{(n)} = \, $
$\left \{(\phi_i^{(n)}, \zeta_i^{(n)}),\,  i\, =\, 1, \, \cdots,\,  n \right \}$ 
are the stationary singular points of the integrand of $\tilde{\chi}^{(n)}$. Then we can 
construct a stationary singular point of the $\tilde{\chi}^{(n+2)}$ integrand 
as $S^{(n+2)} = \left \{ S^{(n)}, (\phi_j^{(n)}, \zeta_j^{(n)}),
(-\phi_j^{(n)}, -\zeta_j^{(n)}) \right \}$, where $j$ is any one of the $n$ 
values in $S^{(n)}$. Clearly by adding a pair with opposite signs we have guaranteed 
that the phase constraints (\ref{ten}) are satisfied.  
Furthermore, all constraint equations involving only cosines remain unchanged and thus 
satisfied. Finally, the stationary condition
$\sin(\zeta_i)\sin(\phi_j)\,  =\,  \sin(\phi_i)\sin(\zeta_j)$ 
is potentially satisfied under the simultaneous sign change of $\phi_j$ and $\zeta_j$.
Whether this occurs will depend on which Riemann sheet one is on. As an important 
example, and previously discussed in Case 2, on the principal disc for $\vert s \vert \le 1$ 
the definitions (\ref{seven}, \ref{eight}) show that $\sin(\zeta_j)$ is an even 
function of $\phi_j$ and the new stationary condition generated by recursion cannot 
be satisfied.  Specifically, the recursion mechanism {\em explicitly excludes the 
possibility of addition to, or cancellation of, Case 2 ``irreducible" singularities
 on the principal disc and leaves the argument that $\vert s \vert =\, 1$ is a 
natural boundary of $\tilde{\chi}$  secure.}

In conclusion, the Landau conditions for a singularity of $\tilde{\chi}^{(n)}$ are also the 
conditions\footnote[4]{As always, these conditions are necessary but not sufficient.  
The numerical evidence from the singularities of the associated ODE is that they are also 
sufficient and each singularity occurs on some Riemann sheet of the integral.} for that 
singularity in $\chi^{(n+2k)}$, $k \ge 1$. Thus for a complete picture of the 
singularities of $\tilde{\chi}^{(n)}$  it is sufficient to list the polynomials defining the 
``irreducible" singularities for the Cases 2-4.

\section{Singularities \label{Singularities}}

The following is a listing of the polynomials defining the Landau singularities of 
$\tilde{\chi}^{(n)}$ for small $n$ for the three Cases 2-4 discussed in  \ref{appendixBernie}. 
Our notation for the polynomials from $\tilde{\chi}^{(n)}$ is $\, P(^{Case} \, n)$ or 
$\, P(^{Case}\, n_{n-k,k})$ with $k$ defined in eqns. (\ref{sixteen}) and (\ref{twentytwo}).  
We do not include the physical singularities $\, w= \, \pm 1/4$ or the simplest unphysical 
singularities at $\, w=\,0$  or $\, \infty$.  We also do not give the (reducible) Case 5, 
which would just be the list of $\, m>\, 0$,  $\,  \,\tilde{\chi}^{(n-2m)}$ polynomials for any 
given $ \,\tilde{\chi}^{(n)}$. An example of the notation we use in the text in this situation
is $\, P(^57/^3 3_{1,2})$ to indicate a Case 3, $\, n\,-2m\,=\, 3$ contribution to 
$ \,\tilde{\chi}^{(7)}$.

\noindent
Case 2. Circle singularities $\, P(^{2}\, n_{n,0})=\, P(^{2} \, n).$
\begin{eqnarray*}
\fl \quad  P(^{2} \, 3)=(1+2w)(1-w), \\
\fl \quad  P(^{2} \, 4)=(1-4w^2), \\
\fl \quad  P(^{2} \, 5)=(1+w)(1+2w-4w^2)(1-3w+w^2), \\
\fl \quad  P(^{2} \, 6)=(1-4w^2)(1-9w^2)(1-w^2), \\
\fl \quad  P(^{2} \, 7)=(1+2w-8w^2-8w^3)(1+2w-w^2-w^3)(1-5w+6w^2-w^3), \\
\fl \quad  P(^{2} \, 8)=(1-8w^2)(1-2w^2)(1-4w^2)(1-12w^2+4w^4), \\
\fl \quad  P(^{2} \, 9)=(1+2w)(1-w)(1-12w^2+8w^3)(1-6w+9w^2-w^3) \\
\fl \qquad \qquad \times (1-3w^2-w^3)(1+3w-w^3), \\
\fl \quad  P(^{2} \, 10)=(1-w^2)(1-5w^2)(1-12w^2+16w^4) \\
\fl \qquad \qquad \times  (1-7w^2+w^4)(1-15w^2+25w^4).
\end{eqnarray*}

\noindent	
Case 3. $f_i=\, 0$ or $\cos\phi_+=\, 1/2w+1,$ $\cos\phi_-=\, 1/2w-1.$
\begin{eqnarray*}
\fl \quad P(^{3}\, 3_{1,2})=(1+3w+4w^2), \\
\fl \quad P(^{3}\, 4_{2,2})=1, \\
\fl \quad P(^{3}\, 5_{3,2})=(1-7w+5w^2-4w^3), \\
\fl \quad P(^{3}\, 5_{1,4})=(1+8w+20w^2+15w^3+4w^4), \\
\fl \quad P(^{3}\, 6_{4,2})=(1-25w^2)(1-w^2+16w^4), \\
\fl \quad P(^{3}\, 7_{5,2})=(1-10w+35w^2-51w^3+21w^4-4w^5), \\
\fl \quad P(^{3}\, 7_{3,4})=(1+7w+26w^2+7w^3+4w^4), \\
\fl \quad P(^{3}\, 7_{1,6})=(1+12w+54w^2+112w^3+105w^4+35w^5+4w^6), \\
\fl \quad P(^{3}\, 8_{6,2})=(1-16w^2+100w^4)(1-20w^2+16w^4-16w^6), \\
\fl \quad P(^{3}\, 8_{4,4})=(1+2w^2).
\end{eqnarray*}

\noindent
Case 4. $f_i \ne 0,$  $\cos\phi_a$ and  $\cos\phi_b$ distinct.
	
\begin{eqnarray*}
\fl \quad P(^{4}\, 3_{2,1})=1, \qquad P(^{4}\, 4_{3,1})=1, \qquad  P(^{4}\, 4_{2,2})=1, \\
\fl \quad P(^{4}\, 5_{4,1})=(1-w-3w^2+4w^3), \\
\fl \quad P(^{4}\, 5_{3,2})=(1+4w+8w^2), \\
\fl \quad P(^{4}\, 6_{5,1})=(1-10w^2+29w^4), \\
\fl \quad P(^{4}\, 6_{4,2})=1, \\
\fl \quad P(^{4}\, 6_{3,3})=1, \\
\fl \quad P(^{4}\, 7_{6,1})=(1-3w-10w^2+35w^3+5w^4-62w^5+17w^6+32w^7-16w^8), \\
\fl \quad P(^{4}\, 7_{5,2})=(1+8w+15w^2-21w^3-60w^4+16w^5+96w^6+64w^7), \\
\fl \quad P(^{4}\, 7_{4,3})=(1-4w-16w^2-48w^3+32w^4-128w^5), \\
\fl \quad P(^{4}\, 8_{7,1})=(1-26w^2+242w^4-960w^6+1685w^8-1138w^{10}), \\
\fl \quad P(^{4}\, 8_{6,2})=(1-10w^2+32w^4), \\
\fl \quad P(^{4}\, 8_{5,3})=(1-30w^2+56w^4-1312w^6), \\
\fl \quad P(^{4}\, 8_{4,4})=1.
\end{eqnarray*}

\section{The one term Fermionic toy model \label{app:ToyModel}}

Here we consider the question of which integral singularities might arise
from combinations of $x_i x_j =\, 1$, $f_i=\, 0$ and $\prod x_i =\, 1$.

The analysis of $\, \tilde{\chi}^{(n)}$ with the full Fermionic factor $(G^{(n)})^2$ is 
complicated in large part because of the very many different $\, x_i\, x_j =\, 1$
combinations to consider.  Thus we are motivated to look at simpler integrals 
intermediate between $\, \Phi_H^{(n)}$ and  $\, \tilde{\chi}^{(n)},$ but which lead to 
the same singularities seen in the ODE for  $\, \tilde{\chi}^{(5)}$ and  $\, \tilde{\chi}^{(6)}$. 
The simplest situation is where the full Fermionic factor is replaced, for 
 $n$ even, by $\, (h_{12})^2,$ and for $n$ odd by $\, (f_{12})^2,$ 
where~\cite{nickel-99}:
\begin{eqnarray}
\label{(D.1)}
	f_{12}& =& \,
 {{1}\over {2}} \, (\sin\phi_1\, -\sin\phi_2)(1 \, +x_1\, x_2)/(1\, -x_1\, x_2).	
\end{eqnarray}
The integrands in these cases are simple enough that long series can be derived with 
only slightly more effort than for $\, \Phi_H^{(n)}.$  We find, for example, by a mod 
prime analysis of the ODE for this ``one term'' reduction of $\, \tilde{\chi}^{(5)}$, that 
the head polynomial has the $\, (1-2w)$ factor that is in the $\, \tilde{\chi}^{(5)}$ ODE 
and in addition has the factor $\, (1-8w^2)$. The latter also shows up in the ``one-term'' 
reduction of  $\, \tilde{\chi}^{(6)}$, where it might have been expected since that 
it is what we found for the $\, \tilde{\chi}^{(6)}$ ODE by diff-Pad\'e analysis. Explicit 
solution of the ODE shows the associated singular functions have, as leading terms, 
$\, (1-2w)^{1/2}$, $\, \ln(1-\sqrt{8} \, w)$ and $\, \ln(1\, +\sqrt{8} \, w)$  for the 
``one term'' $\, \tilde{\chi}^{(5)}$ and $\, (1-8w^2) \ln(1\, -8\, w^2)$ for the ``one term'' 
$\, \tilde{\chi}^{(6)}$. The observed singularities of the ODE make the simple ``one term'' 
reductions of $\, \tilde{\chi}^{(n)}$ ideal integrals for a Landau singularity analysis and 
the details of this analysis is the content of the rest of this appendix. As in 
\ref{appendixBernie}, we allow for various\footnote{``Various'' includes all possible sign 
combinations of the square root factors in the integrand and that means we are considering 
every possible local environment in the Landau analysis.  The same local conditions can 
differ in the global behaviour of the contour distortions used in the integrals but this 
does not affect the singularity conditions. In this sense our singularity search is exhaustive.} 
analytic continuations of the integrands. 
Yet in spite of this we find that the Landau conditions are never satisfied at either 
$\,1\,-2w=\, 0$ or $\, 1\,-8w^2\,=\,  0$. Since the Landau conditions are necessary 
for the integral to be singular we must conclude that the linear ODE generated from 
the series have additional singularities not possessed by the integrals. 

The addition of the new singularity $x_1 x_2 =\, 1$ requires an analysis that can 
be broken into two parts.  The simplest is the determination of the integral 
singularities arising out of just $x_1 x_2 =\, 1$ and $f_i=\, 0$. The more involved 
investigation is for the combination of $f_i=\,0$ with $\, x_1 \, x_2\, =\, 1$ and 
$\prod x_i =\, 1$ simultaneously satisfied and with the normals to the latter two 
hypersurfaces constrained to be parallel\footnote[1]{We need not consider
the third possibility, $x_1 x_2=1$ and $\prod x_i =\,1$, satisfied with 
$x_1 x_2$  and $\prod x_i$ simultaneously stationary. The constraints when $x_1 x_2$  
and $\prod x_i$ are treated as independent are either compatible with each other 
or not.  If the constraints are incompatible solutions will have to be dropped, 
otherwise they can be kept. In either case there is no possibility of new solutions 
being generated.}.

For the simple case of $x_1 x_2 =1$ and $f_i=0$ for all $i$ we first observe that the 
$f_i=0$ pinch conditions lead to the Case 3 singularities (\ref{sixteen}) and the only 
new investigation to be done is to determine whether the presence of different integrand 
factors changes the even $k$ condition in (\ref{sixteen}). The previous proof that the 
presence of $\left( G^{(n)}\right)^2$ did not change the even $k$ condition in 
(\ref{sixteen}) relied on the fact that $G^{(n)}$ is a product of $h_{ij}$ factors 
and thus was in essence also a proof for any single $(h_{12})^2$. Essentially the same 
proof will apply to $(f_{12})^2$ provided we take for $f_{ij}$ the original
$f_{ij}=\, 1/2 \, \left( \sin(\phi_i)-\sin(\phi_j) \right)(1+x_i x_j)/(1-x_i x_j)$
(see (4) in~\cite{nickel-99}) because this form can be reduced to a Taylor expansion 
with terms $(\zeta_i)^{p_i} (\zeta_j)^{p_j}$  with $\, p_i+p_j\, $ odd exactly as in 
the argument for $h_{ij}$ following (\ref{seventeen}). The cumulant reduced form 
$\,f_{ij}\,=\, \left( \sin(\phi_i)\,-\sin(\phi_j) \right)x_i x_j/(1-x_i x_j)$,
which  is more convenient for calculating the full $\tilde{\chi}^{(2n+1)}$, when expanded has 
terms of both odd and even parity and thus will also generate odd $k$ singularities 
(\ref{sixteen}). This has been  observed and is an additional confirmation of the 
argument that it is {\em the very special  nature of the integrand of the Ising
$\tilde{\chi}^{(n)}$ that is responsible for the even $k$ condition in} (\ref{sixteen}).  
It is to be understood in the following that we will be using the original $f_{ij}$
and thus that (\ref{sixteen}) will not be supplemented with odd $k$ terms.

{\em We must also deal with the van Hove singularities arising from $x_1 x_2=\, 1$
with $x_1 x_2$ stationary.}  The product $x_1 x_2 =1$ requires $\zeta_1 =\, -\zeta_2$ 
and hence $\cos(\zeta_1)=\, \cos(\zeta_2)$.  This in turn, because of  
(\ref{thirteen}), gives $\cos(\phi_1)=\, \cos(\phi_2)$ and $\phi_1= \pm \phi_2$.  
The stationary condition is $\sin(\phi_1)=\sin(\phi_2)=0,$ which combined with the 
preceding $\phi_1 = \pm \phi_2$ allows as possible $\phi_1$, $\phi_2$ combinations 
$0,0$ and $\pi, \pi$. The corresponding $\cos(\zeta_1)$ and $\cos(\zeta_2)$ are both 
either $1/(2w)-1$ or $1/(2w)+1$ so that the equality $\cos(\zeta_1)=\cos(\zeta_2)$
required for $x_1 x_2 =1$ is automatic and gives no constraint on $w$. The absence of 
a first order stationary constraint requires that we go to second order so that in 
addition to $\zeta_1'=-\sin(\phi_1)/\sin(\zeta_1)=0$ we demand 
$\zeta_1''=-\left( \cos(\phi_1)\sin^2(\zeta_1)
\, +\cos(\zeta_1)\sin^2(\phi_1) \right)/\sin^3(\zeta_1)=0$
and similarly for $\zeta_2''$.  The expression in braces factorizes into 
$\left( \cos(\phi_1)+\cos(\zeta_1) \right)\left(1-\cos(\phi_1)\cos(\zeta_1)\right)$, 
which, given (\ref{thirteen}) and the possible values $\cos(\phi_1)=\, \pm 1$, reduces 
to $1/w(1 \pm 1/(4w))$. Thus from $\zeta_1''=\, 0$ we get as the only possible Landau
singularities $w=\,  \pm \, 1/4$ or $\infty$, the Case 1 singularities we are not 
considering.

\vskip 0.1cm

We begin the analysis where $x_1 x_2 =1$ and $\prod x_i =1$ are 
simultaneously satisfied and these hypersurfaces touch tangentially, by 
deducing the necessary constraint conditions. As before we treat 
$\phi_i$, $i=\, 1,\,  \cdots,\,  n-1$ as independent giving 
$\zeta_i=\, \zeta_i(\phi_i)$, $i=\, 1,\,  \cdots,\,  n-1$, and 
$\zeta_n=\, \zeta_n(\phi_n)=\zeta_n \left( 2 \pi k-\phi_1 - \cdots \phi_{n-1} \right)$.
Parallel normals requires 
$\alpha \partial \left( \zeta_1 + \cdots + \zeta_n \right)/\partial \phi_i=
\beta \partial (\zeta_1+\zeta_2)/\partial \phi_i$,  
$i=1, \cdots, n-1$, with both $\alpha$ and $\beta$ non-zero.  Explicitly,
\begin{eqnarray}
\label{twentythreea}
&& \alpha \cdot  \left( \zeta_1'\, - \zeta_n' \right)
\, =\,\, \beta \cdot \zeta_1', \qquad 
\alpha \cdot  \left( \zeta_2' \,- \zeta_n' \right)
 \,=\,\, \beta \cdot \zeta_2',   \\
\label{twentythreeb}
&&  \alpha\cdot  \left( \zeta_i' - \zeta_n' \right)=0, 
\qquad \qquad 3 \, \le\, i \,\le \,n
\end{eqnarray}
where again $\zeta'=\, \partial \zeta /\partial \phi$ $=\, -\sin(\phi)/\sin(\zeta)$.
Conditions (\ref{twentythreea}) imply either $\zeta_n'=\, 0$ or $\zeta_1'=\, \zeta_2'$ 
while conditions (\ref{twentythreeb}) reduce to the previous stationary conditions 
(\ref{eleven}) except that $i$ and $j$ are restricted by $i,j >2$.  
The new structure of the stationary conditions has implications for the pinch 
singularities also. Whereas in the discussion at the end of Case 3 
we noted that {\em pinch and van Hove type singularity conditions could not mix},
here we have the possibility that with $m=2$ in that discussion 
$\alpha \partial \left( \zeta_1 + \cdots + \zeta_n \right)/\partial \zeta_1\, $
$=\, \beta \partial (\zeta_1+\zeta_2)/\partial \zeta_1$ and   
$\alpha \partial \left( \zeta_1 + \cdots + \zeta_n \right)/\partial \zeta_2\, $
$=\, \beta \partial (\zeta_1+\zeta_2)/\partial \zeta_2$ 
can both be satisfied with $\alpha\,  =\, \beta$. Thus it is possible to have 
$\sin(\zeta_1)=\, \sin(\zeta_2)=\,0$ in conjunction with the conditions 
(\ref{twentythreeb}) that arise out of the remaining derivatives with respect 
to $\phi_i$, $i>2$. The reverse situation in which $\, \sin(\zeta_1)\,  \ne\,  0$, 
$\sin(\zeta_2)\,  \ne\,  0$ and $\sin(\zeta_i)=0$, $i>2$, is not possible because
there is no analogue of the $\alpha=\, \beta$ solution in this case.  In summary, 
all possibilities covered by (\ref{twentythreea}, \ref{twentythreeb}) plus the 
allowed pinch situations are given by
\begin{eqnarray}
\label{twentyfoura}
&& \sin(\phi_n) \cdot 
\left( \sin(\zeta_1) \sin(\phi_2) -\sin(\phi_1) \sin(\zeta_2) \right)
\, =\,\, 0, \\
\label{twentyfourb}
&& \sin(\zeta_i) \cdot \sin(\phi_j)\,  = \, \, \sin(\phi_i) \cdot \sin( \zeta_j), \\
&&\qquad   {\rm \,\,} i \ne j >2, \,\,\,\qquad 
(\sin(\zeta_i) \ne 0, i>2)              \nonumber
\end{eqnarray}

Details for the various alternatives are given below with the major categories a) 
through d) being the distinctions allowed by (\ref{twentyfoura}).

a) Pinch case: $\sin(\zeta_1)\,=\,\sin(\zeta_2)=\,0.$ \\
Allowed combinations of $\zeta_1$, $\zeta_2$ satisfying 
$\sin(\zeta_1)\,=\, \sin(\zeta_2)\,=\,\,0$  and $x_1 x_2 =\,1$ are $0,0$ and $\pi, \pi$. 
The corresponding $\cos(\phi_1)$ and $\cos(\phi_2)$ are either both 
$\cos(\phi^{(-)})\,=\,1/(2w)-1$ or both $\cos(\phi^{(+)})=\, 1/(2w)+1$.  
If $\phi_1\,=\,-\phi_2$ then conditions (\ref{twentyfourb}) together with (\ref{ten}) 
are just the conditions for the singularities of $\,\tilde{\chi}^{(n-2)}$ in 
the absence of the Fermionic factor $\left( G^{(n)} \right)^2$  and have already been 
discussed. If $\phi_1=\,\phi_2$ the sum on phases $\phi_i$, $i>2$, is no longer 0, but 
$\,\pm 2 \phi_1$ mod $2 \pi$ with $\,\phi_1$ either $\phi^{(-)}$ or $\phi^{(+)}$.
For the new singularities that arise we introduce further subdivisions depending on 
the form taken by (\ref{twentyfourb}).

For $i>\, 2$, all $\zeta_i$ equal and all $\phi_i$ equal as the analogue of Case 2.
The first new situation arises for $\,\tilde{\chi}^{(5)}$.  Here we set $\zeta_i=\,2\pi/3$ and 
solve $\,T_2(\phi^{(\pm)})\,=\,\,T_3(1/(2w)+1/2)$. For $\tilde{\chi}^{(6)}$ set $\zeta_i=\pi/2$ 
and solve $\,T_2(\phi^{(\pm)})=T_4(1/(2w))$. These are the only possibilities for $n<7$ 
and the resulting polynomial equations are
\begin{eqnarray}
&& (1 + 2w + 4w^2 - 4w^3)(1 + 2w - 4w^2 - 4w^3)\,=\,\,0,
 \quad {\rm for} \quad \tilde{\chi}^{(5)}, \quad {\rm and} \nonumber \\
\label{twentyfive}
&& (1 - w^2)(1 - 9w^2 + 16w^4)\,=\,\,0, \qquad \qquad  {\rm for} \quad \tilde{\chi}^{(6)}.
\end{eqnarray}

For $i>2$, $\zeta_i=\,\zeta_a$ or $\zeta_b$ and $\phi_i=\,\phi_a$ or $\phi_b$ as
the analogue of Case 4. The reduction of the singularity conditions to polynomial form is
similar to that described in the Case 4 analysis and will not be described further.  
The only possibilities for $n<7$ are
\begin{eqnarray}
\fl  \qquad   (1 - 2w - 4w^2 + 12w^3 + 16w^4)(1 - w + 9w^2 - 24w^3 + 16w^4) \nonumber \\
\fl  \qquad  (1 - w + w^2 + 80w^3 + 352w^4 + 512w^5 + 256w^6) \nonumber \\
\fl  \qquad  (1 + 2w - 20w^2 - 68w^3 - 32w^4 - 64w^5) = 0,  
  \qquad \qquad {\rm for} \quad \tilde{\chi}^{(5)}, \quad {\rm and} \nonumber \\
 \label{twentysix} \\
\fl  \qquad     (1 - w^2 + 16w^4)(1 - 41w^2 + 640w^4 - 4096w^6)
 \nonumber  \\
\fl  \qquad   (16 - 664w^2 + 11273w^4 - 68290w^6 + 141889w^8 - 16896w^{10} + 65536w^{12}) 
 \nonumber \\
\fl  \qquad   (1 - 77w^2 + 1898w^4 - 20282w^6 + 107013w^8 - 160553w^{10} + 198432w^{12}  
 \nonumber \\
\fl   \qquad        - 87776w^{14} - 7680w^{16} + 20224w^{18} + 4096w^{20}) = 0, 
 \qquad {\rm for} \quad \tilde{\chi}^{(6)}. \nonumber 
\end{eqnarray}

Two of the singularity polynomials for $\tilde{\chi}^{(6)}$, $1-w^2$ and $1-w^2+16w^4$, 
are present also in the absence of $\left( G^{(n)} \right)^2$. All other polynomials 
in (\ref{twentyfive}) and (\ref{twentysix}) {\it are new but have not been seen} in 
any ODE analysis.  Because the integrand singularity surfaces $\,x_1\, x_2 \,=\,1$ 
and $\prod x_i =\,1$ involve the common factor $x_1 x_2$ {\it it is conceivable that}
the $\phi_1$ and $\phi_2$ integration contours {\it might always be constrained to 
lie on the same side of the two surfaces and never be pinched in between. This would 
explain the absence of these singularities} but to resolve such a complicated 
topological question is not something we have attempted. On the other hand, the 
{\it absence} of these singularities is confirmed by the $\zeta \leftrightarrow \phi$ 
interchange symmetry discussed in the Case 3 section and the following b) results.

b) van Hove case: $\sin(\phi_1)=\, \sin(\phi_2)=\,\,0$. \\
Only the $\phi_1$, $\phi_2$  combinations $0,\,0$ and $\pi, \pi$ are allowed since 
the $x_1 x_2=\,1$ constraint requires $\zeta_1=-\zeta_2$ and this is only possible with 
$\cos(\zeta_1)$ and $\cos(\zeta_2)$ either both $1/(2w)-1$ or $1/(2w)+1$.  
This is analogous to the $\phi_1=-\phi_2$ situation of case a) which is that 
the remaining conditions (\ref{twentyfourb}) and (\ref{ten}) just yield the singularities 
of $\tilde{\chi}^{(n-2)}$ without $\left( G^{(n)} \right)^2$. The difference between what we 
have here and in a) is that there are no other possibilities: singularities such as 
(\ref{twentyfive}, \ref{twentysix}) are not generated here and, by inference 
from the $\zeta \leftrightarrow \phi$ interchange symmetry, are not present in a).

c) The combination $\phi_1=-\phi_2$ and $\zeta_1=-\zeta_2$ with $\sin(\phi_1) \ne 0$, 
$\sin(\zeta_1) \ne 0$. \\
The $\zeta_1=-\zeta_2$ condition is required by $x_1 x_2 =1$. 
The remaining $\phi_1=\, -\phi_2$ by
 $\, \sin(\zeta_1)\cdot \sin(\phi_2)\, -\sin(\phi_1) \cdot \sin(\zeta_2)=0$ 
in (\ref{twentyfoura}).  The conditions (\ref{twentyfourb}) and (\ref{ten}) give the 
singularities of $\tilde{\chi}^{(n-2)}$ without $\left(  G^{(n)} \right)^2$ as in b).

d) The case $\sin(\phi_n)=0$. \\
The conditions (\ref{twentyfourb}) require, for all $i>\,2$, $\sin(\phi_i)=\,0$
and thus $\phi_i=0$ or $\pi$. The corresponding $\cos(\zeta_i)$ are 
$\cos(\zeta^{(\pm)})=\,1/(2w) \pm 1$. The condition $x_1 x_2 =1$ requires 
$\zeta_1=\,-\zeta_2$  or $\cos(\zeta_1)=\,\cos(\zeta_2)$ and, then, from the 
definition (\ref{thirteen}) that $\cos(\phi_1) =\,\cos(\phi_2),$ or 
$\phi_1= \,\pm \phi_2$. Note that $\phi_1=\phi_2$ is allowed because (\ref{twentyfoura}) 
is no longer a constraint on $\phi_1$ or $\phi_2$ once $\sin(\phi_n)=\,0$.  
In the case that $\phi_1=\,-\phi_2$ the phase constraint $\sum \phi_i =\,0$  mod $2\pi$ 
is satisfied if there are an even number of $\phi_i=\,\pi$, $i>2$, terms.  In the 
$\phi_1=\phi_2$ case we can set $\,\phi_1=\,\phi_2=\,\pi/2$  and thus accommodate an 
odd number of $\phi_i=\,\pi$, $i>2$, terms as well.  The final result is that when 
$\sin(\phi_n)=\,0$, the van Hove singularities are the Case 3 singularities (\ref{sixteen}) 
for $\tilde{\chi}^{(n-2)}$ but with odd $k$ allowed and when supplemented by Case 5 are:
\begin{eqnarray}
\label{twentyseven}
&&T_k\left( 1/2w +1 \right)  \,\, =\,
\,\, \, T_{n-2m-k}\left( 1/2w -1 \right), \,\,\, \\
&& 0 \, \le\,  k \, \le\,  n-2m, \quad 
 \, \,\,\,m\, =\,\, 1,\,  \cdots, \,\, [n/2]. \nonumber
\end{eqnarray}

The situation here has parallels to that in a) and b). First, the odd $k$ polynomials 
in (\ref{twentyseven}) {\it are not seen} in any ODE analysis and the absence of the 
odd $k$ singularities (\ref{twentyseven}) might well have the same explanation as that 
suggested for the absence of (\ref{twentyfive}) in a). Second, 
recall the remarks just preceding (\ref{twentyfoura}) that the mixed van Hove/pinch 
situation in which $\,\sin(\zeta_1)\, \ne\, 0$, $\, \sin(\zeta_2) \ne 0$ 
and $\sin(\zeta_i)=\,0$, $i\,>\,2$, is not possible and hence yields no singularities.  
But with the interchange $\zeta \leftrightarrow \phi$, we get exactly the situation 
described by (\ref{twentyseven}) for $k$ odd and {\it thus the absence of these 
singularities is also confirmed.}

The four cases a)--d) exhaust all possibilities for singularities of the 
toy $\tilde{\chi}^{(n)}$ integrals where a single $(f_{12})^2$ or $(h_{12})^2$ replaces 
the Fermionic $\left(  G^{(n)} \right)^2$ factor in the integrand.  
Furthermore, by performing the Landau analysis of the $(n-1)$ dimensional integral 
analogs of (\ref{chi3tild}) rather that the $(2n-2)$ dimensional integrals as
in~\cite{bo-ha-ma-ze-07b} we gain a powerful symmetry related tool that enables us to 
compare results obtained in two different ways. Without this tool we can only conclude 
that the Landau singularity conditions (\ref{twentyfive}-\ref{twentyseven}) are 
necessary but not sufficient. With this tool we can eliminate 
(\ref{twentyfive}-\ref{twentyseven}) as singularities of the integrals.  
This proves the toy $\tilde{\chi}^{(n)}$  integrals with $(f_{12})^2$ or $(h_{12})^2$  have 
the same singularities as those without these integrand factors.

We have not investigated systematically what happens when more terms from
$\left(  G^{(n)} \right)^2$  are included but such calculations will have many 
similarities to what has been done in the case of just $(f_{12})^2$ or $(h_{12})^2$.
Furthermore, we have not found any obvious candidates for new singularities.  
Thus we conjecture that the singularities of the integrals $\tilde{\chi}^{(n)}$
are exactly those of the integrals without the Fermionic
$\left(  G^{(n)} \right)^2$ factor and the complete list of these is the
list we have given in~\ref{Singularities}.

\section{Singularity exponents at Landau singularities \label{app:Exponents}}

To determine the singular behaviour of $\tilde{\chi}^{(n)}$  at Landau singularities
requires only a local analysis and calculations similar to those already
done for the Case 2 circle singularities~\cite{nickel-99, nickel-00}.
If only leading exponent information is of interest, the calculations simplify 
dramatically and essentially reduce to a power counting argument. We sketch in 
this appendix the calculations for the cases most analogous to 
Case 2  and derive the formula for the exponent
\begin{equation}
\label{eqE1}
p \,= \, \, [(n-m)^2 + m^2 - 3]/2, \,\,\, 0 \le m \le n, 
\end{equation}
applicable in many situations. Here $n$ is either the $\tilde{\chi}^{(n)}$  index in the
irreducible Cases 2-4 or the corresponding index of the $\tilde{\chi}$ subgraph in
the reducible Case 5. The $m$ in (\ref{eqE1}) is either the $k$ or $n-k$ as defined
in (\ref{sixteen}) and (\ref{twentytwo}). These indices are given as subscripts in our
polynomial labeling scheme (cf. \ref{Singularities}) to allow a direct calculation
of $p$ using (\ref{eqE1}). For example, $p=(3^2+2^2-3)/2=5$ for polynomial $(^35_{3,2})$
or any of the reducible polynomials $(^5N/^35_{3,2})$ with $N =5+2i, \,\,\, i>0.$ Note
that Case 2 polynomials $(^2n)=(^2n_{n,0})$ have $m=\,0$
 so that (\ref{eqE1}) becomes $p=\,(n^2-3)/2,$
the formula for the circle singularities derived in~\cite{nickel-99,nickel-00} as
exponents satisfying (\ref{eqE1}), which we will refer to as ``normal'' exponents. We also
choose to call the principal disc physical exponents $p=-1$ at the
ferromagnetic point and $p=0$ at the anti-ferromagnetic point ``normal''.
There are other situations, possibly arising from the cancellation in
certain expressions after analytic continuation onto different Riemann
sheets. Because of our lack of understanding in most of these cases
we will simply call the other exponents ``anomalous''. The following is
then to be understood as a very tentative approach to the singularity
exponent problem given that we do not have simple general criteria necessary
for the ``normal'' situation to occur. We are in this respect ultimately
guided by the agreement or disagreement with the numerical work of Sections~\ref{sec:analysis}
and \ref{sec:DPn7}. In this regard note that (\ref{eqE1}) represents the leading exponent at
a singularity and the derivation below does not exclude exponents greater
than $p.$ On the other hand, if an exponent less than $p$ is observed a
different mechanism is required. An alternative explanation is that
the observed exponent is associated with a solution of the ODE that is
not the integral $\tilde{\chi}^{(n)}.$ 

The power counting analysis requires that we distinguish between
the two cases $\sin(\zeta_i) =\, \rmi\cdot f_i =\,  0$ or $\ne \, 0$.
 We begin with the latter and determine
first the singularities of $\tilde{\chi}^{(n)}$  without the Fermionic factor. The
singularities in this case come from the zeros of
 $1-\prod x_i =\,  1-\exp(\rmi\sum \zeta_i).$
For the behaviour of the integrand in the vicinity of a zero we define
$\epsilon$ as the deviation of $\omega-1$ from the Landau 
singularity value and $\delta \phi_i$ as
the deviation of the corresponding phase values. A Taylor expansion
of $\sum \zeta_i$ yields
 $\sum \zeta_i = \,2\pi m+\sum \delta \phi_i(\partial\zeta_i/\partial \phi_i)-\epsilon/2
\sum1/\sin(\zeta_i)-1/(4\omega)
\sum \delta \phi_i^2(1-\cos \zeta_i \cos \phi_i)/\sin^3 \zeta_i + \ldots$
 and the stationary condition
(\ref{eleven}) that $\partial\zeta_i/\partial \phi_i$ is the 
same for all $i=\, 1 \ldots \, n$ implies that the linear term
is proportional to $\sum\delta \phi_i$ and thus vanishes because of the phase constraint
in~(\ref{they}). We conclude that the leading terms
 in $1-\prod  x_i$ are of the form $A\epsilon+\sum a_i \delta \phi_i^2$
with $A$ and $a_i$ constants. The rescaling 
$\delta \phi_i = \, \delta \psi_i \sqrt \epsilon$ puts this denominator
factor in the $\tilde{\chi}^{(n)}$  integral into homogeneous
 form proportional to $\epsilon$. The Jacobian of
the transformation from $n-1$ variables $\delta \phi$ 
to $\delta \psi$ yields another overall
factor of $\epsilon^{(n-1)/2}$ so that the final scaling of the integral singularity
is $\epsilon^{(n-3)/2}$, that is to say, a singularity
 power $p=\, (n-3)/2$ which is exactly that observed
in all ODE analyses of these integrals.

The inclusion of the Fermionic factor leads to extra powers of $\epsilon$
that can be determined by counting as follows. In the Case 2 situation
treated in~\cite{nickel-99,nickel-00}, there are $\,n(n-1)/2$
 factors of $h_{ij}^2,$ each of which has
an O$(1)$ denominator and a numerator proportional to 
$(\delta \phi_i - \delta \phi_j)^2$ or 
$\epsilon(\delta \psi_i - \delta \psi_j)^2$and thus O$(\epsilon)$.
 The total singularity power then becomes
$p=\, (n-3)/2+n(n-1)/2=(n^2-3)/2.$ For Case 4, to which the present argument
applies, there are $k$ phases $\phi_a$ and $(n-k)$ phases
 $\phi_b$ at a Landau singularity.
This implies that there are now $k(n-k)$ factors of $h_{ij}^2$ with numerators close
to $\sin^2((\phi_a-\phi_b)/2)$ and these are O$(1)$
 and not O$(\epsilon).$ Only the $k(k-1)/2$
and $(n-k)(n-k-1)/2$ factors within each $\phi_a$ and $\phi_b$ set respectively are
O$(\epsilon).$ This gives the singularity power now as
$p=\, (n-3)/2+k(k-1)/2+(n-k)(n-k-1)/2$ which is (\ref{eqE1}) with $m=k.$

For the reducible Case 5 a new situation arises in the Fermionic
factor. Suppose, for example, in our singularity list of $(\phi,\zeta)$ values
there are $n-k$ pairs of type $(\phi_b,\zeta_b)$ but that of the remaining $k$ pairs,
one is $(-\phi_a,-\zeta_a)$ and only $k-1$ are
 $(\phi_a,\zeta_a)$. Then we know there will be
$k-1$ factors of $h_{ij}^2$ involving a 
$(-\phi_a,-\zeta_a),(\phi_a,\zeta_a)$ combination. The
numerator in each of these factors will be close to $\, \sin^2(\phi_a)$ and thus
O$(1)$ and not proportional to $(\delta \phi_i - \delta \phi_j)^2$
 and O$(\epsilon).$ The denominator on the
other hand is 
$(1-x_i x_j)^2 = [1-\exp(i(\zeta_a+\delta \zeta_i)+i(-\zeta_a+\delta \zeta_j))]^2 
\approx -(\delta \zeta_i+\delta \zeta_j)^2 = {\rm O}(\epsilon).$
In summary, there are $k-1$ Fermionic factor terms each O$(1/\epsilon).$ Furthermore,
of the original $k(k-1)/2$ factors involving
factors of $\phi\,$ of type $\phi_a$ only the remaining
$k(k-1)/2-(k-1)=(k-1)(k-2)/2$ are O$(\epsilon)$ as before. Factors involving only
type $\phi_b$ or type $\pm \phi_a$ and $\phi_b$ combinations are also unaffected. Counting
all powers of $\epsilon$ listed above shows that the singularity exponent will be
$(n-3)/2-(k-1)+(k-1)(k-2)/2+(n-k)(n-k-1)/2 = [(k-2)^2+(n-k)^2-3]/2$ which
is again (E.1) but with $m$ and $n-m$ identified with the irreducible subgraph
values $k-2$ and $n-k.$ If there is more than one sign reversed pair in
the reducible Case 5, then the above counting argument can be repeated
and leads to the conclusion that (\ref{eqE1}) is the general result.

The arguments when $\sin(\zeta_i)=\, \rmi\cdot f_i = 0$ at the Landau singularity are
very similar to what is described above and show that (\ref{eqE1}) still applies,
in particular for Case 3. One technical difference and partial result
is worth noting. Each $f_i$ is of the 
form $\sqrt{(\epsilon+a_i\delta \phi_i)}$ and
thus the appropriate rescaling to make this singular function homogeneous
in $\epsilon$ is $\delta \phi_i = \delta \psi_i \cdot \epsilon$.
 The singularity power counting in the absence of
the Fermionic factor then gives an $\epsilon^{n-1}$ 
from the Jacobian of the transformation,
an $\epsilon^{-n/2}$ from the $n$ denominator factors $f_i$ 
and another $\epsilon^{-1/2}$ from the denominator
$1-\prod x_i$ which reduces, in leading order, to $\sum f_i$.
 The product of these three
$\epsilon$ factors is $\epsilon^{(n-3)/2}$ and thus we have 
as the singularity exponent in the
absence of the Fermionic factor the value $p=\, (n-3)/2$ exactly as in the
$\sin(\zeta_i)=\, \rmi\cdot f_i \ne 0$ situation.

One set of comparisons of (\ref{eqE1}) with observed exponent values can
be found in Table~\ref{Ta:C5expo}. While there are many cases of agreement, it is
also clear that there are ``anomalous'' exponents that we cannot account
for. These include the $-2,$ $-7/4,$ $-3/2$ and $-5/4$ at the ferromagnetic
point, and more relevant for the present discussion, the powers $2,$ $5/2$
and $0$ for the polynomials $(^55/^23)$ and $(^55/^33_{1,2}).$ The understanding of
these cases is left as a challenge for the future.

\end{document}